\documentclass[letterpaper,12pt]{article}

\usepackage{geometry}
\usepackage{lscape}

\usepackage{mathptmx}

\usepackage{graphics}
\usepackage{graphicx}
\usepackage[numbers,sort&compress]{natbib}

\usepackage{amssymb}

\usepackage{setspace}

\begin{document}
\begin{center}
\LARGE
Decomposition of the Total Effect for Two Mediators: A Natural Counterfactual Interaction Effect Framework
\end{center}


\begin{center}
Xin Gao\footnote{\label{a1}Department of Mathematics and Statistics, University of New Mexico, Albuquerque, NM, 87131, USA}$^,$\footnote{\label{a2}Comprehensive Cancer Center, University of New Mexico, Albuquerque, NM, 87131, USA}, Li Li$^1$, Li Luo$^{2,}$\footnote{\label{a3}Department of Internal Medicine, University of New Mexico, Albuquerque, NM, 87131, USA}\renewcommand{\thefootnote}{\fnsymbol{footnote}}\footnote[1]{Corresponding to: {\ttfamily LLuo@salud.unm.edu}}
\end{center}
\smallskip
\begin{abstract}
Mediation analysis has been used in many disciplines to explain the mechanism or process that underlies an observed relationship between an exposure variable and an outcome variable via the inclusion of mediators. Decompositions of the total causal effect of an exposure variable into effects characterizing mediation pathways and interactions have gained an increasing amount of interest in the last decade. In this work, we develop decompositions for scenarios where the two mediators are causally sequential or non-sequential. Current developments in this area have primarily focused on either decompositions without interaction components or with interactions but assuming no causally sequential order between the mediators. We propose a new concept called natural counterfactual interaction effect that captures the two-way and three-way interactions for both scenarios that extend the two-way mediated interactions in literature. We develop a unified approach for decomposing the total effect into the effects that are due to mediation only, interaction only, both mediation and interaction, neither mediation nor interaction within the counterfactual framework. Finally, we illustrate the proposed decomposition method using a real data analysis where the two mediators are causally sequential.
\end{abstract}

\doublespacing
\section{Introduction}
\pagenumbering{arabic}

Mediation analysis has become the technique of choice to identify and explain the mechanism that underlies an observed relationship between an exposure or treatment variable and an outcome variable via the inclusion of intermediate variables, known as mediators. Decompositions of the total effect of the exposure into effects characterizing mediation pathways and interactions help researchers understand the effects through different mechanisms and have gained much attention in literature and application in the last decade \cite{vmul,vpre,d,s,m,v3,v4,vbook,b}. In our motivating example, we are interested in the effects of drinking alcohol on Systolic Blood Pressure (SBP) via the mediators, Body Mass Index (BMI) and Gamma Glutamyl Transferase (GGT), and their interaction effects. Besides, the mediator BMI is previously reported to affect GGT and not vice versa, and hence the two mediators are causally sequential. Current developments in this area for scenarios considering two mediators have primarily focused on either decomposition without interaction components, or decomposition allowing interactions but assuming no causally sequential order between the mediators. Daniel \cite{d} and Steen et al. \cite{s} discussed the decompositions in a general framework with causally sequential mediators; however, their decompositions do not include interaction components. Bellavia and Valeri \cite{b} proposed a decomposition with components describing interactions, but they assumed these mediators are causally non-sequential.

In this work, we develop decomposition methods for the scenarios when the two mediators are causally sequential and extensive interaction effects exist where existing decomposition methods are limited. Our approach also applies to the non-sequential mediators' scenario. We present a unified approach for decomposing the total effect into the components that are due to mediation only, interaction only, both mediation and interaction, neither mediation nor interaction within the counterfactual framework. Our decomposition methods are motivated by VanderWeele's four-way decomposition \cite{v4} of the total effect with one mediator, where the interaction effects include a reference interaction effect for interaction only and a mediated interaction effect for both mediation and interaction. VanderWeele \cite{v4} emphasized that these interaction terms are often considered of the greatest public health importance \cite{Rothman1,Rothman2,Hosmer}. We also propose a new concept called natural counterfactual interaction effect for describing the two-way and three-way interactions in the two-mediator scenarios that extend the mediated interaction from VanderWeele's work \cite{v4}. Since the causal structures are more complex with two mediators, the decompositions have multiple terms for mediation only, interaction only, and both mediation and interaction. More importantly, we find that the terms for interaction only are all identifiable at the individual level when the two mediators are causally non-sequential, but some of them are no longer identifiable when the two mediators are causally sequential. When the two mediators are casually non-sequential, our decomposition uses a different approach from what was proposed by Bellavia and Valeri \cite{b}. For example, their population-averaged mediated interaction effect between $A$ and $M_1$ is evaluated by controlling $M_2$ at a fixed level while our natural counterfactual interaction effect is essentially a weighted mediated interaction effect where the weights are determined by the distributions of both mediators in the population.

The rest of the paper is organized as follows: Section 2 reviews VanderWeele's four-way decomposition; Section 3 presents decompositions of total effect for two-mediator scenarios; Section 4 lays out identification assumptions and gives the empirical formulas for computing each component in the decomposition with two causally sequential mediators; Section 5 presents our real data analysis; Section 6 concludes the paper with discussions.

\section{Decomposition of the total effect in a single-mediator scenario}

\subsection{Counterfactual definitions}
\begin{figure}[!h]
\centering
\scalebox{0.5}[0.5]{\includegraphics{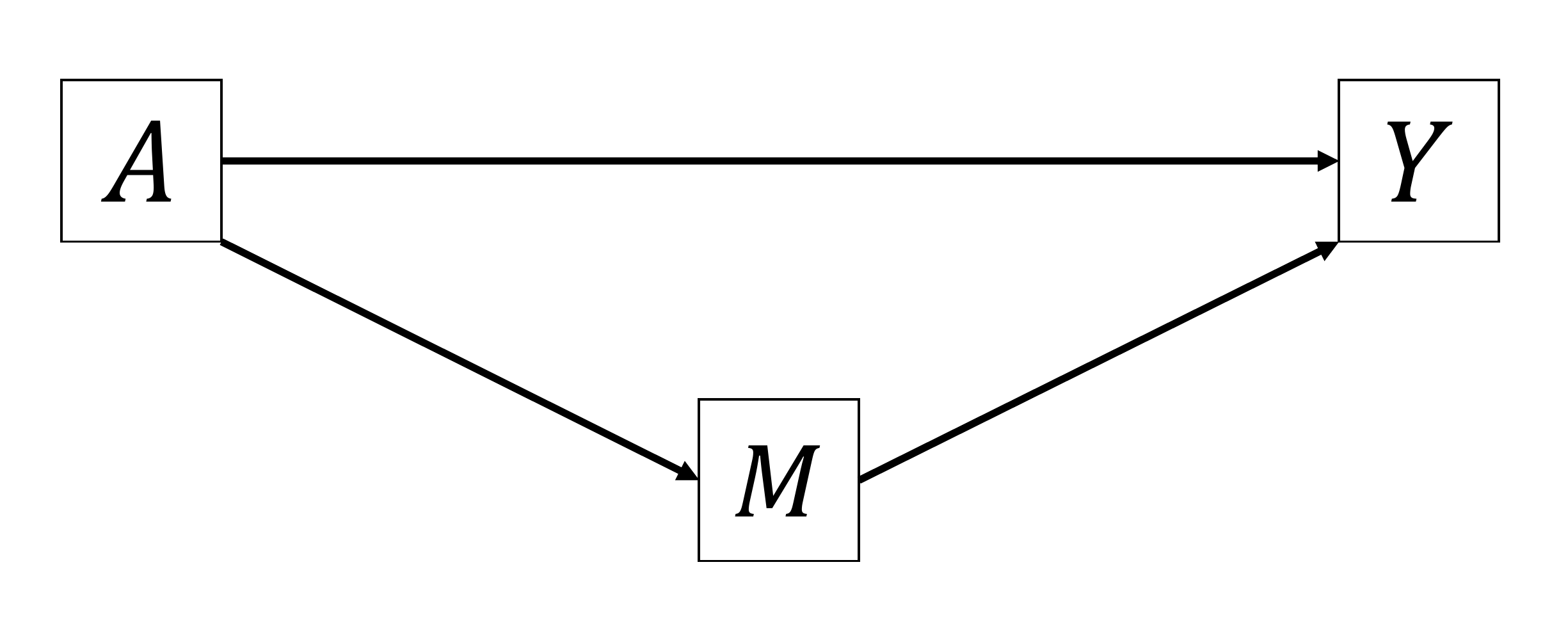}}
\caption{Directed acyclic graph of a single-mediator scenario.}
\label{fig1}
\end{figure}

\begin{figure}[!h]
\centering
\scalebox{0.5}[0.5]{\includegraphics{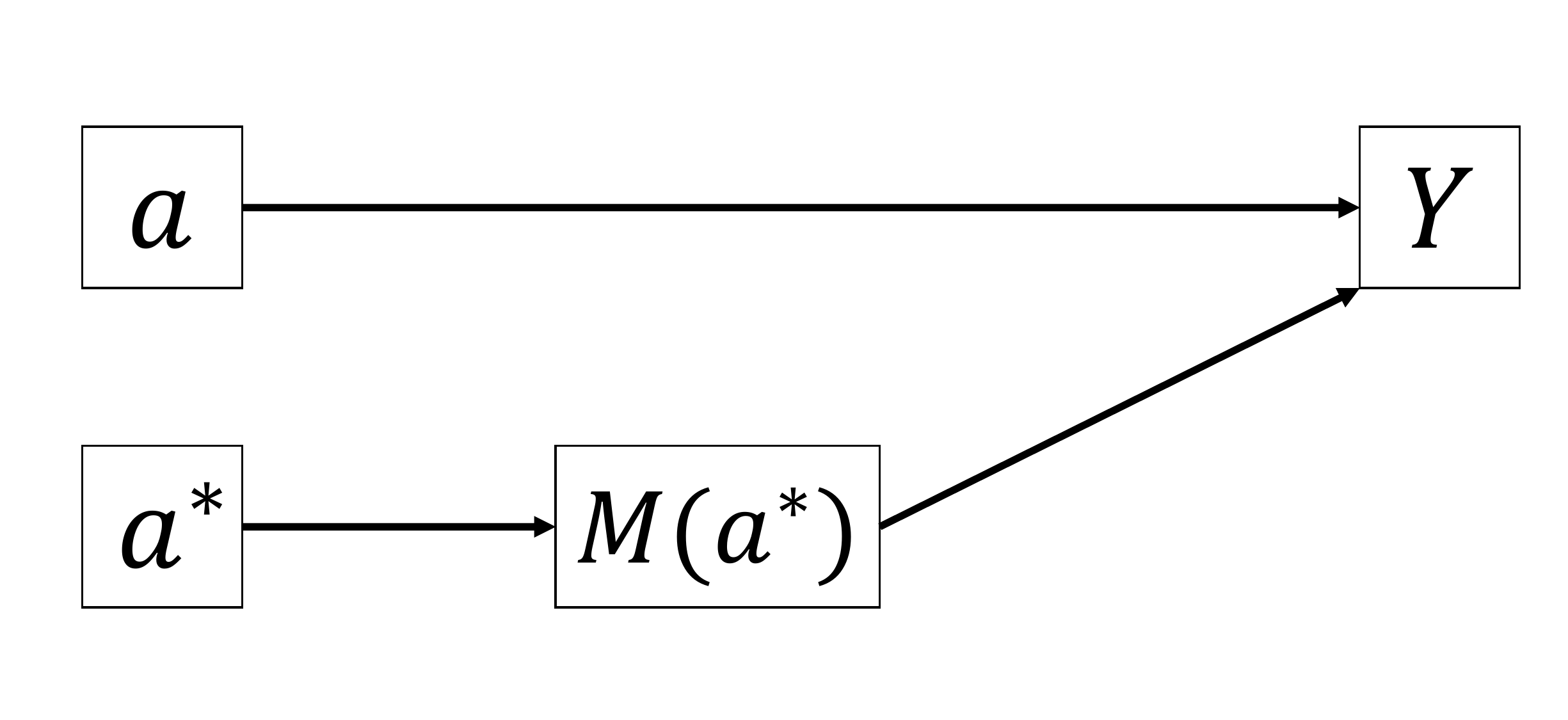}}
\caption{Nested counterfactual formula $Y(a, M_1(a^\ast))$.}
\label{fig2}
\end{figure}

Consider the single-mediator scenario in Figure \ref{fig1}. Counterfactual formulas give the potential value of outcome $Y$ or mediator $M$ that would have been observed if the exposure $A$ or mediator $M$ were fixed at a certain level \cite{vbook,riden,p01}. Let $Y(a)$ denote the potential value of $Y$ that would have been observed if the exposure $A$ were fixed at a constant level $a$ \cite{vbook}. Similarly, $M(a)$ denotes the potential value of $M$ that would have been observed if $A$ were fixed at $a$ and $Y(a,m)$ denotes the potential value of $Y$ that would have been observed if $A$ and $M$ were fixed at $a$ and $m$, respectively \cite{vbook}. A nested counterfactual formula $Y(a, M(a^{\ast}))$ denotes the potential value of $Y$ that would have been observed if the exposure were fixed at $a$ and the mediator $M$ were set to what would have been observed or potential value when the exposure were fixed at $a^{\ast}$ (Figure \ref{fig2}) \cite{vbook}.\\

\subsection{Two-way decomposition}
The total effect ($TE$) of the exposure $A$ for an individual is defined as the difference between $Y(a)$ and $Y(a^{\ast})$ \cite{vbook}, where $a$ and $a^{\ast}$ are the treatment and reference level of the exposure $A$, respectively. The classical decomposition of the total effect has two components: natural direct effect ($NDE$) and natural indirect effect ($NIE$) \cite{vbook,p01,rsem}. $NDE$ represents the causal effect along the direct path from $A$ to $Y$ and $NIE$ represents the causal effect along the indirect path from $A$ through $M$ to $Y$. The effects are defined using the following formulas: 
\begin{eqnarray*}
  TE & = & Y(a)-Y(a^\ast)\\
  & = & Y(a,M(a)) - Y(a^{\ast},M(a^\ast))\\
  & = & Y(a,M(a)) - Y(a,M(a^\ast)) + Y(a,M(a^\ast)) - Y(a^{\ast},M(a^\ast)),\\
  \\
  NDE & = & Y(a,M(a^\ast)) - Y(a^\ast,M(a^\ast)),\\
  \\
  NIE & = & Y(a,M(a)) - Y(a,M(a^\ast)).
\end{eqnarray*}

The second equality of $TE$ follows by the composition axiom \cite{vbook,a} and the third equality of $TE$ follows by subtracting and adding the same counterfactual formula $Y(a,M(a^\ast))$. $NDE$ is the difference in the potential value of outcome when $A$ goes from $a^\ast$ to $a$ and $M$ is at its potential value $M(a^\ast)$. $NIE$ is the difference in the potential value of outcome had $M$ goes from $M(a^\ast)$ to $M(a)$ while $A$ is at its treatment level $a$. In literature, $NDE$ and $NIE$ are also referred to as pure direct effect ($PDE$) \cite{riden} and total indirect effect ($TDE$) \cite{riden}, respectively. Furthermore, $NDE$ also corresponds to a path-specific effect proposed by Pearl \cite{p01}. 

\subsection{Four-way decomposition with interactions}
VanderWeele \cite{v4} proposed a four-way decomposition in a single-mediator scenario where the exposure interacts with the mediator. The total effect of the exposure on the outcome is decomposed into components due to mediation only, to interaction only, to both mediation and interaction, and to neither mediation nor interaction. These four components are termed as pure indirect effect ($PIE$), reference interaction effect ($INT_{ref}(m^\ast)$), mediated interaction effect ($INT_{med}$) and controlled direct effect ($CDE(m^\ast)$), respectively, where $m^\ast$ is an arbitrarily chosen fixed reference level of the mediator $M$. At the individual level, the four components are expressed in general forms \cite{v4}:
\begin{eqnarray*}
  CDE(m^\ast) & = & Y(a,m^\ast)-Y(a^\ast,m^\ast),\\
  \\
  INT_{ref}(m^\ast) & = & \sum_{m} [Y(a,m)-Y(a^\ast,m)-Y(a,m^\ast)+Y(a^\ast,m^\ast)]\times I(M(a^\ast)=m),\\
  \\
  INT_{med} & = & \sum_{m} [Y(a,m)-Y(a^\ast,m)-Y(a,m^\ast)+Y(a^\ast,m^\ast)]\\
  & & \times [I(M(a)=m)-I(M(a^\ast)=m)],\\
  \\
  PIE & = & \sum_{m}[Y(a^\ast,m)-Y(a^\ast,m^\ast)]\times [I(M(a)=m)-I(M(a^\ast)=m)].
\end{eqnarray*}

The reference and mediated interaction effects can also be expressed in the form of the counterfactual formulas in our view:
\begin{eqnarray*}
  INT_{ref}(m^\ast) & = & Y(a,M(a^\ast))- Y(a^\ast,M(a^\ast))-Y(a,m^\ast)+Y(a^\ast,m^\ast),\\
  \\
  INT_{med} & = & Y(a,M(a))-Y(a^\ast,M(a))-Y(a,M(a^\ast))+Y(a^\ast,M(a^\ast)).
\end{eqnarray*}

$CDE$ measures the effect of $A$ had $M$ be fixed at level $m^\ast$. $INT_{ref}(m^\ast)$ measures the change in the effect of $A$ had $M$ go from $m^\ast$ to $M({a^\ast})$. If $M({a^\ast}) = m^\ast$, $INT_{ref}(m^\ast)$ for the individual considered is reduced to zero. $INT_{med}$ describes the change in the effect of $A$ had $M$ go from $M({a^\ast})$ to $M({a})$. When $A$ has no effect on the mediator, $M({a^\ast})=M({a})$, and $INT_{med}$ becomes zero. $PIE$ describes the effect of $M$ when $A$ is set at $a^\ast$ and $M$ goes from $M({a^\ast})$ to $M({a})$.

When $A$ and $M$ are both binary with the conditions $a=1$, $a^\ast=0$ and $m^\ast=0$, the counterfactual definitions of the components become \cite{v4}: 
\begin{eqnarray*}
CDE(0) & = & Y(1,0)-Y(0,0),\\
\\
INT_{ref}(0) & = & [Y(1,1)-Y(1,0)-Y(0,1)+Y(0,0)]\times M(0),\\
\\
INT_{med} & = & [Y(1,1)-Y(1,0)-Y(0,1)+Y(0,0)]\times[M(1)-M(0)],\\ 
\\
PIE & = & [Y(0,1)-Y(0,0)]\times[M(1)-M(0)],
\end{eqnarray*}
where $1$ is the treatment level and $0$ is the reference level \cite{v4}. \\

Both $INT_{ref}$ and $INT_{med}$ have an additive interaction $[Y(1,1)-Y(1,0)-Y(0,1)+Y(0,0)]$ term which will be non-zero for an individual if the joint effect of having both the exposure and the mediator present differs from the sum of the effects of having only the exposure or mediator present. The additive interaction effect is generally considered of great public health importance \cite{Rothman1,Rothman2,Hosmer}. The difference between $INT_{ref}$ and $INT_{med}$ is that  $INT_{ref}$ is non-zero only if the mediator is present in the absence of exposure (i.e. $M(0) = 1$) whereas  $INT_{med}$ is non-zero only if the exposure has an effect on the mediator (i.e. $M(1)-M(0) \neq 0$).

Based on the counterfactual formula form of mediated interaction $INT_{med}$, we propose the natural counterfactual interaction effect and provide the following definition. The mediated interaction effect and natural counterfactual interaction effect are mathematically equivalent in the single mediator scenario, we define it from a different perspective only for building up the concepts for scenarios with two mediators in section 3.\\

\textbf{Definition 1}. 
Natural counterfactual interaction effect of $A$ and $M$ in a single-mediator scenario:
\begin{eqnarray*}
NatINT_{AM} := Y(a,M(a)) - Y(a^\ast,M(a)) - Y(a,M(a^\ast)) + Y(a^{\ast},M(a^\ast)),
\end{eqnarray*}
where $M(a^\ast)$ and $M(a)$ denote the potential values of $M$ that would have occurred if $A$ were fixed at $a^\ast$ and $a$, respectively. 

\section{Decomposition of the total effect in two-mediator scenarios}

When two mediators are considered, two-way interaction of the two mediators and three-way interaction of the exposure and the two mediators are likely to exist \cite{v4,b,vbook}. There may also be a causal sequence between the two mediators, i.e. there is a direct causal link between the two mediators (Figure \ref{fig4}). There is limited research on how to define interactions when the two mediators are causally sequential. We aim to develop interpretable interactions concepts and decomposition approaches for the two-mediator scenarios.

\subsection{Mediators causally non-sequential}
We first consider the scenario when the two mediators are causally non-sequential, i.e., there is no direct causal link between the two mediators, which is shown in Figure \ref{fig3}. Below, we define two-way natural counterfactual interaction effects of $A$ and $M_1$, $A$ and $M_2$, $M_1$ and $M_2$, and a three-way natural counterfactual interaction effect of $A, M_1$ and $M_2$.
\\
\\
\textbf{Definition 2}. Natural counterfactual interaction effects in a causally non-sequential two-mediator scenario:
\begin{eqnarray*}
  NatINT_{AM_1} & := & Y\left(a,M_1(a),M_2(a^\ast)\right)-Y\left(a^\ast,M_1(a),M_2(a^\ast)\right)\\
  & & - Y\left(a,M_1(a^\ast),M_2(a^\ast)\right)+Y\left(a^\ast,M_1(a^\ast),M_2(a^\ast)\right),\\
  \\
  NatINT_{AM_2} & := & Y\left(a,M_1(a^\ast),M_2(a)\right)- Y\left(a^\ast,M_1(a^\ast),M_2(a)\right)\\
  & & -Y\left(a,M_1(a^\ast),M_2(a^\ast)\right)+Y\left(a^\ast,M_1(a^\ast),M_2(a^\ast)\right),\\
  \\
 NatINT_{M_1M_2} & := & Y\left(a^\ast,M_1(a),M_2(a)\right)-Y\left(a^\ast,M_1(a^\ast),M_2(a)\right)\\
  & & -Y\left(a^\ast,M_1(a),M_2(a^\ast)\right)+Y\left(a^\ast,M_1(a^\ast),M_2(a^\ast)\right), \\
  \\
   NatINT_{AM_1M_2} & := & Y\left(a,M_1(a),M_2(a)\right)-Y\left(a^\ast,M_1(a),M_2(a)\right)\\
  & & - Y\left(a,M_1(a^\ast),M_2(a)\right)+Y\left(a^\ast,M_1(a^\ast),M_2(a)\right)\\
  & & -Y\left(a,M_1(a),M_2(a^\ast)\right)+Y\left(a^\ast,M_1(a),M_2(a^\ast)\right)\\
  & & +Y\left(a,M_1(a^\ast),M_2(a^\ast)\right)-Y\left(a^\ast,M_1(a^\ast),M_2(a^\ast)\right).
\end{eqnarray*}

$NatINT_{AM_1}$, $NatINT_{AM_2}$, and $NatINT_{AM_1M_2}$ are components that capture the effects due to both mediation and interaction with the exposure.  $ NatINT_{M_1M_2}$ describes the effect due to mediation and interaction between the two mediators. When measuring the interaction between $A$ and $M_1$, $M_2$ is not fixed but takes its potential value $M_2(a^\ast)$ for each individual had the exposure been the reference level. Similarly, when measuring the interaction between $A$ and $M_2$, $M_1$ is not fixed but takes its potential value $M_1(a^\ast)$ for the individual. The three-way interaction $NatINT_{AM_1M_2}$ is similar to a three-way additive interaction. To demonstrate the similarity, we consider that $A$ is binary with the conditions $a=1$ and $a^\ast =0$; $NatINT_{AM_1M_2} $ becomes 
\begin{eqnarray*}&& Y\left(1,M_1(1),M_2(1)\right)-Y\left(0,M_1(1),M_2(1)\right)
- Y\left(1,M_1(0),M_2(1)\right)+Y\left(0,M_1(0),M_2(1)\right)\\
\\
&&-Y\left(1,M_1(1),M_2(0)\right)+Y\left(0,M_1(1),M_2(0)\right)
+Y\left(1,M_1(0),M_2(0)\right)-Y\left(0,M_1(0),M_2(0)\right).
\end{eqnarray*}

The above three-way interaction measures the change in the two-way interaction between $A$ and $M_1$ when $M_2$ goes from $M_2(0)$ to $M_2(1)$. It also measures the change in the interaction between $A$ and $M_2$ when $M_1$ goes from $M_1(0)$ to $M_1(1)$ or the change in the interaction between $M_1$ and $M_2$ when $A$ goes from $0$ to $1$.

In Appendix A, we show that the total effect can be decomposed into ten components at the individual level: 
\begin{eqnarray*}
  TE & = & CDE(m_1^\ast,m_2^\ast)+INT_{ref\mbox{-}AM_1}(m_1^\ast,m_2^\ast)+INT_{ref\mbox{-}AM_2}(m_1^\ast,m_2^\ast)\\
  & & +INT_{ref\mbox{-}AM_1M_2}(m_1^\ast,m_2^\ast)+ NatINT_{AM_1} + NatINT_{AM_2}+ NatINT_{AM_1M_2}\\
  & & + NatINT_{M_1M_2} + PIE_{M_1} + PIE_{M_2},
\end{eqnarray*}
where $m_1^\ast$ and $m_2^\ast$ are fixed reference levels for $M_1$ and $M_2$, respectively, 
\begin{eqnarray*}
  CDE(m_1^\ast,m_2^\ast) & = & Y(a,m_1^\ast,m_2^\ast)-Y(a^\ast,m_1^\ast,m_2^\ast),\\
  \\
INT_{ref\mbox{-}AM_1}(m_1^\ast,m_2^\ast) & = & Y(a, M_1(a^\ast),m_2^\ast)-Y(a^\ast,M_1(a^\ast),m_2^\ast)-Y(a,m_1^\ast,m_2^\ast)+Y(a^\ast,m_1^\ast,m_2^\ast),\\
\\
INT_{ref\mbox{-}AM_2}(m_1^\ast,m_2^\ast) & = &  Y(a, m_1^\ast, M_2(a^\ast))-Y(a^\ast, m_1^\ast, M_2(a^\ast))-Y(a,m_1^\ast,m_2^\ast)+Y(a^\ast,m_1^\ast,m_2^\ast),\\
\\
INT_{ref\mbox{-}AM_1M_2}(m_1^\ast,m_2^\ast) & = & Y(a,M_1(a^\ast),M_2(a^\ast))-Y(a^\ast,M_1(a^\ast),M_2(a^\ast))-Y(a,m_1^\ast,M_2(a^\ast)) \\
&& + Y(a^\ast,m_1^\ast,M_2(a^\ast)) -Y(a,M_1(a^\ast),m_2^\ast)+Y(a^\ast,M_1(a^\ast),m_2^\ast) \\
&& + Y(a,m_1^\ast,m_2^\ast)-Y(a^\ast,m_1^\ast,m_2^\ast), \\
\\
PIE_{M_1} &= & Y(a^\ast,M_1(a),M_2(a^\ast))-Y(a^\ast,M_1(a^\ast),M_2(a^\ast)), \\
\\
PIE_{M_2} & = & Y(a^\ast,M_1(a^\ast),M_2(a))-Y(a^\ast,M_1(a^\ast),M_2(a^\ast)).
\end{eqnarray*}

Similar to the four-way decomposition, $CDE$ denotes controlled direct effect due to neither mediation nor interaction, $INT_{ref}$'s denote reference interaction effects due to interactions only, and $PIE$'s denote pure indirect effects due to mediation only \cite{v4}. $NatINT_{M_1M_2}$ can be interpreted as the effect due to the mediation through both $M_1$ and $M_2$, and the interaction between $M_1$ and $M_2$. Since the interaction is not involved with the change in exposure $A$, the interpretation can be simply put as the effect due to the mediation through both $M_1$ and $M_2$ only. These ten components are displayed in Table 1 assuming that $A$, $M_1$ and $M_2$ are binary with $a=1$, $a^\ast=0$, $m_1^\ast=0$ and $m_2^\ast=0$.

Bellavia and Valeri \cite{b} proposed a ten-component decomposition for the same directed acyclic graph in Figure \ref{fig3}. We show in Appendix B that their decomposition is a special case of our proposed decomposition under the extra condition $M_1(0)=M_2(0)=0$. Their $CDE$ and $INT_{ref}$'s have corresponding terms in our decomposition but their mediated interaction effects and pure natural indirect effects are different from our natural counterfactual interactions and pure indirect effects. The top panel in Figure \ref{fig4} illustrates their mediated interaction effect between $A$ and $M_1$ where $M_2$ is assigned a fixed value at $m_2^\ast=0$. The bottom panel in Figure \ref{fig4} illustrates the natural counterfactual interaction effect between $A$ and $M_1$, where both $M_1$ and $M_2$ take their potential values. 

Our natural counterfactual interaction effects account for the distributions of $M_1(0)$ and $M_2(0)$. If the population distribution of $M_2(0)$ has probability of $1$ taking the value $0$, the $NatINT_{AM_1}$ is consistent with the mediated interaction effect between $A$ and $M_1$ as proposed by Bellavia and Valeri. However, if the population distribution of $M_2(0)$ does not have probability of $1$ taking the value $0$, the natural counterfactual interaction effects are more suitable to describe the population average of the counterfactual interaction effects. Table 1 lists the specific decomposition components. Table 2 presents the results under the extra condition $M_1(0)=M_2(0)=0$, which are reduced to those proposed by Bellavia and Valeri \cite{b}.

\subsection{Mediators causally sequential}

In this section, we consider the scenario where the two mediators are causally sequential, i.e., there is a direct causal link from mediator $M_1$ to $M_{2}$ (Figure \ref{fig5}). Let $M_2(a^\ast, M_{1}(a))$ be the potential value of $M_2$ if $A$ were fixed at $a^\ast$ and $M_1$ were at its potential value had $A$ been set at $a$. Similarly, $M_2(a^\ast, M_{1}(a^\ast))$ denotes the potential value of $M_2$ if $A$ were fixed at $a^\ast$ and $M_1$ were at its potential value had $A$ been set at $a^\ast$. Counterfactual values for $Y$ are expressed using nested formulas but not all of them are identifiable. For example, $Y\left(a,M_1(a),M_2(a,M_1(a^\ast))\right)$ is not identifiable since it has two distinct counterfactual values of mediator $M_1$, i.e., $M_1(a)$ and $M_1(a^\ast)$, which means $M_1$ is activated by two different values of $A$ at the same time. Avin et al. \cite{a} showed that such counterfactual formulas are not identifiable. We present identifiable decomposition components only for those identifiable counterfactual formulas of $Y$. \\
\\
\textbf{Definition 3}. Natural counterfactual interaction effects in a causally sequential two-mediator scenario:
\begin{eqnarray*}
  NatINT_{AM_1} & := & Y\left(a,M_1(a),M_2(a^\ast,M_1(a))\right)-Y\left(a^\ast,M_1(a),M_2(a^\ast,M_1(a))\right)\\
  & & - Y\left(a,M_1(a^\ast),M_2(a^\ast,M_1(a^\ast))\right)+Y\left(a^\ast,M_1(a^\ast),M_2(a^\ast,M_1(a^\ast))\right),\\
  \\
  NatINT_{AM_2} & := & Y\left(a,M_1(a^\ast),M_2(a,M_1(a^\ast))\right)-Y\left(a^\ast,M_1(a^\ast),M_2(a,M_1(a^\ast))\right)\\
  & & -Y\left(a,M_1(a^\ast),M_2(a^\ast,M_1(a^\ast))\right)+Y\left(a^\ast,M_1(a^\ast),M_2(a^\ast,M_1(a^\ast))\right),\\
  \\
    NatINT_{M_1M_2} & := & Y\left(a^\ast,M_1(a),M_2(a,M_1(a))\right)-Y\left(a^\ast,M_1(a^\ast),M_2(a,M_1(a^\ast))\right)\\
  & & -Y\left(a^\ast,M_1(a),M_2(a^\ast,M_1(a))\right)+Y\left(a^\ast,M_1(a^\ast),M_2(a^\ast,M_1(a^\ast))\right),\\
  \\
NatINT_{AM_1M_2} & := & Y\left(a,M_1(a),M_2(a,M_1(a))\right)-Y\left(a^\ast,M_1(a),M_2(a,M_1(a))\right)\\
  & & - Y\left(a,M_1(a^\ast),M_2(a,M_1(a^\ast))\right)+Y\left(a^\ast,M_1(a^\ast),M_2(a,M_1(a^\ast))\right)\\
  & & -Y\left(a,M_1(a),M_2(a^\ast,M_1(a))\right)+Y\left(a^\ast,M_1(a),M_2(a^\ast,M_1(a))\right)\\
  & & +Y\left(a,M_1(a^\ast),M_2(a^\ast,M_1(a^\ast))\right)-Y\left(a^\ast,M_1(a^\ast),M_2(a^\ast,M_1(a^\ast))\right).
\end{eqnarray*}

These interaction terms are similar to those in Definition 2 except that $M_2$ has an extra input from $M_1$. In $NatINT_{AM_1} $, $M_2$ is neither fixed nor set at a level independent of $M_1$; rather, $M_2$  changes whenever $M_1$ changes. Therefore, $NatINT_{AM_1} $ captures the change in the total effect of $M_1$ (going from $M_1(a^\ast)$ to $M_1(a)$) on the response when $A$ goes from $a^\ast$ to $a$. In $NatINT_{M_1M_2}$,  $M_2$ would still partially depend on the level of $M_1$. Hence this component describes the interaction between $M_1$ and $M_2$ had $M_2$ only change its exposure input. Similarly, the three-way interaction $NatINT_{AM_1M_2}$ can be interpreted as the change in the interaction between $A$ and $M_1$ when $M_2$ has its exposure input going from $a^\ast$ to $a$.

We show in Appendix C that the total effect can be decomposed into 9 components at the individual level: 
\begin{eqnarray*}
  TE & = & CDE(m_1^\ast,m_2^\ast)+INT_{ref\mbox{-}AM_1}(m_1^\ast,m_2^\ast)+INT_{ref\mbox{-}AM_2+AM_1M_2}(m_2^\ast)\\
  & & + NatINT_{AM_1} + NatINT_{AM_2}+ NatINT_{AM_1M_2}+ NatINT_{M_1M_2}\\
  & & + PIE_{M_1} + PIE_{M_2},
\end{eqnarray*}
where 
\begin{eqnarray*}
CDE(m_1^\ast,m_2^\ast) & = & Y(a,m_1^\ast,m_2^\ast)-Y(a^\ast,m_1^\ast,m_2^\ast),\\
\\
INT_{ref\mbox{-}AM_1}(m_1^\ast,m_2^\ast) & = & Y(a,M_1(a^\ast),m_2^\ast)-Y(a^\ast,M_1(a^\ast),m_2^\ast)-Y(a,m_1^\ast,m_2^\ast)+Y(a^\ast,m_1^\ast,m_2^\ast),\\
  \\
  INT_{ref\mbox{-}AM_2+AM_1M_2}(m_2^\ast) & = & Y(a,M_1(a^\ast),M_2(a^\ast,M_1(a^\ast)))-Y(a,M_1(a^\ast),m_2^\ast)\\
  & & - Y(a^\ast,M_1(a^\ast),M_2(a^\ast,M_1(a^\ast)))+Y(a^\ast,M_1(a^\ast),m_2^\ast),\\
  \\
PIE_{M_1} &= & Y(a^\ast,M_1(a),M_2(a^\ast,M_1(a)))-Y(a^\ast,M_1(a^\ast),M_2(a^\ast,M_1(a^\ast))), \\
\\
PIE_{M_2} & = & Y(a^\ast,M_1(a^\ast),M_2(a,M_1(a^\ast)))-Y(a^\ast,M_1(a^\ast),M_2(a^\ast,M_1(a^\ast))).
\end{eqnarray*}

Compared to the decomposition in Section 3.1, reference interaction effects in the above case have fewer terms. $INT_{ref\mbox{-}AM_2}(m_1^\ast,m_2^\ast)$ and $INT_{ref\mbox{-}AM_1M_2}(m_1^\ast,m_2^\ast)$ are summed into $INT_{ref\mbox{-}AM_2+AM_1M_2}(m_2^\ast)$ to have identifiable effects. We show the detailed proof in Appendix D. $INT_{ref\mbox{-}AM_2+AM_1M_2}(m_2^\ast)$ can be interpreted as the effect due to the interaction between $A$ and $M_2$ only, conditioning on the potential value of $M_1$ at the reference level $a^\ast$. Because of the direct causal link between the two mediators, $M_2$ possesses two types of mediation, $M_2(1,1)-M_2(0,1)\neq 0$ and $M_2(1,0)-M_2(0,0)\neq 0$. They collectively contribute to $NatINT_{AM_1M_2}$ and $NatINT_{M_1M_2}$ (Appendix C). These nine components and their interpretations are shown in Table 3 for the special case when $A$, $M_1$ and $M_2$ are binary with the conditions $a=1$, $a^\ast=0$, $m_1^\ast=0$ and $m_2^\ast=0$.\\

\section{Identification assumptions and empirical formulas}
The decompositions for one- and two-mediator scenarios thus far have been primarily conceptual. The individual-level effects in the decompositions cannot be identified from data, but under certain assumptions on confounding the population-averages of those components can be identified from data \cite{v3}.

\subsection{Identification assumptions}
We first consider a single-mediator scenario. Four identification assumptions are required \cite{vcon}, which are listed below as ($A^\prime 1$) -- ($A^\prime 4$): 
\begin{eqnarray*}
  & & Y(a,m) \perp A|C     \hspace{2.4cm}    (A^\prime 1)\\  
  & & Y(a,m) \perp M|\{A,C\}     \hspace{1.5cm}    (A^\prime 2)\\ 
  & & M(a) \perp A|C     \hspace{2.8cm}    (A^\prime 3)\\
  & & Y(a,m) \perp M(a^\ast)|C,     \hspace{1.5cm}    (A^\prime 4)
\end{eqnarray*}
where $C$ is a set of covariates. The assumptions above state that given a covariate set $C$ or $\{A,C\}$, there exist no unmeasured variables confounding the association between exposure $A$ and outcome $Y$ ($A^\prime 1$); no unmeasured variables confounding the association between mediator $M$ and outcome $Y$ ($A^\prime 2$) and no unmeasured variables confounding the association between exposure $A$ and mediator $M$ ($A^\prime 3$) \cite{vbook}. ($A^\prime 4$) is a strong assumption and a few researchers published their works on this topic \cite{v4,s,ralt}. It could be interpreted as there exist no variables that are causal descendants of exposure $A$, and in the meantime, that confound the association between mediator $M$ and outcome $Y$ \cite{s,p01}. \\

The analogs of ($A^\prime 1$) -- ($A^\prime 4$) for a directed acyclic graph with two sequential mediators can be found by first considering $M_1$ and $M_2$ as a set \cite{s}. Namely, we have four corresponding identification assumptions ($A1$) -- ($A4$):
\begin{eqnarray*}
  & & Y(a,m_1,m_2) \perp A|C     \hspace{5.1cm}    (A1)\\  
  & & Y(a,m_1,m_2) \perp \{M_1,M_2\}|\{A,C\}     \hspace{2.9cm}    (A2)\\ 
  & & \{M_1(a),M_2(a,m_1)\} \perp A|C     \hspace{3.85cm}    (A3)\\
  & & Y(a,m_1,m_2) \perp \{M_1(a^\ast),M_2(a^\ast,m_1)\}|C.     \hspace{1.55cm}    (A4)\\
\end{eqnarray*}

Similarly, the assumptions above state that given a covariate set $C$ or $\{A,C\}$, there exists no unmeasured variables confounding the association between exposure $A$ and outcome $Y$ ($A1$), no unmeasured variables confounding the association between the mediator set $\{M_1,M_2\}$ and outcome $Y$ ($A2$), no unmeasured variables confounding the association between exposure $A$ and the mediator set $\{M_1,M_2\}$ ($A3$) and no unmeasured variables that are causal descendants of exposure $A$, and in the meantime, that confound the association between the mediator set $\{M_1,M_2\}$ and outcome $Y$ ($A4$) \cite{s, vcon}. 

In order to account for the confounding between $M_1$ and $M_2$, two more assumptions are required: 
\begin{eqnarray*}
  & & M_2(a,m_1) \perp M_1|\{A,C\}     \hspace{2.2cm}    (A5)\\  
  & & M_2(a,m_1) \perp M_1(a^\ast)|C,     \hspace{2.2cm}    (A6)
\end{eqnarray*}
where ($A5$) and ($A6$) state, respectively, that there exists no unmeasured variables confounding the association between $M_1$ and $M_2$ given $\{A,C\}$, and no unmeasured variables that are causal descendants of exposure $A$, and in the meantime, are confounding the association between $M_1$ and $M_2$ \cite{s}. 

Steen et al \cite{s} presented comprehensive identification assumptions for the causal structures with multiple mediators and pointed out that weaker identification assumptions than ($A1$) -- ($A6$) can be considered under certain decompositions. 

\subsection{Empirical formulas}

Suppose a set of covariates $C$ satisfies the assumptions on confounding for a decomposition. We can obtain the expected value of each component in the decomposition using the iterated conditional expectation rule. We focus on the scenario with two causally sequential mediators. Suppose $M_1$ and $M_2$ are categorical and let $p_{am_1m_2}=E[Y|A=a,M_1=m_1,M_2=m_2, C = c]$. The following formulas can be obtained: 
\begin{eqnarray*}
E\left[CDE(m_1^\ast,m_2^\ast)\right]& = & p_{am_1^\ast m_2^\ast}-p_{a^\ast m_1^\ast m_2^\ast}\\
\\
E[INT_{ref\mbox{-}AM_1}(m_1^\ast,m_2^\ast)] & = & \sum_{m_1}(p_{am_1 m_2^\ast}-p_{a m_1^\ast m_2^\ast}-p_{a^\ast m_1 m_2^\ast}+p_{a^\ast m_1^\ast m_2^\ast})\\
& & \times Pr(M_1=m_1|a^\ast, c)\\
\\
E[INT_{ref\mbox{-}AM_2+AM_1M_2}(m_2^\ast)] & = & \sum_{m_2}\sum_{m_1}(p_{am_1 m_2}-p_{a m_1 m_2^\ast}-p_{a^\ast m_1 m_2}+p_{a^\ast m_1 m_2^\ast})\\
& & \times Pr(M_1=m_1|a^\ast, c)\\
& & \times Pr(M_2=m_2|a^\ast,m_1, c)\\
\\
E[NatINT_{AM_1}] & = & \sum_{m_2}\sum_{m_1}(p_{am_1 m_2}-p_{a^\ast m_1 m_2})\\
& & \times Pr(M_2=m_2|a^\ast,m_1, c)\\
& & \times [Pr(M_1=m_1|a, c)-Pr(M_1=m_1|a^\ast, c)]\\
\\
E[NatINT_{AM_2}] & = & \sum_{m_2}\sum_{m_1}(p_{am_1 m_2}-p_{a^\ast m_1 m_2})\\
& & \times Pr(M_1=m_1|a^\ast, c)\\
& & \times [Pr(M_2=m_2|a,m_1, c)-Pr(M_2=m_2|a^\ast,m_1, c)]\\
\\
E[NatINT_{AM_1M_2}] & = & \sum_{m_2}\sum_{m_1}(p_{am_1 m_2}-p_{a^\ast m_1 m_2})\\
& & \times [Pr(M_1=m_1|a, c)-Pr(M_1=m_1|a^\ast, c)]\\
& & \times [Pr(M_2=m_2|a,M_1=m_1, c)-Pr(M_2=m_2|a^\ast,M_1=m_1, c)]\\
\\
E[NatINT_{M_1M_2}] & = & \sum_{m_2}\sum_{m_1}p_{a^\ast m_1 m_2}\\
& & \times [Pr(M_1=m_1|a, c)-Pr(M_1=m_1|a^\ast, c)]\\
& & \times [Pr(M_2=m_2|a,m_1, c)-Pr(M_2=m_2|a^\ast,m_1, c)]\\
\\
E[PIE_{M_1}] & = & \sum_{m_2}\sum_{m_1}p_{a^\ast m_1 m_2}\\
& & \times Pr(M_2=m_2|a^\ast,m_1, c)\\
& & \times [Pr(M_1=m_1|a, c)-Pr(M_1=m_1|a^\ast, c)]\\
\\
E[PIE_{M_2}] & = & \sum_{m_2}\sum_{m_1}p_{a^\ast m_1 m_2}\\
& & \times Pr(M_1=m_1|a^\ast, c)\\
& & \times [Pr(M_2=m_2|a,m_1, c)-Pr(M_2=m_2|a^\ast,m_1, c)].
\end{eqnarray*}

When $M_1$ and $M_2$ are continuous, empirical formulas can be obtained by replacing the sums by integrations and the conditional probabilities by conditional densities.

\subsection{Relations to linear models}
Suppose $Y$, $M_1$, and $M_2$ are continuous. For the scenario with two causally sequential mediators, we assume that the following regression models for $Y$, $M_1$, and $M_2$ are specified:
\begin{eqnarray*}
E[Y|A,M_1,M_2,C] & = & \theta_0 + \theta_1A + \theta_2M_1 + \theta_3M_2 + \theta_4AM_1 + \theta_5AM_2 + \theta_6M_1M_2\\ 
& & + \theta_7AM_1M_2 + \theta_8^\prime C\\
E[M_2|A,M_1,C] & = & \beta_0 + \beta_1A + \beta_2M_1 + \beta_3AM_1 + \beta_4^\prime C\\
E[M_1|A,C] & = & \gamma_0 + \gamma_1A + \gamma_2^\prime C, 
\end{eqnarray*}
the results on the effect components are given in Appendix E.

For the scenario with two causally non-sequential mediators, assume that a set of covariates $C$ satisfies the identification assumptions for the decomposition and assume that the following regression models for $Y$, $M_1$, and $M_2$ are specified:
\begin{eqnarray*}
E[Y|A,M_1,M_2,C] & = & \theta_0 + \theta_1A + \theta_2M_1 + \theta_3M_2 + \theta_4AM_1 + \theta_5AM_2 + \theta_6M_1M_2\\ 
& & + \theta_7AM_1M_2 + \theta_8^\prime C\\
E[M_2|A,C] & = & \beta_0 + \beta_1A  + \beta_4^\prime C\\
E[M_1|A,C] & = & \gamma_0 + \gamma_1A + \gamma_2^\prime C, 
\end{eqnarray*}
the results can be obtained as a special case of those derived from the scenario with two causally sequential mediators by setting parameters $\beta_2$ and $\beta_3$ to zero. \\

\section{Illustration with real data}
To illustrate the concept of natural counterfactual interaction effect and the decomposition methods, we used the 2015-2016 data from the National Health and Nutrition Examination Survey on the hazard of drinking alcohol as a contribution to the abnormal pattern in mortality \cite{d,l}. The dataset was downloaded from {\ttfamily http://www.cdc.gov/nhanes}. Exposure $A$ is alcohol drinking, mediator $M_1$ is Body Mass Index (BMI), mediator $M_2$ is the log-transformed Gamma Glutamyl Transferase (GGT), and outcome $Y$ is Systolic Blood Pressure (SBP). Sex and Age are considered a sufficient set satisfying the assumption on confounding. In addition, BMI is known to affect GGT. The hypothetical causal diagram is shown in Figure \ref{fig6}. 

Log transformation was performed for $M_2$ due to the skewness of the data. The fixed reference levels of $M_1$ and $\log(M_2)$ were chosen at their corresponding mean levels where $m_1^\ast=29.5$ and ${\log{(m_2)}}^\ast=3.05$. Three linear models were fit for $Y$, $\log(M_2)$ and $M_1$. The 95\% confidence intervals were obtained by using a bootstrap method \cite{v}. 

Table \ref{tab::data1} presents the decomposition of total effect conditional on males and the mean level of age at $48.3$. The controlled direct effect is $0.238$ (95\% C.I. = $-0.969$ to $1.429$); the reference interaction effect between $A$ and $M_1$ is $-0.059$ ($-0.203$ to $0.039$); the sum of two reference interaction effect is $-0.115$ ($-0.516$ to $0.219$); the natural counterfactual interaction effect between $A$ and $M_1$ is $-0.018$ ($-0.125$ to $0.056$); the natural counterfactual interaction effect between $A$ and $\log(M_2)$ is $-0.026$ ($-0.194$ to $0.095$); the natural counterfactual interaction effect among $A$, $M_1$ and $\log(M_2)$ is $0.000386$ ($-0.0059$ to $0.0082$); the natural counterfactual interaction effect between $M_1$ and $\log(M_2)$ is $0.000873$ ($-0.0094$ to $0.0123$); the pure direct effect is $0.0636$ ($-1.226$ to $1.317$); the pure indirect effect through $M_1$ is $-0.0409$ ($-0.206$ to $0.109$); the pure indirect effect through $\log(M_2)$ is $0.143$ ($0.00803$ to $0.363$); the total effect is $0.123$ ($-1.178$ to $1.396$). The results of the decomposition of the total effect conditional on females and the mean level of age are shown in Table \ref{tab::data2}. It can be seen that the pure indirect effect through $\log(M_2)$ is the only significant effect contributing to the outcome for both females and males.

\section{Conclusion}

In this work, we develop decompositions for scenarios where the two mediators are causally sequential or non-sequential. We propose a unified approach for decomposing the total effect into components that are due to mediation only, interaction only, both mediation and interaction, and neither mediation nor interaction within the counterfactual framework. The decomposition was implemented via a new concept called natural counterfactual interaction effect that we proposed to describe the two-way and three-way interactions for both scenarios that extend the two-way mediated interactions in existing literature. To estimate the components of our proposed decompositions, we lay out the identification assumptions. We also derive the formulas when the response is assumed to be continuous with a linear model. 

We believe that our proposed new concept of natural counterfactual interaction effects and the decomposition methods for the causal framework with two sequential or non-sequential mediators provide a powerful tool to decipher the refined path effects while appropriately account for the interaction effects among the exposure and mediators. The counterfactual interaction effects evaluate the interaction terms that involve mediators by treating them at the natural levels. There is a gap in existing research of decomposing total effect into mediation and interaction effects for the scenario of two sequential mediators, and our proposed methods have the potential to fill in the gap. The proposed work provides the foundation to generalize into decomposition of total effect for more complicated causal structures involving more than two sequential mediators, which we will explore in the future work. 

\clearpage
\begin{landscape}
\begin{table}[!h]
\centering
\caption{Decomposition of the Total Effect in a Two Non-sequential Mediators Scenario When $A$, $M_1$ and $M_2$ are Binary with $a=1$, $a^\ast=0$, $m_1^\ast=0$ and $m_2^\ast=0$}
\begin{tabular}{lll}
\hline
\textbf{Effect}        & \textbf{Definition}    & \textbf{Interpretation}\\
\hline
$ CDE(0,0)$      & $Y(1,0,0)-Y(0,0,0)$      & Due to neither mediation nor interaction\\
& &\\
$INT_{ref\mbox{-}AM_1}(0,0)$       & $[Y(1,1,0)-Y(0,1,0)-Y(1,0,0)+Y(0,0,0)]\times M_1(0)$    & Due to the interaction between $A$ and $M_1$ only \\
& & \\
$INT_{ref\mbox{-}AM_2}(0,0) $  & $[Y(1,0,1)-Y(0,0,1)-Y(1,0,0)+Y(0,0,0)]\times M_2(0)$ & Due to the interaction between $A$ and $M_2$ only \\
& & \\
$INT_{ref\mbox{-}AM_1M_2}(0,0)$ & $[Y(1,1,1)-Y(0,1,1)-Y(1,0,1)+Y(0,0,1)$ & Due to the interaction between $A$, $M_1$ and $M_2$ only
\\& $-Y(1,1,0)+Y(0,1,0)+Y(1,0,0)-Y(0,0,0)]$ &
\\& $\times M_1(0)\times M_2(0)$   & \\
& & \\
$NatINT_{AM_1}$      & $\sum_{m_2}[Y(1,1,m_2)I(M_2(0)=m_2)-Y(0,1,m_2)I(M_2(0)=m_2)$      & Due to the mediation through $M_1$ and the interaction\\
 & $- Y(1,0,m_2)I(M_2(0)=m_2)+Y(0,0,m_2)I(M_2(0)=m_2)]$ & between $A$ and $M_1$ conditioning on the potential value \\
& $\times[M_1(1)-M_1(0)]$ & of $M_2$ with the fixed reference level $a^\ast=0$\\
& & \\
$NatINT_{AM_2}$      & $\sum_{m_1}[Y(1,m_1,1)I(M_1(0)=m_1)-Y(0,m_1,1)I(M_1(0)=m_1)$      & Due to the mediation through $M_2$ and the interaction\\
& $- Y(1,m_1,0)I(M_1(0)=m_1)+Y(0,m_1,0)I(M_1(0)=m_1)]$ & between $A$ and $M_2$ conditioning on the potential value\\
& $\times[M_2(1)-M_2(0)]$ & of $M_1$ with the fixed reference level $a^\ast=0$\\
& & \\
$ NatINT_{AM_1M_2}$      & $[Y(1,1,1)-Y(0,1,1)-Y(1,0,1)+Y(0,0,1)$      & Due to the mediation through both $M_1$ and $M_2$ and the\\
& $- Y(1,1,0)+Y(0,1,0)+Y(1,0,0)-Y(0,0,0)]$ &  interaction between $A$, $M_1$ and $M_2$\\
& $ \times[M_1(1)-M_1(0)]\times[M_2(1)-M_2(0)]$ & \\
& & \\
$NatINT_{M_1M_2}$      & $[Y(0,1,1)-Y(0,0,1)-Y(0,1,0)+Y(0,0,0)]$      & Due to the mediation through both $M_1$ and $M_2$ only\\
&$\times[M_1(1)-M_1(0)]\times[M_2(1)-M_2(0)]$ &\\
& & \\
$PIE_{M_1}$      & $\sum_{m_2}[Y(0,1,m_2)I(M_2(0)=m_2)-Y(0,0,m_2)I(M_2(0)=m_2)]$      &Due to the mediation through $M_1$ only conditioning on \\
&$\times[M_1(1)-M_1(0)]$ &the potential value of $M_2$ with the fixed reference level $a^\ast=0$\\
& &\\
$PIE_{M_2}$      & $\sum_{m_1}[Y(0,m_1,1)I(M_1(0)=m_1)-Y(0,m_1,0)I(M_1(0)=m_1)]$      & Due to the mediation through $M_2$ only conditioning on\\
&$\times[M_2(1)-M_2(0)]$ & the potential value of $M_1$ with the fixed reference level $a^\ast=0$\\
\hline
\end{tabular}
\end{table}
\end{landscape}

\clearpage
\begin{landscape}
\begin{table}[!h]
\centering
\caption{Proposed Interaction and Pure Indirect Effects for Non-Sequential Two Mediators Scenario with Binary $A$, $M_1$ and $M_2$ under the Extra Conditions $M_1(0)=M_2(0)=0$}
\begin{tabular}{lll}
\hline
\textbf{Effect}        & \textbf{Definition}    & \textbf{Interpretation}\\
\hline
$NatINT_{AM_1}$      & $[Y(1,1,0)-Y(0,1,0)-Y(1,0,0)+Y(0,0,0)]$      & Due to the mediation through $M_1$ and the interaction\\
 & $\times[M_1(1)-M_1(0)]$ & between $A$ and $M_1$ conditioning on $M_2(0)=0$ \\
& & \\
$NatINT_{AM_2}$ & $[Y(1,0,1)-Y(0,0,1)-Y(1,0,0)+Y(0,0,0)]$      & Due to the mediation through $M_2$ and the interaction\\
& $\times[M_2(1)-M_2(0)]$ & between $A$ and $M_2$ conditioning on $M_1(0)=0$\\
& & \\
$ NatINT_{AM_1M_2}$      & $[Y(1,1,1)-Y(0,1,1)-Y(1,0,1)+Y(0,0,1)$      & Due to the mediation through both $M_1$ and $M_2$ and the\\
& $- Y(1,1,0)+Y(0,1,0)+Y(1,0,0)-Y(0,0,0)]$ &  interaction between $A$, $M_1$ and $M_2$ conditioning on \\
& $ \times[M_1(1)M_2(1)-M_1(0)M_2(0)]$ & $M_1(0)=M_2(0)=0$\\
& & \\
$NatINT_{M_1M_2}$      & $[Y(0,1,1)-Y(0,0,1)-Y(0,1,0)+Y(0,0,0)]$      & Due to the mediation through both $M_1$ and $M_2$ only\\
&$\times [M_1(1)M_2(1)-M_1(0)M_2(0)]$ & conditioning on $M_1(0)=M_2(0)=0$\\
& & \\
$PIE_{M_1}$      & $[Y(0,1,0)-Y(0,0,0)]\times[M_1(1)-M_1(0)]$      &Due to the mediation through $M_1$ only conditioning on \\
& & $M_2(0)=0$\\
& &\\
$PIE_{M_2}$      & $[Y(0,0,1)-Y(0,0,0)]\times[M_2(1)-M_2(0)]$      & Due to the mediation through $M_2$ only conditioning on\\
& & $M_1(0)=0$\\
\hline
\end{tabular}
\end{table}
\end{landscape}

\clearpage
\begin{landscape}
\begin{table}[!h]
\centering
\caption{Decomposition of the Total Effect in a Two Sequential Mediators Scenario When $A$, $M_1$ and $M_2$ are Binary with $a=1$, $a^\ast=0$, $m_1^\ast=0$ and $m_2^\ast=0$}
\scalebox{0.95}{
\begin{tabular}{lll}
\hline
\textbf{Effect}        & \textbf{Definition}    & \textbf{Interpretation}\\
\hline
$ CDE(0,0)$      & $Y(1,0,0)-Y(0,0,0)$      & Due to neither mediation nor interaction\\
& &\\
$INT_{ref\mbox{-}AM_1}(0,0)$       & $[Y(1,1,0)-Y(0,1,0)-Y(1,0,0)+Y(0,0,0)]\times M_1(0)$    & Due to the interaction between $A$ and $M_1$ only \\
& & \\
$INT_{ref\mbox{-}AM_2+AM_1M_2}(0) $  & $[Y(1,0,1)-Y(0,0,1)-Y(1,0,0)+Y(0,0,0)]$ & Due to the interaction between $A$ and $M_2$ only \\
& $\times [1-M_1(0)]\times M_2(0,0)$ & conditioning on the potential value of $M_1$ with\\

&$+[Y(1,1,1)-Y(0,1,1)-Y(1,1,0)+Y(0,1,0)]$ & the fixed reference level $a^\ast=0$\\
& $\times M_1(0)\times M_2(0,1)$ & \\
& & \\
$NatINT_{AM_1}$      & $\sum_{m_2}[Y(1,1,m_2)I(M_2(0,1)=m_2)-Y(0,1,m_2)I(M_2(0,1)=m_2)$      & Due to the mediation through $M_1$ and the interaction\\
 & $- Y(1,0,m_2)I(M_2(0,0)=m_2)+Y(0,0,m_2)I(M_2(0,0)=m_2)]$ & between $A$ and $M_1$ conditioning on the potential values \\
& $\times[M_1(1)-M_1(0)]$ & of $M_2$ with the fixed reference level $a^\ast=0$\\
& & \\
$NatINT_{AM_2}$      & $\sum_{m_1}[Y(1,m_1,1)I(M_1(0)=m_1)-Y(0,m_1,1)I(M_1(0)=m_1)$      & Due to the mediation through $M_2$ and the interaction\\
& $- Y(1,m_1,0)I(M_1(0)=m_1)+Y(0,m_1,0)I(M_1(0)=m_1)]$ & between $A$ and $M_2$ conditioning on the potential value\\
& $\times[M_2(1,m_1)-M_2(0,m_1)]$&of $M_1$ with the fixed reference level $a^\ast=0$\\
& & \\

$ NatINT_{AM_1M_2}$      & $[Y(1,1,1)-Y(0,1,1)- Y(1,1,0)+Y(0,1,0)]$      & Due to the mediation through both $M_1$ and $M_2$ and the\\
& $\times[M_1(1)-M_1(0)]\times[M_2(1,1)-M_2(0,1)]$ &  interaction between $A$, $M_1$ and $M_2$\\
& $+ [-Y(1,0,1)+Y(0,0,1)+Y(1,0,0)-Y(0,0,0)]$ & \\
&$\times[M_1(1)-M_1(0)]\times[M_2(1,0)-M_2(0,0)]$ &\\
& & \\

$NatINT_{M_1M_2}$      & $[Y(0,1,1)-Y(0,1,0)]\times [M_1(1)-M_1(0)]$      & Due to the mediation through both $M_1$ and $M_2$ only\\
&$\times [M_2(1,1)-M_2(0,1)]$ &\\
&$+ [-Y(0,0,1)+Y(0,0,0)] \times [M_1(1)-M_1(0)]$ &\\
&$\times [M_2(1,0)-M_2(0,0)]$ & \\
& & \\
$PIE_{M_1}$      & $\sum_{m_2} [Y(0,1,m_2)I(M_2(0,1)=m_2)-Y(0,0,m_2)I(M_2(0,0)=m_2)]$      &Due to the mediation through $M_1$ only conditioning on \\
&$\times [M_1(1)-M_1(0)]$ &the potential values of $M_2$ with the fixed reference level $a^\ast=0$\\
& &\\
$PIE_{M_2}$      & $\sum_{m_1}[Y(0,m_1,1)\times I(M_1(0)=m_1)-Y(0,m_1,0)\times I(M_1(0)=m_1)]$      & Due to the mediation through $M_2$ only conditioning on\\
&$\times [M_2(1,m_1)-M_2(0,m_1)]$ & the potential value of $M_1$ with the fixed reference level $a^\ast=0$\\
\hline
\end{tabular}
}
\end{table}
\end{landscape}

\clearpage
\begin{table}[!h]
\centering
\caption{Illustration with Real Data: Decomposition of Total Effect Conditional on Males and the Mean Age. \label{tab::data1}}
\begin{tabular}{lll}
\hline
Component        & Estimate    & 95\% C.I. \\
\hline
$ CDE(m_1^\ast,\log(m_2)^\ast)$      & $0.238$      & $-0.969, 1.429$\\
$INT_{ref\mbox{-}AM_1}(m_1^\ast,\log(m_2)^\ast)$       & $-0.059$    &$ -0.203, 0.039$\\
$INT_{ref\mbox{-}A\log(M_2)+AM_1\log(M_2)}(\log(m_2)^\ast)$  & $-0.115$ & $-0.516,0.219$ \\
$NatINT_{AM_1}$      & $-0.018$      & $-0.125,0.056$\\
$NatINT_{A\log(M_2)}$      & $-0.026$      & $-0.194,0.095$\\
$NatINT_{AM_1\log(M_2)}$      & $0.000386$      & $-0.0059,0.0082$\\
$NatINT_{M_1\log(M_2)}$      & $0.000873$      & $-0.0094,0.0123$\\
$PDE$      & $0.0636$      & $-1.226,1.317$\\
$PIE_{M_1}$      & $-0.0409$      & $-0.206,0.109$\\
$PIE_{\log(M_2)}$      & $0.143$      & $0.00803,0.363$\\
$TE$      & $0.123$      & $-1.178,1.396$\\
\hline
\end{tabular}
\end{table}

\clearpage
\begin{table}[!h]
\centering
\caption{Illustration with Real Data: Decomposition of Total Effect Conditional on Females and the Mean Age. \label{tab::data2}}
\begin{tabular}{lll}
\hline
Component        & Estimate    & 95\% C.I. \\
\hline
$ CDE(m_1^\ast,\log(m_2)^\ast)$      & $0.238$      & $-0.969, 1.429$\\
$INT_{ref\mbox{-}AM_1}(m_1^\ast,\log(m_2)^\ast)$       & $0.087$    &$ -0.0359, 0.263$\\
$INT_{ref\mbox{-}A\log(M_2)+AM_1\log(M_2)}(\log(m_2)^\ast)$  & $0.0658$ & $-0.395,0.533$ \\
$NatINT_{AM_1}$      & $-0.0207$      & $-0.135,0.060$\\
$NatINT_{A\log(M_2)}$      & $-0.0286$      & $-0.206,0.0896$\\
$NatINT_{AM_1\log(M_2)}$      & $0.000377$      & $-0.00586,0.00863$\\
$NatINT_{M_1\log(M_2)}$      & $0.000860$      & $-0.00936,0.0117$\\
$PDE$      & $0.391$      & $-0.828,1.581$\\
$PIE_{M_1}$      & $-0.0448$      & $-0.219,0.114$\\
$PIE_{\log(M_2)}$      & $0.137$      & $0.00752,0.353$\\
$TE$      & $0.435$      & $-0.788,1.629$\\
\hline
\end{tabular}
\end{table}

\clearpage

\clearpage

\clearpage
\begin{figure}[!h]
\centering
\scalebox{0.5}[0.5]{\includegraphics{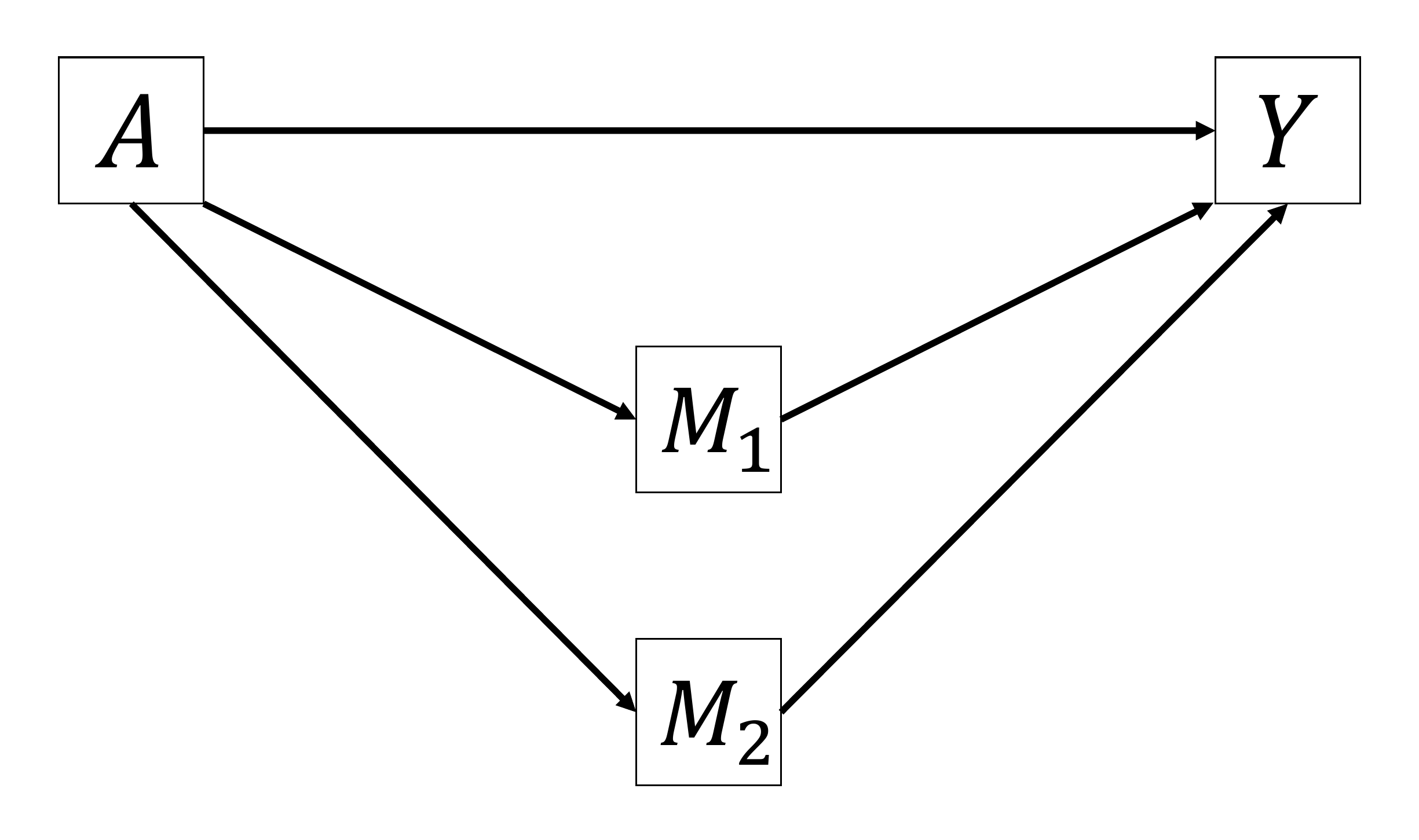}}
\caption{Directed acyclic graph with two non-sequential mediators.}
\label{fig3}
\end{figure}

\clearpage
\begin{figure}[!h]
\centering
\scalebox{0.5}[0.5]{\includegraphics{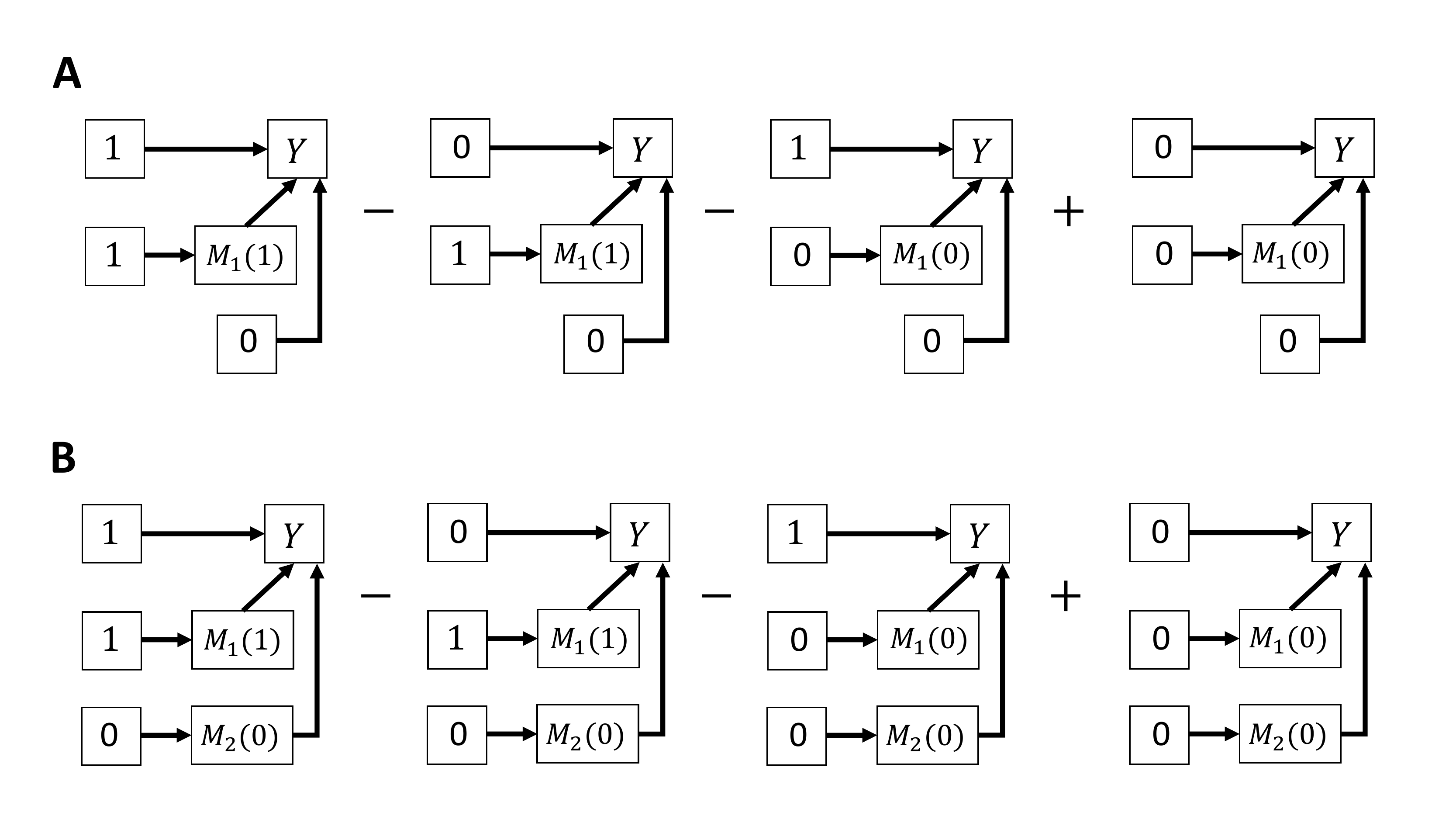}}
\caption{A comparison between the mediated interaction effect and the natural counterfactual interaction effect between $A$ and $M_1$ in a non-sequential two-mediator scenario.}
\label{fig4}
\end{figure}

\clearpage
\begin{figure}[!h]
\centering
\scalebox{0.5}[0.5]{\includegraphics{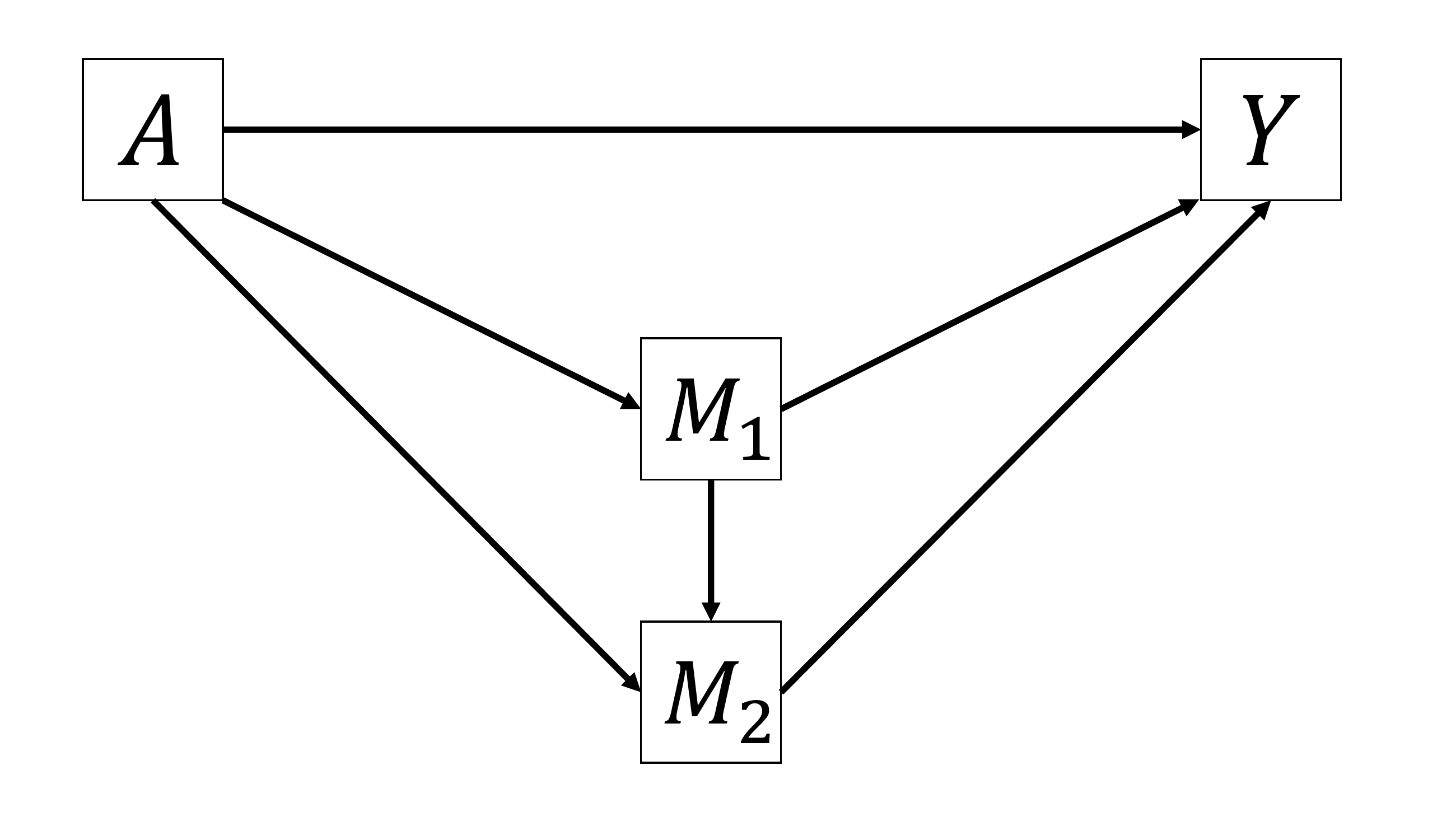}}
\caption{Directed acyclic graph with two sequential mediators.}
\label{fig5}
\end{figure}

\clearpage
\begin{figure}[!h]
\centering
\scalebox{0.5}[0.5]{\includegraphics{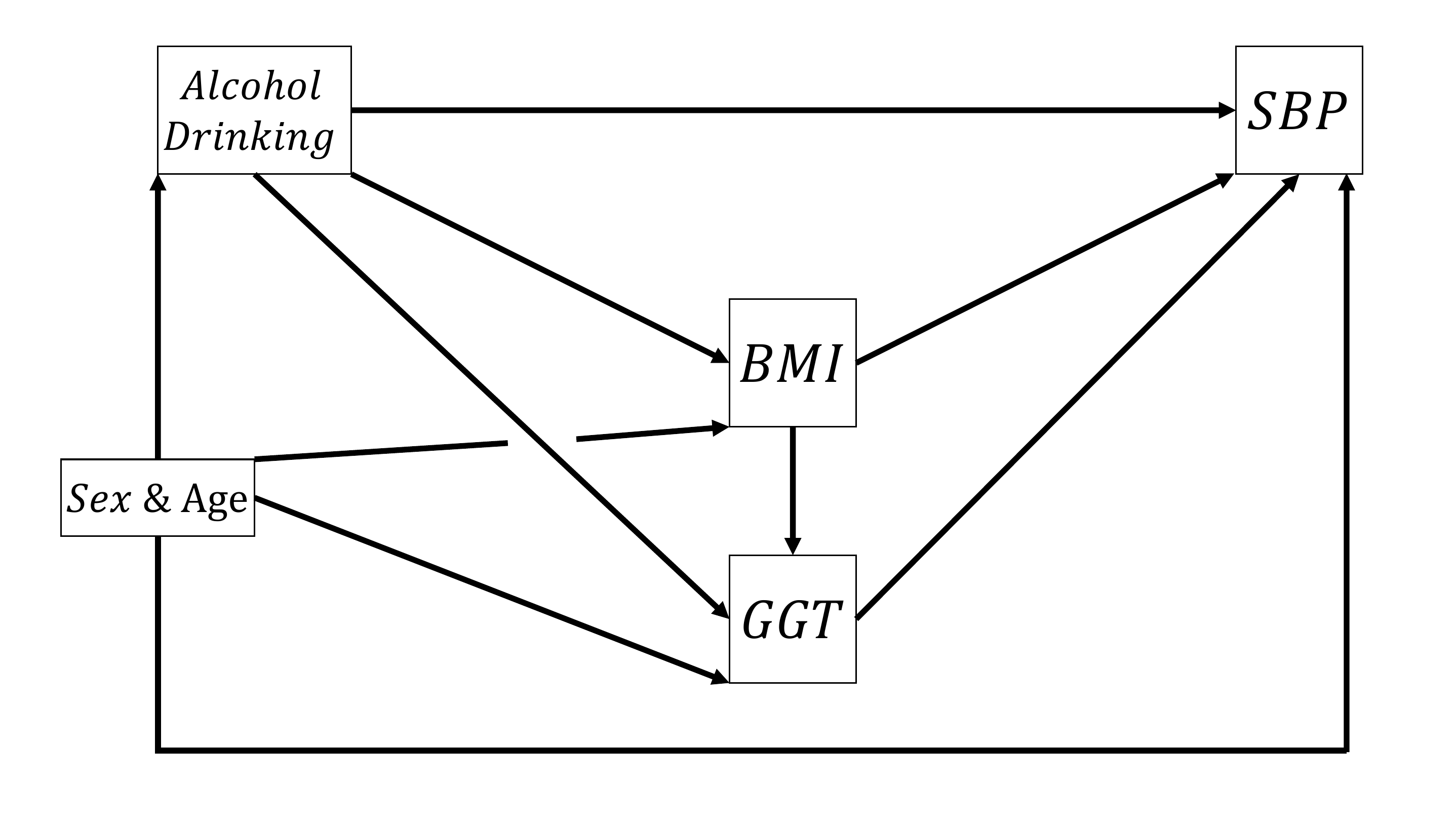}}
\caption{The directed acyclic graph for the study on hazard of drinking alcohol.}
\label{fig6}
\end{figure}

\clearpage
\section*{Acknowledgments}
This research was partially supported by UNM Comprehensive Cancer Center Support Grant NCI P30CA118100, the Biostatistics shared resource and UNM METALS Superfund Research Center (1P42ES025589).

\clearpage

\clearpage
\section*{Appendix A. Decomposition of total effect with the notion of natural counterfactual interaction effect in a non-sequential two-mediator scenario and the corresponding interpretations}
Suppose we have a directed acyclic graph as shown in Figure \ref{fig3}. We show in the following that the total effect can be decomposed into the following 10 components at the individual level:
\begin{eqnarray*}
  TE & = & CDE(m_1^\ast,m_2^\ast)+INT_{ref\mbox{-}AM_1}(m_1^\ast,m_2^\ast)+INT_{ref\mbox{-}AM_2}(m_1^\ast,m_2^\ast)\\
  & & +INT_{ref\mbox{-}AM_1M_2}(m_1^\ast,m_2^\ast)+ NatINT_{AM_1} + NatINT_{AM_2}+ NatINT_{AM_1M_2}\\
  & & + NatINT_{M_1M_2} + PIE_{M_1} + PIE_{M_2},
\end{eqnarray*}
where the natural counterfactual interaction effects are listed in Definition 2. We also give the corresponding interpretation for each component.\\

\noindent
\emph{Proof}: 

We first decompose the total effect into total direct effect ($TDE$) \cite{riden}, seminatural indirect effect through $M_1$ ($SIE_{M_1}$) \cite{p14} and pure indirect effect (path-specific effect) through $M_2$ ($PIE_{M_2}$) \cite{riden,p01}.
\begin{eqnarray*}
  TE & = & Y(a)-Y(a^\ast)\\
     \\
     & = & Y(a,M_1(a),M_2(a))-Y(a^\ast,M_1(a^\ast),M_2(a^\ast))\\
     \\
     & = & Y(a,M_1(a),M_2(a))-Y(a^\ast,M_1(a),M_2(a))\\
     & & +Y(a^\ast,M_1(a),M_2(a))-Y(a^\ast,M_1(a^\ast),M_2(a))\\
     & & +Y(a^\ast,M_1(a^\ast),M_2(a))-Y(a^\ast,M_1(a^\ast),M_2(a^\ast)),
\end{eqnarray*}
where the second equality follows the composition axiom \cite{vbook,a} and the third equality follows by adding and subtracting the same counterfactual formulas. 

The formulas of $TDE$, $SIE_{M_1}$ and $PIE_{M_2}$ are presented as follows: 
\begin{eqnarray*}
  TDE & = & Y(a,M_1(a),M_2(a))-Y(a^\ast,M_1(a),M_2(a))\\
     \\
  SIE_{M_1} & = & Y(a^\ast,M_1(a),M_2(a))-Y(a^\ast,M_1(a^\ast),M_2(a))\\
     \\
  PIE_{M_2} & = & Y(a^\ast,M_1(a^\ast),M_2(a))-Y(a^\ast,M_1(a^\ast),M_2(a^\ast)),
\end{eqnarray*}
where $TE = TDE+SIE_{M_1}+PIE_{M_2}$.

We focus on $TDE$ in the next step and decompose it into natural counterfactual interaction effects and pure direct effect ($PDE$) \cite{riden,p01} by subtracting $PDE$ from $TDE$, where $PDE$ satisfies the definition of a path-specific effect \cite{p01} and equals the following contrast of two counterfactual formulas:
\begin{eqnarray*}
  PDE & = & Y(a,M_1(a^\ast),M_2(a^\ast))-Y(a^\ast,M_1(a^\ast),M_2(a^\ast)).
\end{eqnarray*}

We have the following results:
\begin{eqnarray*}
  TDE-PDE & = & Y(a,M_1(a),M_2(a))-Y(a^\ast,M_1(a),M_2(a))\\
  & & -Y(a,M_1(a^\ast),M_2(a^\ast))+Y(a^\ast,M_1(a^\ast),M_2(a^\ast))\\
  \\
  & = & Y(a,M_1(a),M_2(a))-Y(a^\ast,M_1(a),M_2(a))\\
  & & -Y(a,M_1(a^\ast),M_2(a^\ast))+Y(a^\ast,M_1(a^\ast),M_2(a^\ast))\\
  & & +Y(a^\ast,M_1(a^\ast),M_2(a^\ast))-Y(a^\ast,M_1(a^\ast),M_2(a^\ast))\\
  & & +Y(a^\ast,M_1(a^\ast),M_2(a))-Y(a^\ast,M_1(a^\ast),M_2(a))\\
  & & +Y(a^\ast,M_1(a),M_2(a^\ast))-Y(a^\ast,M_1(a),M_2(a^\ast))\\
  & & +Y(a,M_1(a^\ast),M_2(a^\ast))-Y(a,M_1(a^\ast),M_2(a^\ast))\\
  & & +Y(a,M_1(a^\ast),M_2(a))-Y(a,M_1(a^\ast),M_2(a))\\
  & & +Y(a,M_1(a),M_2(a^\ast))-Y(a,M_1(a),M_2(a^\ast))\\
  \\
  & = & Y(a,M_1(a),M_2(a^\ast))-Y(a^\ast,M_1(a),M_2(a^\ast))\\
  & & - Y(a,M_1(a^\ast),M_2(a^\ast))+Y(a^\ast,M_1(a^\ast),M_2(a^\ast))\\
  & & + Y(a,M_1(a^\ast),M_2(a))- Y(a^\ast,M_1(a^\ast),M_2(a))\\
  & & -Y(a,M_1(a^\ast),M_2(a^\ast))+Y(a^\ast,M_1(a^\ast),M_2(a^\ast))\\
  & & + Y(a,M_1(a),M_2(a))-Y(a^\ast,M_1(a),M_2(a))\\
  & & - Y(a,M_1(a^\ast),M_2(a))+Y(a^\ast,M_1(a^\ast),M_2(a))\\
  & & -Y(a,M_1(a),M_2(a^\ast))+Y(a^\ast,M_1(a),M_2(a^\ast))\\
  & & +Y(a,M_1(a^\ast),M_2(a^\ast))-Y(a^\ast,M_1(a^\ast),M_2(a^\ast)),
\end{eqnarray*}
where the second equality follows by adding and subtracting the same counterfactual formulas, and the third equality follows by rearranging all the terms. 

Therefore, we have the following formulas satisfying Definition 2:
\begin{eqnarray*}
  NatINT_{AM_1} & = & Y(a,M_1(a),M_2(a^\ast))-Y(a^\ast,M_1(a),M_2(a^\ast))\\
  & & - Y(a,M_1(a^\ast),M_2(a^\ast))+Y(a^\ast,M_1(a^\ast),M_2(a^\ast))\\
  \\
  NatINT_{AM_2} & = & Y(a,M_1(a^\ast),M_2(a))-Y(a^\ast,M_1(a^\ast),M_2(a))\\
  & & -Y(a,M_1(a^\ast),M_2(a^\ast))+Y(a^\ast,M_1(a^\ast),M_2(a^\ast))\\
  \\
  NatINT_{AM_1M_2} & = & Y(a,M_1(a),M_2(a))-Y(a^\ast,M_1(a),M_2(a))\\
  & & - Y(a,M_1(a^\ast),M_2(a))+Y(a^\ast,M_1(a^\ast),M_2(a))\\
  & & -Y(a,M_1(a),M_2(a^\ast))+Y(a^\ast,M_1(a),M_2(a^\ast))\\
  & & +Y(a,M_1(a^\ast),M_2(a^\ast))-Y(a^\ast,M_1(a^\ast),M_2(a^\ast)).
\end{eqnarray*}

Accordingly, $TDE$ can be decomposed into the following components:
\begin{eqnarray*}
TDE &=& PDE + NatINT_{AM_1}+NatINT_{AM_2}+NatINT_{AM_1M_2}.
\end{eqnarray*}

We next focus on $PDE$ (path-specific effect) and decompose it into $CDE$ and reference interaction effects \cite{v4, b}:
\begin{eqnarray*}
PDE & = & Y(a,M_1(a^\ast),M_2(a^\ast))-Y(a^\ast,M_1(a^\ast),M_2(a^\ast))\\
\\
    & = & \sum_{m_2}\sum_{m_1}Y(a,m_1,m_2)\times I(M_1(a^\ast)=m_1)\times I(M_2(a^\ast)=m_2)\\
    & & - \sum_{m_2}\sum_{m_1}Y(a^\ast,m_1,m_2)\times I(M_1(a^\ast)=m_1)\times I(M_2(a^\ast)=m_2)\\
    \\
    & = & \sum_{m_2}\sum_{m_1}[Y(a,m_1,m_2)-Y(a^\ast,m_1,m_2)]\times I(M_1(a^\ast)=m_1)\times I(M_2(a^\ast)=m_2)\\
    \\
    & = & \sum_{m_2}\sum_{m_1}[Y(a,m_1,m_2)-Y(a^\ast,m_1,m_2)-Y(a,m_1^\ast,m_2^\ast)+Y(a^\ast,m_1^\ast,m_2^\ast)]\\
    & & \times I(M_1(a^\ast)=m_1)\times I(M_2(a^\ast)=m_2)\\
    & & + \sum_{m_2}\sum_{m_1}[Y(a,m_1^\ast,m_2^\ast)-Y(a^\ast,m_1^\ast,m_2^\ast)]\times I(M_1(a^\ast)=m_1)\times I(M_2(a^\ast)=m_2)\\
    \\
    & = & \sum_{m_2}\sum_{m_1}[Y(a,m_1,m_2)-Y(a^\ast,m_1,m_2)-Y(a,m_1^\ast,m_2^\ast)+Y(a^\ast,m_1^\ast,m_2^\ast)]\\
    & & \times I(M_1(a^\ast)=m_1)\times I(M_2(a^\ast)=m_2)\\
    & & + Y(a,m_1^\ast,m_2^\ast)-Y(a^\ast,m_1^\ast,m_2^\ast)\\
    \\
    & = & \sum_{m_2}\sum_{m_1}[Y(a,m_1,m_2)-Y(a^\ast,m_1,m_2)-Y(a,m_1^\ast,m_2^\ast)+Y(a^\ast,m_1^\ast,m_2^\ast)\\
    & & +Y(a^\ast,m_1^\ast,m_2^\ast)-Y(a^\ast,m_1^\ast,m_2^\ast)+Y(a^\ast,m_1^\ast,m_2)-Y(a^\ast,m_1^\ast,m_2)\\
    & & +Y(a^\ast,m_1,m_2^\ast)-Y(a^\ast,m_1,m_2^\ast)+Y(a,m_1^\ast,m_2^\ast)-Y(a,m_1^\ast,m_2^\ast)\\
    & & +Y(a,m_1^\ast,m_2)-Y(a,m_1^\ast,m_2)+Y(a,m_1,m_2^\ast)-Y(a,m_1,m_2^\ast)]\\
    & & \times I(M_1(a^\ast)=m_1)\times I(M_2(a^\ast)=m_2)\\
    & & + Y(a,m_1^\ast,m_2^\ast)-Y(a^\ast,m_1^\ast,m_2^\ast)\\
    \\
    & = & \sum_{m_2}\sum_{m_1}[Y(a,m_1,m_2^\ast)-Y(a^\ast,m_1,m_2^\ast)-Y(a,m_1^\ast,m_2^\ast)+Y(a^\ast,m_1^\ast,m_2^\ast)]\\
    & & \times I(M_1(a^\ast)=m_1)\times I(M_2(a^\ast)=m_2)\\
    & & + \sum_{m_2}\sum_{m_1}[Y(a,m_1^\ast,m_2)-Y(a^\ast,m_1^\ast,m_2)-Y(a,m_1^\ast,m_2^\ast)+Y(a^\ast,m_1^\ast,m_2^\ast)]\\
    & & \times I(M_1(a^\ast)=m_1)\times I(M_2(a^\ast)=m_2)\\
    & & + \sum_{m_2}\sum_{m_1}[Y(a,m_1,m_2)-Y(a^\ast,m_1,m_2)-Y(a,m_1^\ast,m_2)+Y(a^\ast,m_1^\ast,m_2)\\
    & & -Y(a,m_1,m_2^\ast)+Y(a^\ast,m_1,m_2^\ast)+Y(a,m_1^\ast,m_2^\ast)-Y(a^\ast,m_1^\ast,m_2^\ast)]\\
    & & \times I(M_1(a^\ast)=m_1)\times I(M_2(a^\ast)=m_2)\\
    & & + Y(a,m_1^\ast,m_2^\ast)-Y(a^\ast,m_1^\ast,m_2^\ast)\\
    \\
    & = & \sum_{m_1}[Y(a,m_1,m_2^\ast)-Y(a^\ast,m_1,m_2^\ast)-Y(a,m_1^\ast,m_2^\ast)+Y(a^\ast,m_1^\ast,m_2^\ast)]\\
    & & \times I(M_1(a^\ast)=m_1)\\
    & & + \sum_{m_2}[Y(a,m_1^\ast,m_2)-Y(a^\ast,m_1^\ast,m_2)-Y(a,m_1^\ast,m_2^\ast)+Y(a^\ast,m_1^\ast,m_2^\ast)]\\
    & & \times I(M_2(a^\ast)=m_2)\\
    & & + \sum_{m_2}\sum_{m_1}[Y(a,m_1,m_2)-Y(a^\ast,m_1,m_2)-Y(a,m_1^\ast,m_2)+Y(a^\ast,m_1^\ast,m_2)\\
    & & -Y(a,m_1,m_2^\ast)+Y(a^\ast,m_1,m_2^\ast)+Y(a,m_1^\ast,m_2^\ast)-Y(a^\ast,m_1^\ast,m_2^\ast)]\\
    & & \times I(M_1(a^\ast)=m_1)\times I(M_2(a^\ast)=m_2)\\
    & & + Y(a,m_1^\ast,m_2^\ast)-Y(a^\ast,m_1^\ast,m_2^\ast).
\end{eqnarray*}

According to the derivation above, the following formulas can be obtained:
\begin{eqnarray*}
CDE(m_1^\ast,m_2^\ast) & = & Y(a,m_1^\ast,m_2^\ast)-Y(a^\ast,m_1^\ast,m_2^\ast)\\
\\
INT_{ref\mbox{-}AM_1}(m_1^\ast,m_2^\ast) & = & \sum_{m_1}[Y(a,m_1,m_2^\ast)-Y(a^\ast,m_1,m_2^\ast)-Y(a,m_1^\ast,m_2^\ast)+Y(a^\ast,m_1^\ast,m_2^\ast)]\\
& & \times I(M_1(a^\ast)=m_1)\\
\\
INT_{ref\mbox{-}AM_2}(m_1^\ast,m_2^\ast) & = & \sum_{m_2}[Y(a,m_1^\ast,m_2)-Y(a^\ast,m_1^\ast,m_2)-Y(a,m_1^\ast,m_2^\ast)+Y(a^\ast,m_1^\ast,m_2^\ast)]\\
& & \times I(M_2(a^\ast)=m_2)\\
\\
INT_{ref\mbox{-}AM_1M_2}(m_1^\ast,m_2^\ast) & = & \sum_{m_2}\sum_{m_1}[Y(a,m_1,m_2)-Y(a^\ast,m_1,m_2)-Y(a,m_1^\ast,m_2)+Y(a^\ast,m_1^\ast,m_2)\\
    & & -Y(a,m_1,m_2^\ast)+Y(a^\ast,m_1,m_2^\ast)+Y(a,m_1^\ast,m_2^\ast)-Y(a^\ast,m_1^\ast,m_2^\ast)]\\
    & & \times I(M_1(a^\ast)=m_1)\times I(M_2(a^\ast)=m_2).
\end{eqnarray*}

With a little mathematical derivation, $INT_{ref\mbox{-}AM_1}$, $INT_{ref\mbox{-}AM_2}$ and $INT_{ref\mbox{-}AM_1M_2}$ can be expressed in the form of the counterfactual formula:
\begin{eqnarray*}
INT_{ref\mbox{-}AM_1}(m_1^\ast,m_2^\ast) & = & Y(a,M_1(a^\ast),m_2^\ast)-Y(a^\ast,M_1(a^\ast),m_2^\ast)-Y(a,m_1^\ast,m_2^\ast)+Y(a^\ast,m_1^\ast,m_2^\ast)\\
\\
INT_{ref\mbox{-}AM_2}(m_1^\ast,m_2^\ast) & = & Y(a,m_1^\ast,M_2(a^\ast))-Y(a^\ast,m_1^\ast,M_2(a^\ast))-Y(a,m_1^\ast,m_2^\ast)+Y(a^\ast,m_1^\ast,m_2^\ast)\\
\\
INT_{ref\mbox{-}AM_1M_2}(m_1^\ast,m_2^\ast) & = & Y(a,M_1(a^\ast),M_2(a^\ast))-Y(a^\ast,M_1(a^\ast),M_2(a^\ast))\\
& & -Y(a,m_1^\ast,M_2(a^\ast))+Y(a^\ast,m_1^\ast,M_2(a^\ast))\\
& & -Y(a,M_1(a^\ast),m_2^\ast)+Y(a^\ast,M_1(a^\ast),m_2^\ast)\\
& & +Y(a,m_1^\ast,m_2^\ast)-Y(a^\ast,m_1^\ast,m_2^\ast).
\end{eqnarray*}

Therefore, $PDE$ can be decomposed into the following components:
\begin{eqnarray*}
PDE & = & CDE(m_1^\ast,m_2^\ast)+INT_{ref\mbox{-}AM_1}(m_1^\ast,m_2^\ast)+INT_{ref\mbox{-}AM_2}(m_1^\ast,m_2^\ast)+INT_{ref\mbox{-}AM_1M_2}(m_1^\ast,m_2^\ast).
\end{eqnarray*}

We next focus on $SIE_{M_1}$ and try to decompose it into $PIE_{M_1}$ and $NatINT_{M_1M_2}$ by subtracting $PIE_{M_1}$ from $SIE_{M_1}$:
\begin{eqnarray*}
SIE_{M_1}-PIE_{M_1} & = & Y(a^\ast,M_1(a),M_2(a))-Y(a^\ast,M_1(a^\ast),M_2(a))\\
& & -Y(a^\ast,M_1(a),M_2(a^\ast))+Y(a^\ast,M_1(a^\ast),M_2(a^\ast))\\
\\
& = & NatINT_{M_1M_2},
\end{eqnarray*}
where $NatINT_{M_1M_2}$ satisfies Definition 2.

Therefore, $SIE_{M_1}$ can be decomposed into the following components:
\begin{eqnarray*}
SIE_{M_1} & = & PIE_{M_1}+ NatINT_{M_1M_2}.
\end{eqnarray*}

Combining all the derivations above, we have the decomposition of total effect as follows:
\begin{eqnarray*}
  TE & = & CDE(m_1^\ast,m_2^\ast)+INT_{ref\mbox{-}AM_1}(m_1^\ast,m_2^\ast)+INT_{ref\mbox{-}AM_2}(m_1^\ast,m_2^\ast)\\
  & & +INT_{ref\mbox{-}AM_1M_2}(m_1^\ast,m_2^\ast)+ NatINT_{AM_1} + NatINT_{AM_2}+ NatINT_{AM_1M_2}\\
  & & + NatINT_{M_1M_2} + PIE_{M_1} + PIE_{M_2}.
\end{eqnarray*}

We next present the interpretation for each component assuming binary $A$, $M_1$ and $M_2$ with the conditions $a=1$, $a^\ast=0$, $m_1^\ast=0$ and $m_2^\ast=0$ for illustration purpose. While other interpretations were proposed in the literature \cite{v4,b}, our work represent a different and more flexible interpretation from the perspective of population averages which accounts for the distribution of the mediators in the causal structure.
\subsection*{controlled direct effect}
With the specified conditions, the controlled direct effect can be written as:
\begin{eqnarray*}
CDE(m_1^\ast,m_2^\ast) & = & Y(a,m_1^\ast,m_2^\ast)-Y(a^\ast,m_1^\ast,m_2^\ast)\\
\\
\Rightarrow \quad\quad CDE(0,0) & = & Y(1,0,0)-Y(0,0,0).
\end{eqnarray*}

$CDE(m_1^\ast,m_2^\ast)$ can be interpreted as the effect due to neither mediation nor interaction.
\subsection*{reference interaction effects}
With the specified conditions, the reference interaction effect between $A$ and $M_1$ can be written as:
\begin{eqnarray*}
INT_{ref\mbox{-}AM_1}(m_1^\ast,m_2^\ast) & = & \sum_{m_1}[Y(a,m_1,m_2^\ast)-Y(a^\ast,m_1,m_2^\ast)-Y(a,m_1^\ast,m_2^\ast)+Y(a^\ast,m_1^\ast,m_2^\ast)]\\
& & \times I(M_1(a^\ast)=m_1)\\
\\
\Rightarrow INT_{ref\mbox{-}AM_1}(0,0) & = & \sum_{m_1}[Y(1,m_1,0)-Y(0,m_1,0)-Y(1,0,0)+Y(0,0,0)]\times I(M_1(0)=m_1)\\
\\
& = & [Y(1,0,0)-Y(0,0,0)-Y(1,0,0)+Y(0,0,0)]\times I(M_1(0)=0)\\
& & + [Y(1,1,0)-Y(0,1,0)-Y(1,0,0)+Y(0,0,0)]\times I(M_1(0)=1)\\
\\
& = & [Y(1,1,0)-Y(0,1,0)-Y(1,0,0)+Y(0,0,0)]\times I(M_1(0)=1)\\
\\
& = & [Y(1,1,0)-Y(0,1,0)-Y(1,0,0)+Y(0,0,0)]\times M_1(0).
\end{eqnarray*}

$INT_{ref\mbox{-}AM_1}(m_1^\ast,m_2^\ast)$ can be interpreted as the effect due to the interaction between $A$ and $M_1$ only.\\

The reference interaction effect between $A$ and $M_2$ can be written as:
\begin{eqnarray*}
INT_{ref\mbox{-}AM_2}(m_1^\ast,m_2^\ast) & = & \sum_{m_2}[Y(a,m_1^\ast,m_2)-Y(a^\ast,m_1^\ast,m_2)-Y(a,m_1^\ast,m_2^\ast)+Y(a^\ast,m_1^\ast,m_2^\ast)]\\
& & \times I(M_2(a^\ast)=m_2)\\
\\
\Rightarrow INT_{ref\mbox{-}AM_2}(0,0) & = & \sum_{m_2}[Y(1,0,m_2)-Y(0,0,m_2)-Y(1,0,0)+Y(0,0,0)]\times I(M_2(0)=m_2)\\
\\
& = & [Y(1,0,0)-Y(0,0,0)-Y(1,0,0)+Y(0,0,0)]\times I(M_2(0)=0)\\
& & + [Y(1,0,1)-Y(0,0,1)-Y(1,0,0)+Y(0,0,0)]\times I(M_2(0)=1)\\
\\
& = & [Y(1,0,1)-Y(0,0,1)-Y(1,0,0)+Y(0,0,0)]\times I(M_2(0)=1)\\
\\
& = & [Y(1,0,1)-Y(0,0,1)-Y(1,0,0)+Y(0,0,0)]\times M_2(0).
\end{eqnarray*}

$INT_{ref\mbox{-}AM_2}(m_1^\ast,m_2^\ast)$ can be interpreted as the effect due to the interaction between $A$ and $M_2$ only.\\

The reference interaction effect between $A$, $M_1$ and $M_2$ can be written as:
\begin{eqnarray*}
INT_{ref\mbox{-}AM_1M_2}(m_1^\ast,m_2^\ast) & = & \sum_{m_2}\sum_{m_1}[Y(a,m_1,m_2)-Y(a^\ast,m_1,m_2)-Y(a,m_1^\ast,m_2)+Y(a^\ast,m_1^\ast,m_2)\\
    & & -Y(a,m_1,m_2^\ast)+Y(a^\ast,m_1,m_2^\ast)+Y(a,m_1^\ast,m_2^\ast)-Y(a^\ast,m_1^\ast,m_2^\ast)]\\
    & & \times I(M_1(a^\ast)=m_1)\times I(M_2(a^\ast)=m_2)\\
\\
\Rightarrow INT_{ref\mbox{-}AM_1M_2}(0,0) & = & \sum_{m_2}\sum_{m_1}[Y(1,m_1,m_2)-Y(0,m_1,m_2)-Y(1,0,m_2)+Y(0,0,m_2)\\
    & & -Y(1,m_1,0)+Y(0,m_1,0)+Y(1,0,0)-Y(0,0,0)]\\
    & & \times I(M_1(0)=m_1)\times I(M_2(0)=m_2)\\
\\
& = & \sum_{m_2}[Y(1,0,m_2)-Y(0,0,m_2)-Y(1,0,m_2)+Y(0,0,m_2)\\
    & & -Y(1,0,0)+Y(0,0,0)+Y(1,0,0)-Y(0,0,0)]\\
    & & \times I(M_1(0)=0)\times I(M_2(0)=m_2)\\
& & + \sum_{m_2}[Y(1,1,m_2)-Y(0,1,m_2)-Y(1,0,m_2)+Y(0,0,m_2)\\
    & & -Y(1,1,0)+Y(0,1,0)+Y(1,0,0)-Y(0,0,0)]\\
    & & \times I(M_1(0)=1)\times I(M_2(0)=m_2)\\
\\
& = & [Y(1,1,0)-Y(0,1,0)-Y(1,0,0)+Y(0,0,0)\\
    & & -Y(1,1,0)+Y(0,1,0)+Y(1,0,0)-Y(0,0,0)]\\
    & & \times I(M_1(0)=1)\times I(M_2(0)=0)\\
& & + [Y(1,1,1)-Y(0,1,1)-Y(1,0,1)+Y(0,0,1)\\
    & & -Y(1,1,0)+Y(0,1,0)+Y(1,0,0)-Y(0,0,0)]\\
    & & \times I(M_1(0)=1)\times I(M_2(0)=1)\\
\\
& = & [Y(1,1,1)-Y(0,1,1)-Y(1,0,1)+Y(0,0,1)\\
    & & -Y(1,1,0)+Y(0,1,0)+Y(1,0,0)-Y(0,0,0)]\\
    & & \times M_1(0)\times M_2(0).
\end{eqnarray*}

$INT_{ref\mbox{-}AM_1M_2}(m_1^\ast,m_2^\ast)$ can be interpreted as the effect due to the interaction between $A$, $M_1$ and $M_2$ only.
\subsection*{natural counterfactual interaction effects}
The natural counterfactual interaction effect between $A$ and $M_1$ can be rewritten as: 
\begin{eqnarray*}
  NatINT_{AM_1} & = & Y(a,M_1(a),M_2(a^\ast))-Y(a^\ast,M_1(a),M_2(a^\ast))\\
  & & - Y(a,M_1(a^\ast),M_2(a^\ast))+Y(a^\ast,M_1(a^\ast),M_2(a^\ast))\\
\\
& = & \sum_{m_2}\sum_{m_1}Y(a,m_1,m_2)I(M_1(a)=m_1)I(M_2(a^\ast)=m_2)\\
& & - \sum_{m_2}\sum_{m_1}Y(a^\ast,m_1,m_2)I(M_1(a)=m_1)I(M_2(a^\ast)=m_2)\\
& & - \sum_{m_2}\sum_{m_1}Y(a,m_1,m_2)I(M_1(a^\ast)=m_1)I(M_2(a^\ast)=m_2)\\
& & + \sum_{m_2}\sum_{m_1}Y(a^\ast,m_1,m_2)I(M_1(a^\ast)=m_1)I(M_2(a^\ast)=m_2)\\
\\
& = & \sum_{m_2}\sum_{m_1}[Y(a,m_1,m_2)-Y(a^\ast,m_1,m_2)]I(M_1(a)=m_1)I(M_2(a^\ast)=m_2)\\
& & - \sum_{m_2}\sum_{m_1}[Y(a,m_1,m_2)-Y(a^\ast,m_1,m_2)]I(M_1(a^\ast)=m_1)I(M_2(a^\ast)=m_2)\\
\\
& = & \sum_{m_2}\sum_{m_1}[Y(a,m_1,m_2)-Y(a^\ast,m_1,m_2)]I(M_2(a^\ast)=m_2)[I(M_1(a)=m_1)-I(M_1(a^\ast)=m_1)]\\
\\
& = & \sum_{m_2}\sum_{m_1}[Y(a,m_1,m_2)I(M_2(a^\ast)=m_2)-Y(a^\ast,m_1,m_2)I(M_2(a^\ast)=m_2)]\\
& & \times[I(M_1(a)=m_1)-I(M_1(a^\ast)=m_1)]\\
\\
& = & \sum_{m_2}\sum_{m_1}[Y(a,m_1,m_2)I(M_2(a^\ast)=m_2)-Y(a^\ast,m_1,m_2)I(M_2(a^\ast)=m_2)\\
& & - Y(a,m_1^\ast,m_2)I(M_2(a^\ast)=m_2)+Y(a^\ast,m_1^\ast,m_2)I(M_2(a^\ast)=m_2)]\\
& & \times[I(M_1(a)=m_1)-I(M_1(a^\ast)=m_1)],
\end{eqnarray*}
where the sixth equation follows by adding two extra terms which do not change the value of $NatINT_{AM_1}$.\\

With the specified conditions, $NatINT_{AM_1}$ can be written as:
\begin{eqnarray*}
  NatINT_{AM_1} & = & \sum_{m_2}\sum_{m_1}[Y(1,m_1,m_2)I(M_2(0)=m_2)-Y(0,m_1,m_2)I(M_2(0)=m_2)\\
& & - Y(1,0,m_2)I(M_2(0)=m_2)+Y(0,0,m_2)I(M_2(0)=m_2)]\\
& & \times[I(M_1(1)=m_1)-I(M_1(0)=m_1)]\\
\\
& = & \sum_{m_2}[Y(1,0,m_2)I(M_2(0)=m_2)-Y(0,0,m_2)I(M_2(0)=m_2)\\
& & - Y(1,0,m_2)I(M_2(0)=m_2)+Y(0,0,m_2)I(M_2(0)=m_2)]\\
& & \times[I(M_1(1)=0)-I(M_1(0)=0)]\\
& & + \sum_{m_2}[Y(1,1,m_2)I(M_2(0)=m_2)-Y(0,1,m_2)I(M_2(0)=m_2)\\
& & - Y(1,0,m_2)I(M_2(0)=m_2)+Y(0,0,m_2)I(M_2(0)=m_2)]\\
& & \times[I(M_1(1)=1)-I(M_1(0)=1)]\\
\\
& = & \sum_{m_2}[Y(1,1,m_2)I(M_2(0)=m_2)-Y(0,1,m_2)I(M_2(0)=m_2)\\
& & - Y(1,0,m_2)I(M_2(0)=m_2)+Y(0,0,m_2)I(M_2(0)=m_2)]\\
& & \times[I(M_1(1)=1)-I(M_1(0)=1)]\\
\\
& = & \sum_{m_2}[Y(1,1,m_2)I(M_2(0)=m_2)-Y(0,1,m_2)I(M_2(0)=m_2)\\
& & - Y(1,0,m_2)I(M_2(0)=m_2)+Y(0,0,m_2)I(M_2(0)=m_2)]\\
& & \times[M_1(1)-M_1(0)],
\end{eqnarray*}
where the indicator function $I(M_2(0)=m_2)$ indicates that $M_2$ is at its potential value $M_2(0)$ which may vary with respect to different individuals. \\

$NatINT_{AM_1}$ can be interpreted as the effect due to the mediation through $M_1$ and the interaction between $A$ and $M_1$ conditioning on the potential value of $M_2$ with the fixed reference level $a^\ast$.\\

The natural counterfactual interaction effect between $A$ and $M_2$ can be rewritten as: 
\begin{eqnarray*}  
  NatINT_{AM_2} & = & Y(a,M_1(a^\ast),M_2(a))-Y(a^\ast,M_1(a^\ast),M_2(a))\\
  & & -Y(a,M_1(a^\ast),M_2(a^\ast))+Y(a^\ast,M_1(a^\ast),M_2(a^\ast))\\
\\
& = & \sum_{m_2}\sum_{m_1}Y(a,m_1,m_2)I(M_1(a^\ast)=m_1)I(M_2(a)=m_2)\\
& & - \sum_{m_2}\sum_{m_1}Y(a^\ast,m_1,m_2)I(M_1(a^\ast)=m_1)I(M_2(a)=m_2)\\
& & - \sum_{m_2}\sum_{m_1}Y(a,m_1,m_2)I(M_1(a^\ast)=m_1)I(M_2(a^\ast)=m_2)\\
& & + \sum_{m_2}\sum_{m_1}Y(a^\ast,m_1,m_2)I(M_1(a^\ast)=m_1)I(M_2(a^\ast)=m_2)\\
\\
& = & \sum_{m_2}\sum_{m_1}[Y(a,m_1,m_2)-Y(a^\ast,m_1,m_2)]I(M_1(a^\ast)=m_1)I(M_2(a)=m_2)\\
& & - \sum_{m_2}\sum_{m_1}[Y(a,m_1,m_2)-Y(a^\ast,m_1,m_2)]I(M_1(a^\ast)=m_1)I(M_2(a^\ast)=m_2)\\
\\
& = & \sum_{m_2}\sum_{m_1}[Y(a,m_1,m_2)-Y(a^\ast,m_1,m_2)]I(M_1(a^\ast)=m_1)[I(M_2(a)=m_2)-I(M_2(a^\ast)=m_2)]\\
\\
& = & \sum_{m_2}\sum_{m_1}[Y(a,m_1,m_2)I(M_1(a^\ast)=m_1)-Y(a^\ast,m_1,m_2)I(M_1(a^\ast)=m_1)]\\
& & \times[I(M_2(a)=m_2)-I(M_2(a^\ast)=m_2)]\\
\\
& = & \sum_{m_2}\sum_{m_1}[Y(a,m_1,m_2)I(M_1(a^\ast)=m_1)-Y(a^\ast,m_1,m_2)I(M_1(a^\ast)=m_1)\\
& & - Y(a,m_1,m_2^\ast)I(M_1(a^\ast)=m_1) + Y(a^\ast,m_1,m_2^\ast)I(M_1(a^\ast)=m_1)]\\
& & \times[I(M_2(a)=m_2)-I(M_2(a^\ast)=m_2)],
\end{eqnarray*}
where the sixth equation follows by adding two extra terms which do not change the value of $NatINT_{AM_2}$.\\

With the specified conditions, $NatINT_{AM_2}$ can be written as:
\begin{eqnarray*}
  NatINT_{AM_2} & = & \sum_{m_2}\sum_{m_1}[Y(1,m_1,m_2)I(M_1(0)=m_1)-Y(0,m_1,m_2)I(M_1(0)=m_1)\\
& & - Y(1,m_1,0)I(M_1(0)=m_1)+Y(0,m_1,0)I(M_1(0)=m_1)]\\
& & \times[I(M_2(1)=m_2)-I(M_2(0)=m_2)]\\
\\
& = & \sum_{m_1}[Y(1,m_1,0)I(M_1(0)=m_1)-Y(0,m_1,0)I(M_1(0)=m_1)\\
& & - Y(1,m_1,0)I(M_1(0)=m_1)+Y(0,m_1,0)I(M_1(0)=m_1)]\\
& & \times[I(M_2(1)=0)-I(M_2(0)=0)]\\
& & + \sum_{m_1}[Y(1,m_1,1)I(M_1(0)=m_1)-Y(0,m_1,1)I(M_1(0)=m_1)\\
& & - Y(1,m_1,0)I(M_1(0)=m_1)+Y(0,m_1,0)I(M_1(0)=m_1)]\\
& & \times[I(M_2(1)=1)-I(M_2(0)=1)]\\
\\
& = & \sum_{m_1}[Y(1,m_1,1)I(M_1(0)=m_1)-Y(0,m_1,1)I(M_1(0)=m_1)\\
& & - Y(1,m_1,0)I(M_1(0)=m_1)+Y(0,m_1,0)I(M_1(0)=m_1)]\\
& & \times[I(M_2(1)=1)-I(M_2(0)=1)]\\
\\
& = & \sum_{m_1}[Y(1,m_1,1)I(M_1(0)=m_1)-Y(0,m_1,1)I(M_1(0)=m_1)\\
& & - Y(1,m_1,0)I(M_1(0)=m_1)+Y(0,m_1,0)I(M_1(0)=m_1)]\\
& & \times[M_2(1)-M_2(0)],
\end{eqnarray*}
where the indicator function $I(M_1(0)=m_1)$ indicates that $M_1$ is at its potential value $M_1(0)$ which may vary with respect to different individuals. \\

$NatINT_{AM_2}$ can be interpreted as the effect due to the mediation through $M_2$ and the interaction between $A$ and $M_2$ conditioning on the potential value of $M_1$ with the fixed reference level $a^\ast$.\\

The natural counterfactual interaction effect between $A$, $M_1$ and $M_2$ can be rewritten as:
\begin{eqnarray*}
  NatINT_{AM_1M_2} & = & Y(a,M_1(a),M_2(a))-Y(a^\ast,M_1(a),M_2(a))\\
  & & - Y(a,M_1(a^\ast),M_2(a))+Y(a^\ast,M_1(a^\ast),M_2(a))\\
  & & -Y(a,M_1(a),M_2(a^\ast))+Y(a^\ast,M_1(a),M_2(a^\ast))\\
  & & +Y(a,M_1(a^\ast),M_2(a^\ast))-Y(a^\ast,M_1(a^\ast),M_2(a^\ast))\\
\\
& = & \sum_{m_2}\sum_{m_1}Y(a,m_1,m_2)I(M_1(a)=m_1)I(M_2(a)=m_2)\\
& & - \sum_{m_2}\sum_{m_1}Y(a^\ast,m_1,m_2)I(M_1(a)=m_1)I(M_2(a)=m_2)\\
& & - \sum_{m_2}\sum_{m_1}Y(a,m_1,m_2)I(M_1(a^\ast)=m_1)I(M_2(a)=m_2)\\
& & + \sum_{m_2}\sum_{m_1}Y(a^\ast,m_1,m_2)I(M_1(a^\ast)=m_1)I(M_2(a)=m_2)\\
& & - \sum_{m_2}\sum_{m_1}Y(a,m_1,m_2)I(M_1(a)=m_1)I(M_2(a^\ast)=m_2)\\
& & + \sum_{m_2}\sum_{m_1}Y(a^\ast,m_1,m_2)I(M_1(a)=m_1)I(M_2(a^\ast)=m_2)\\
& & + \sum_{m_2}\sum_{m_1}Y(a,m_1,m_2)I(M_1(a^\ast)=m_1)I(M_2(a^\ast)=m_2)\\
& & - \sum_{m_2}\sum_{m_1}Y(a^\ast,m_1,m_2)I(M_1(a^\ast)=m_1)I(M_2(a^\ast)=m_2)\\
\\
& = & \sum_{m_2}\sum_{m_1}[Y(a,m_1,m_2)-Y(a^\ast,m_1,m_2)]I(M_1(a)=m_1)I(M_2(a)=m_2)\\
& & - \sum_{m_2}\sum_{m_1}[Y(a,m_1,m_2)-Y(a^\ast,m_1,m_2)]I(M_1(a^\ast)=m_1)I(M_2(a)=m_2)\\
& & - \sum_{m_2}\sum_{m_1}[Y(a,m_1,m_2)-Y(a^\ast,m_1,m_2)]I(M_1(a)=m_1)I(M_2(a^\ast)=m_2)\\
& & + \sum_{m_2}\sum_{m_1}[Y(a,m_1,m_2)-Y(a^\ast,m_1,m_2)]I(M_1(a^\ast)=m_1)I(M_2(a^\ast)=m_2)\\
\\
& = & \sum_{m_2}\sum_{m_1}[Y(a,m_1,m_2)-Y(a^\ast,m_1,m_2)][I(M_1(a)=m_1)-I(M_1(a^\ast)=m_1)]\\
& & \times[I(M_2(a)=m_2)-I(M_2(a^\ast)=m_2)]\\
\\
& = & \sum_{m_2}\sum_{m_1}[Y(a,m_1,m_2)-Y(a^\ast,m_1,m_2)-Y(a,m_1^\ast,m_2)+Y(a^\ast,m_1^\ast,m_2)\\
& & - Y(a,m_1,m_2^\ast)+Y(a^\ast,m_1,m_2^\ast)+Y(a,m_1^\ast,m_2^\ast)-Y(a^\ast,m_1^\ast,m_2^\ast)]\\
& & \times[I(M_1(a)=m_1)-I(M_1(a^\ast)=m_1)]\\
& & \times[I(M_2(a)=m_2)-I(M_2(a^\ast)=m_2)],
\end{eqnarray*}
where the fifth equation follows by adding six extra terms which do not change the value of $NatINT_{AM_1M_2}$.\\

With the specified conditions, $NatINT_{AM_1M_2}$ can be written as: 
\begin{eqnarray*}
  NatINT_{AM_1M_2} & = & \sum_{m_2}\sum_{m_1}[Y(1,m_1,m_2)-Y(0,m_1,m_2)-Y(1,0,m_2)+Y(0,0,m_2)\\
& & - Y(1,m_1,0)+Y(0,m_1,0)+Y(1,0,0)-Y(0,0,0)]\\
& & \times[I(M_1(1)=m_1)-I(M_1(0)=m_1)]\\
& & \times[I(M_2(1)=m_2)-I(M_2(0)=m_2)]\\
\\
& = & \sum_{m_2}[Y(1,0,m_2)-Y(0,0,m_2)-Y(1,0,m_2)+Y(0,0,m_2)\\
& & - Y(1,0,0)+Y(0,0,0)+Y(1,0,0)-Y(0,0,0)]\\
& & \times[I(M_1(1)=0)-I(M_1(0)=0)]\\
& & \times[I(M_2(1)=m_2)-I(M_2(0)=m_2)]\\
& & + \sum_{m_2}[Y(1,1,m_2)-Y(0,1,m_2)-Y(1,0,m_2)+Y(0,0,m_2)\\
& & - Y(1,1,0)+Y(0,1,0)+Y(1,0,0)-Y(0,0,0)]\\
& & \times[I(M_1(1)=1)-I(M_1(0)=1)]\\
& & \times[I(M_2(1)=m_2)-I(M_2(0)=m_2)]\\
\\
& = & [Y(1,1,0)-Y(0,1,0)-Y(1,0,0)+Y(0,0,0)\\
& & - Y(1,1,0)+Y(0,1,0)+Y(1,0,0)-Y(0,0,0)]\\
& & \times[I(M_1(1)=1)-I(M_1(0)=1)]\\
& & \times[I(M_2(1)=0)-I(M_2(0)=0)]\\
& + & [Y(1,1,1)-Y(0,1,1)-Y(1,0,1)+Y(0,0,1)\\
& & - Y(1,1,0)+Y(0,1,0)+Y(1,0,0)-Y(0,0,0)]\\
& & \times[I(M_1(1)=1)-I(M_1(0)=1)]\\
& & \times[I(M_2(1)=1)-I(M_2(0)=1)]\\
\\
& = & [Y(1,1,1)-Y(0,1,1)-Y(1,0,1)+Y(0,0,1)\\
& & - Y(1,1,0)+Y(0,1,0)+Y(1,0,0)-Y(0,0,0)]\\
& & \times[I(M_1(1)=1)-I(M_1(0)=1)]\\
& & \times[I(M_2(1)=1)-I(M_2(0)=1)]\\
\\
& = & [Y(1,1,1)-Y(0,1,1)-Y(1,0,1)+Y(0,0,1)\\
& & - Y(1,1,0)+Y(0,1,0)+Y(1,0,0)-Y(0,0,0)]\\
& & \times[M_1(1)-M_1(0)]\\
& & \times[M_2(1)-M_2(0)],
\end{eqnarray*}

$NatINT_{AM_1M_2}$ can be interpreted as the effect due to the mediation through both $M_1$ and $M_2$, and the interaction between $A$, $M_1$ and $M_2$.\\

The natural counterfactual interaction effect between $M_1$ and $M_2$ can be rewritten as: 
\begin{eqnarray*}
NatINT_{M_1M_2} & = & Y(a^\ast,M_1(a),M_2(a))-Y(a^\ast,M_1(a^\ast),M_2(a))\\
& & -Y(a^\ast,M_1(a),M_2(a^\ast))+Y(a^\ast,M_1(a^\ast),M_2(a^\ast))\\
\\
& = & \sum_{m_2}\sum_{m_1}Y(a^\ast,m_1,m_2)I(M_1(a)=m_1)I(M_2(a)=m_2)\\
& & - \sum_{m_2}\sum_{m_1}Y(a^\ast,m_1,m_2)I(M_1(a^\ast)=m_1)I(M_2(a)=m_2)\\
& & - \sum_{m_2}\sum_{m_1}Y(a^\ast,m_1,m_2)I(M_1(a)=m_1)I(M_2(a^\ast)=m_2)\\
& & + \sum_{m_2}\sum_{m_1}Y(a^\ast,m_1,m_2)I(M_1(a^\ast)=m_1)I(M_2(a^\ast)=m_2)\\
\\
& = & \sum_{m_2}\sum_{m_1}Y(a^\ast,m_1,m_2)[I(M_1(a)=m_1)-I(M_1(a^\ast)=m_1)]\\
& & \times[I(M_2(a)=m_2)-I(M_2(a^\ast)=m_2)]\\
\\
& = & \sum_{m_2}\sum_{m_1}[Y(a^\ast,m_1,m_2)-Y(a^\ast,m_1^\ast,m_2)-Y(a^\ast,m_1,m_2^\ast)+Y(a^\ast,m_1^\ast,m_2^\ast)]\\& & \times[I(M_1(a)=m_1)-I(M_1(a^\ast)=m_1)]\\
& & \times[I(M_2(a)=m_2)-I(M_2(a^\ast)=m_2)],
\end{eqnarray*}
where the fourth equation follows by adding three extra terms which do not change the value of $NatINT_{M_1M_2}$.\\

With the specified conditions, $NatINT_{M_1M_2}$ can be written as:
\begin{eqnarray*}
NatINT_{M_1M_2} & = & \sum_{m_2}\sum_{m_1}[Y(0,m_1,m_2)-Y(0,0,m_2)-Y(0,m_1,0)+Y(0,0,0)]\\& & \times[I(M_1(1)=m_1)-I(M_1(0)=m_1)]\\
& & \times[I(M_2(1)=m_2)-I(M_2(0)=m_2)]\\
\\
& = &  \sum_{m_2}[Y(0,0,m_2)-Y(0,0,m_2)-Y(0,0,0)+Y(0,0,0)]\\& & \times[I(M_1(1)=0)-I(M_1(0)=0)]\\
& & \times[I(M_2(1)=m_2)-I(M_2(0)=m_2)]\\
& & +  \sum_{m_2}[Y(0,1,m_2)-Y(0,0,m_2)-Y(0,1,0)+Y(0,0,0)]\\& & \times[I(M_1(1)=1)-I(M_1(0)=1)]\\
& & \times[I(M_2(1)=m_2)-I(M_2(0)=m_2)]\\
\\
& = & [Y(0,1,0)-Y(0,0,0)-Y(0,1,0)+Y(0,0,0)]\\
& & \times[I(M_1(1)=1)-I(M_1(0)=1)]\\
& & \times[I(M_2(1)=0)-I(M_2(0)=0)]\\
& & + [Y(0,1,1)-Y(0,0,1)-Y(0,1,0)+Y(0,0,0)]\\
& & \times[I(M_1(1)=1)-I(M_1(0)=1)]\\
& & \times[I(M_2(1)=1)-I(M_2(0)=1)]\\
\\
& = & [Y(0,1,1)-Y(0,0,1)-Y(0,1,0)+Y(0,0,0)]\\
& & \times[I(M_1(1)=1)-I(M_1(0)=1)]\\
& & \times[I(M_2(1)=1)-I(M_2(0)=1)]\\
\\
& = & [Y(0,1,1)-Y(0,0,1)-Y(0,1,0)+Y(0,0,0)]\\
& & \times[M_1(1)-M_1(0)]\\
& & \times[M_2(1)-M_2(0)].
\end{eqnarray*}

$NatINT_{M_1M_2}$ can be interpreted as the effect due to the mediation through both $M_1$ and $M_2$, and the interaction between $M_1$ and $M_2$. Since the interaction is not involved with the change in exposure $A$, the interpretation can be simply put as the effect due to the mediation through both $M_1$ and $M_2$ only.

\subsection*{pure indirect effects}
The pure indirect effect through $M_1$ can be rewritten as:
\begin{eqnarray*}
  PIE_{M_1} & = & Y(a^\ast,M_1(a),M_2(a^\ast))-Y(a^\ast,M_1(a^\ast),M_2(a^\ast))\\
\\
& = & \sum_{m_2}\sum_{m_1}Y(a^\ast,m_1,m_2)I(M_1(a)=m_1)I(M_2(a^\ast)=m_2)\\
& & - \sum_{m_2}\sum_{m_1}Y(a^\ast,m_1,m_2)I(M_1(a^\ast)=m_1)I(M_2(a^\ast)=m_2)\\
\\
& = & \sum_{m_2}\sum_{m_1}Y(a^\ast,m_1,m_2)[I(M_1(a)=m_1)-I(M_1(a^\ast)=m_1)]I(M_2(a^\ast)=m_2).
\end{eqnarray*}

With the specified conditions, $PIE_{M_1}$ can be written as:
\begin{eqnarray*}
  PIE_{M_1} & = & \sum_{m_2}\sum_{m_1}Y(0,m_1,m_2)[I(M_1(1)=m_1)-I(M_1(0)=m_1)]I(M_2(0)=m_2)\\
\\
& = & \sum_{m_2}Y(0,0,m_2)[I(M_1(1)=0)-I(M_1(0)=0)]I(M_2(0)=m_2)\\
& & + \sum_{m_2}Y(0,1,m_2)[I(M_1(1)=1)-I(M_1(0)=1)]I(M_2(0)=m_2)\\
\\
& = & - \sum_{m_2}Y(0,0,m_2)[I(M_1(1)=1)-I(M_1(0)=1)]I(M_2(0)=m_2)\\
& & + \sum_{m_2}Y(0,1,m_2)[I(M_1(1)=1)-I(M_1(0)=1)]I(M_2(0)=m_2)\\
\\
& = & \sum_{m_2}[Y(0,1,m_2)-Y(0,0,m_2)][I(M_1(1)=1)-I(M_1(0)=1)]I(M_2(0)=m_2)\\
\\
& = & \sum_{m_2}[Y(0,1,m_2)I(M_2(0)=m_2)-Y(0,0,m_2)I(M_2(0)=m_2)][M_1(1)-M_1(0)],
\end{eqnarray*}
where the third equation follows by the facts that $I(M_1(1)=0)=1-I(M_1(1)=1)$ and $I(M_1(0)=0)=1-I(M_1(0)=1)$ and the indicator function, $I(M_2(0)=m_2)$, indicates that $M_2$ is at its potential value $M_2(0)$ which may vary with respect to different individuals. \\

$PIE_{M_1}$ can be interpreted as the effect due to the mediation through $M_1$ only, conditioning on the potential value of $M_2$ with the fixed reference level $a^\ast$. \\

The pure indirect effect through $M_2$ can be rewritten as:
\begin{eqnarray*}
  PIE_{M_2} & = & Y(a^\ast,M_1(a^\ast),M_2(a))-Y(a^\ast,M_1(a^\ast),M_2(a^\ast))\\
\\
& = & \sum_{m_2}\sum_{m_1}Y(a^\ast,m_1,m_2)I(M_1(a^\ast)=m_1)I(M_2(a)=m_2)\\
& & - \sum_{m_2}\sum_{m_1}Y(a^\ast,m_1,m_2)I(M_1(a^\ast)=m_1)I(M_2(a^\ast)=m_2)\\
\\
& = & \sum_{m_2}\sum_{m_1}Y(a^\ast,m_1,m_2)I(M_1(a^\ast)=m_1)[I(M_2(a)=m_2)-I(M_2(a^\ast)=m_2)].
\end{eqnarray*}

With the specified conditions, $PIE_{M_2}$ can be written as:
\begin{eqnarray*}
  PIE_{M_2} & = & \sum_{m_2}\sum_{m_1}Y(0,m_1,m_2)I(M_1(0)=m_1)[I(M_2(1)=m_2)-I(M_2(0)=m_2)]\\
\\
& = & \sum_{m_1}Y(0,m_1,0)I(M_1(0)=m_1)[I(M_2(1)=0)-I(M_2(0)=0)]\\
& & + \sum_{m_1}Y(0,m_1,1)I(M_1(0)=m_1)[I(M_2(1)=1)-I(M_2(0)=1)]\\
\\
& = & - \sum_{m_1}Y(0,m_1,0)I(M_1(0)=m_1)[I(M_2(1)=1)-I(M_2(0)=1)]\\
& & + \sum_{m_1}Y(0,m_1,1)I(M_1(0)=m_1)[I(M_2(1)=1)-I(M_2(0)=1)]\\
\\
& = & \sum_{m_1}[Y(0,m_1,1)I(M_1(0)=m_1)-Y(0,m_1,0)I(M_1(0)=m_1)][I(M_2(1)=1)-I(M_2(0)=1)]\\
\\
& = & \sum_{m_1}[Y(0,m_1,1)I(M_1(0)=m_1)-Y(0,m_1,0)I(M_1(0)=m_1)][M_2(1)-M_2(0)],
\end{eqnarray*}
where the third equation follows by the facts that $I(M_2(1)=0)=1-I(M_2(1)=1)$ and $I(M_2(0)=0)=1-I(M_2(0)=1)$ and the indicator function, $I(M_1(0)=m_1)$, indicates that $M_1$ is at its potential value $M_1(0)$ which may vary with respect to different individuals. \\

$PIE_{M_2}$ can be interpreted as the effect due to the mediation through $M_2$ only, conditioning on the potential value of $M_1$ with the fixed reference level $a^\ast$. \\


\clearpage
\section*{Appendix B. The mediated interaction effects in a non-sequential two-mediator scenario}

Suppose we have a directed acyclic graph as shown in Figure \ref{fig3}. We show that the mediated interaction effects proposed by Bellavia and Valeri \cite{b}, if they exist (not equal to zero), are equivalent to the natural counterfactual interaction effects when $A$, $M_1$ and $M_2$ are binary with the conditions $a=1$, $a^\ast=0$, $m_1^\ast=0$, $m_2^\ast=0$ and $M_1(0)=M_2(0)=0$. 

\noindent
\emph{Proof}: 
\subsection*{mediated interaction effect between $A$ and $M_1$}
From Appendix A, we know that with the conditions $a=1$, $a^\ast=0$, $m_1^\ast=0$ and $m_2^\ast=0$ the $NatINT_{AM_1}$ can be written as:
\begin{eqnarray*}
  NatINT_{AM_1} & = & \sum_{m_2}[Y(1,1,m_2)I(M_2(0)=m_2)-Y(0,1,m_2)I(M_2(0)=m_2)\\
& & - Y(1,0,m_2)I(M_2(0)=m_2)+Y(0,0,m_2)I(M_2(0)=m_2)]\\
& & \times[M_1(1)-M_1(0)].
\end{eqnarray*}

If we apply the condition $M_2(0)=0$, the equation can be simplified to the following expression:
\begin{eqnarray*}
  NatINT_{AM_1} & = & [Y(1,1,0)I(M_2(0)=0)-Y(0,1,0)I(M_2(0)=0)\\
& & - Y(1,0,0)I(M_2(0)=0)+Y(0,0,0)I(M_2(0)=0)]\\
& & \times[M_1(1)-M_1(0)]\\
& & + [Y(1,1,1)I(M_2(0)=1)-Y(0,1,1)I(M_2(0)=1)\\
& & - Y(1,0,1)I(M_2(0)=1)+Y(0,0,1)I(M_2(0)=1)]\\
& & \times[M_1(1)-M_1(0)]\\
& = & [Y(1,1,0)-Y(0,1,0) - Y(1,0,0)+Y(0,0,0)]\times[M_1(1)-M_1(0)],
\end{eqnarray*}
where the second equality follows by the condition $M_2(0)=0$. 
This expression is identical to the mediated interaction effect between $A$ and $M_1$ proposed by Bellavia and Valeri \cite{b}.\\

\subsection*{mediated interaction effect between $A$ and $M_2$}
From Appendix A, we know that with the conditions $a=1$, $a^\ast=0$, $m_1^\ast=0$ and $m_2^\ast=0$ the $NatINT_{AM_2}$ can be written as:
\begin{eqnarray*}
  NatINT_{AM_2} & = & \sum_{m_1}[Y(1,m_1,1)I(M_1(0)=m_1)-Y(0,m_1,1)I(M_1(0)=m_1)\\
& & - Y(1,m_1,0)I(M_1(0)=m_1)+Y(0,m_1,0)I(M_1(0)=m_1)]\\
& & \times[M_2(1)-M_2(0)].
\end{eqnarray*}

If we apply the condition $M_1(0)=0$, the equation can be simplified to the following expression:
\begin{eqnarray*}
  NatINT_{AM_2} & = & [Y(1,0,1)I(M_1(0)=0)-Y(0,0,1)I(M_1(0)=0)\\
& & - Y(1,0,0)I(M_1(0)=0)+Y(0,0,0)I(M_1(0)=0)]\\
& & \times[M_2(1)-M_2(0)]\\
& & + [Y(1,1,1)I(M_1(0)=1)-Y(0,1,1)I(M_1(0)=1)\\
& & - Y(1,1,0)I(M_1(0)=1)+Y(0,1,0)I(M_1(0)=1)]\\
& & \times[M_2(1)-M_2(0)]\\
\\
& = & [Y(1,0,1)-Y(0,0,1) - Y(1,0,0) + Y(0,0,0)
\times[M_2(1)-M_2(0)],
\end{eqnarray*}
where the second equality follows by the condition $M_1(0)=0$. 
This expression is identical to the mediated interaction effect between $A$ and $M_2$ proposed by Bellavia and Valeri \cite{b}.
\subsection*{mediated interaction effect between $A$, $M_1$ and $M_2$}
From Appendix A, we know that with the conditions $a=1$, $a^\ast=0$, $m_1^\ast=0$ and $m_2^\ast=0$ the $NatINT_{AM_1M_2}$ can be written as:
\begin{eqnarray*}
  NatINT_{AM_1M_2} & = & [Y(1,1,1)-Y(0,1,1)-Y(1,0,1)+Y(0,0,1)\\
& & - Y(1,1,0)+Y(0,1,0)+Y(1,0,0)-Y(0,0,0)]\\
& & \times[M_1(1)-M_1(0)]\\
& & \times[M_2(1)-M_2(0)].
\end{eqnarray*}

If we apply the conditions $M_1(0)=0$ and $M_2(0)=0$, the equation can be simplified to the following expression:
\begin{eqnarray*}
NatINT_{AM_1M_2} & = & [Y(1,1,1)-Y(0,1,1)-Y(1,0,1)+Y(0,0,1)\\
& & - Y(1,1,0)+Y(0,1,0)+Y(1,0,0)-Y(0,0,0)]\\
& & \times[M_1(1)-0]\\
& & \times[M_2(1)-0]\\
\\
& = & [Y(1,1,1)-Y(0,1,1)-Y(1,0,1)+Y(0,0,1)\\
& & - Y(1,1,0)+Y(0,1,0)+Y(1,0,0)-Y(0,0,0)]\\
& & \times M_1(1)\times M_2(1)\\
\\
& = & [Y(1,1,1)-Y(0,1,1)-Y(1,0,1)+Y(0,0,1)\\
& & - Y(1,1,0)+Y(0,1,0)+Y(1,0,0)-Y(0,0,0)]\\
& & \times [M_1(1)M_2(1)-0]\\
\\
& = & [Y(1,1,1)-Y(0,1,1)-Y(1,0,1)+Y(0,0,1)\\
& & - Y(1,1,0)+Y(0,1,0)+Y(1,0,0)-Y(0,0,0)]\\
& & \times [M_1(1)M_2(1)-M_1(0)M_2(0)],
\end{eqnarray*}
where the last equality is identical to the mediated interaction effect between $A$, $M_1$ and $M_2$ proposed by Bellavia and Valeri \cite{b}.
\subsection*{pure natural indirect effect between $M_1$ and $M_2$ ($PNIE_{M_1M_2}$)}
From Appendix A, we know that with the conditions $a=1$, $a^\ast=0$, $m_1^\ast=0$ and $m_2^\ast=0$ the $NatINT_{M_1M_2}$ can be written as:
\begin{eqnarray*}
NatINT_{M_1M_2} & = & [Y(0,1,1)-Y(0,0,1)-Y(0,1,0)+Y(0,0,0)]\\
& & \times[M_1(1)-M_1(0)]\\
& & \times[M_2(1)-M_2(0)].
\end{eqnarray*}

If we apply the conditions $M_1(0)=0$ and $M_2(0)=0$, the equation can be simplified to the following expression:
\begin{eqnarray*}
NatINT_{M_1M_2} & = & [Y(0,1,1)-Y(0,0,1)-Y(0,1,0)+Y(0,0,0)]\\
& & \times[M_1(1)-0]\\
& & \times[M_2(1)-0]\\
\\
& = & [Y(0,1,1)-Y(0,0,1)-Y(0,1,0)+Y(0,0,0)]\\
& & \times M_1(1) \times M_2(1)\\
\\
& = & [Y(0,1,1)-Y(0,0,1)-Y(0,1,0)+Y(0,0,0)]\\
& & \times [M_1(1)M_2(1)-0]\\
\\
& = & [Y(0,1,1)-Y(0,0,1)-Y(0,1,0)+Y(0,0,0)]\\
& & \times [M_1(1)M_2(1)-M_1(0)M_2(0)],
\end{eqnarray*}
where the last equality is identical to the pure natural indirect effect between $M_1$ and $M_2$ ($PNIE_{M_1M_2}$) proposed by Bellavia and Valeri \cite{b}.
\subsection*{pure indirect effect through $M_1$}
From Appendix A, we know that with the conditions $a=1$, $a^\ast=0$, $m_1^\ast=0$ and $m_2^\ast=0$ the $PIE_{M_1}$ can be written as:
\begin{eqnarray*}
  PIE_{M_1} & = & \sum_{m_2}[Y(0,1,m_2)I(M_2(0)=m_2)-Y(0,0,m_2)I(M_2(0)=m_2)][M_1(1)-M_1(0)].
\end{eqnarray*}

If we apply the condition $M_2(0)=0$, the equation can be simplified to the following expression:
\begin{eqnarray*}
  PIE_{M_1} & = & [Y(0,1,0)-Y(0,0,0)][M_1(1)-M_1(0)],
\end{eqnarray*}
where the equality is identical to the pure natural indirect effect through $M_1$ ($PNIE_{M_1}$) proposed by Bellavia and Valeri \cite{b}.

\subsection*{pure indirect effect through $M_2$}
From Appendix A, we know that with the conditions $a=1$, $a^\ast=0$, $m_1^\ast=0$ and $m_2^\ast=0$ the $PIE_{M_2}$ can be written as:
\begin{eqnarray*}
  PIE_{M_2} & = & \sum_{m_1}[Y(0,m_1,1)I(M_1(0)=m_1)-Y(0,m_1,0)I(M_1(0)=m_1)][M_2(1)-M_2(0)].
\end{eqnarray*}

If we apply the condition $M_1(0)=0$, the equation can be simplified to the following expression:
\begin{eqnarray*}
  PIE_{M_2} & = & [Y(0,0,1)-Y(0,0,0)][M_2(1)-M_2(0)],
\end{eqnarray*}
where the equality is identical to the pure natural indirect effect through $M_2$ ($PNIE_{M_2}$) proposed by Bellavia and Valeri \cite{b}.

\subsection*{graphical comparison between the mediated interaction effect and the natural counterfactual interaction effect between $A$ and $M_1$}
With the conditions $a=1$ and $a^\ast=0$, the natural counterfactual interaction effect can be written as:
\begin{eqnarray*}
  NatINT_{AM_1} & = & Y(1,M_1(1),M_2(0))-Y(0,M_1(1),M_2(0))\\
  & & - Y(1,M_1(0),M_2(0))+Y(0,M_1(0),M_2(0)),
\end{eqnarray*}
which is illustrated in Figure \ref{fig4} B.\\

If we apply the condition $M_2(0)=0$, the natural counterfactual interaction effect will be simplified to the mediated interaction effect between $A$ and $M_1$:
\begin{eqnarray*}
  NatINT_{AM_1} & = & Y(1,M_1(1),0)-Y(0,M_1(1),0))\\
  & & - Y(1,M_1(0),0)+Y(0,M_1(0),0),
\end{eqnarray*}
which is illustrated in Figure \ref{fig4} A.

\clearpage
\section*{Appendix C. Decomposition of total effect in a sequential two-mediator scenario}

Suppose we have a directed acyclic graph as shown in Figure \ref{fig5}. We show that the total effect can be decomposed into the following 9 components at the individual level:
\begin{eqnarray*}
  TE & = & CDE(m_1^\ast,m_2^\ast)+INT_{ref\mbox{-}AM_1}(m_1^\ast,m_2^\ast)+INT_{ref\mbox{-}AM_2+AM_1M_2}(m_2^\ast)\\
  & & + NatINT_{AM_1} + NatINT_{AM_2}+ NatINT_{AM_1M_2}+ NatINT_{M_1M_2}\\
  & & + PIE_{M_1} + PIE_{M_2},
\end{eqnarray*}
where all the natural counterfactual interaction effects are listed in Definition 3. We also give the corresponding interpretation for each component.\\

\noindent
\emph{Proof}: 

We first decompose the total effect into total direct effect ($TDE$) \cite{riden}, seminatural indirect effect through $M_1$ ($SIE_{M_1}$) \cite{p14} and pure indirect effect (path-specific effect) through $M_2$ ($PIE_{M_2}$) \cite{riden,p01}.
\begin{eqnarray*}
  TE & = & Y(a)-Y(a^\ast)\\
  \\
  & = & Y(a,M_1(a),M_2(a,M_1(a)))-Y(a^\ast,M_1(a^\ast),M_2(a^\ast,M_1(a^\ast)))\\
  \\
  & = & Y(a,M_1(a),M_2(a,M_1(a)))-Y(a^\ast,M_1(a),M_2(a,M_1(a)))\\
  & & + Y(a^\ast,M_1(a),M_2(a,M_1(a))) - Y(a^\ast,M_1(a^\ast),M_2(a,M_1(a^\ast)))\\
  & & + Y(a^\ast,M_1(a^\ast),M_2(a,M_1(a^\ast)))-Y(a^\ast,M_1(a^\ast),M_2(a^\ast,M_1(a^\ast))),
\end{eqnarray*}
where the second equality follows by the composition axiom \cite{vbook,a} and the third equality follows by adding and subtracting the same identifiable counterfactual formulas. 

The formulas of $TDE$, $SIE_{M_1}$ and $PIE_{M_2}$ are presented below:
\begin{eqnarray*}
  TDE & = & Y(a,M_1(a),M_2(a,M_1(a)))-Y(a^\ast,M_1(a),M_2(a,M_1(a)))\\
  \\
  SIE_{M_1} & = & Y(a^\ast,M_1(a),M_2(a,M_1(a))) - Y(a^\ast,M_1(a^\ast),M_2(a,M_1(a^\ast)))\\
  \\
  PIE_{M_2} & = & Y(a^\ast,M_1(a^\ast),M_2(a,M_1(a^\ast)))-Y(a^\ast,M_1(a^\ast),M_2(a^\ast,M_1(a^\ast))),
\end{eqnarray*}
where $TE = TDE + SIE_{M_1}+PIE_{M_2}$.

We next focus on $TDE$ and decompose it into natural counterfactual interaction effects and pure direct effect ($PDE$) \cite{riden,p01} by subtracting $PDE$ from $TDE$, where $PDE$ satisfies the definition of a path-specific effect \cite{p01} and equals the following difference of two identifiable counterfactual formulas:
\begin{eqnarray*}
  PDE & = & Y(a,M_1(a^\ast),M_2(a^\ast,M_1(a^\ast)))-Y(a^\ast,M_1(a^\ast),M_2(a^\ast,M_1(a^\ast))).
\end{eqnarray*}

We have the following results:
\begin{eqnarray*}
  TDE-PDE & = & Y(a,M_1(a),M_2(a,M_1(a)))-Y(a^\ast,M_1(a),M_2(a,M_1(a)))\\
  & & -Y(a,M_1(a^\ast),M_2(a^\ast,M_1(a^\ast)))+Y(a^\ast,M_1(a^\ast),M_2(a^\ast,M_1(a^\ast)))\\
  \\
  & = & Y(a,M_1(a),M_2(a,M_1(a)))-Y(a^\ast,M_1(a),M_2(a,M_1(a)))\\
  & & -Y(a,M_1(a^\ast),M_2(a^\ast,M_1(a^\ast)))+Y(a^\ast,M_1(a^\ast),M_2(a^\ast,M_1(a^\ast)))\\
  & & +Y(a^\ast,M_1(a^\ast),M_2(a^\ast,M_1(a^\ast)))-Y(a^\ast,M_1(a^\ast),M_2(a^\ast,M_1(a^\ast)))\\
  & & +Y(a^\ast,M_1(a^\ast),M_2(a,M_1(a^\ast)))-Y(a^\ast,M_1(a^\ast),M_2(a,M_1(a^\ast)))\\
  & & +Y(a^\ast,M_1(a),M_2(a^\ast,M_1(a)))-Y(a^\ast,M_1(a),M_2(a^\ast,M_1(a)))\\
  & & +Y(a,M_1(a^\ast),M_2(a^\ast,M_1(a^\ast)))-Y(a,M_1(a^\ast),M_2(a^\ast,M_1(a^\ast)))\\
  & & +Y(a,M_1(a^\ast),M_2(a,M_1(a^\ast)))-Y(a,M_1(a^\ast),M_2(a,M_1(a^\ast)))\\
  & & +Y(a,M_1(a),M_2(a^\ast,M_1(a)))-Y(a,M_1(a),M_2(a^\ast,M_1(a)))\\
  \\
  & = & Y(a,M_1(a),M_2(a^\ast,M_1(a)))-Y(a^\ast,M_1(a),M_2(a^\ast,M_1(a)))\\
  & & -Y(a,M_1(a^\ast),M_2(a^\ast,M_1(a^\ast)))+Y(a^\ast,M_1(a^\ast),M_2(a^\ast,M_1(a^\ast)))\\
  & & +Y(a,M_1(a^\ast),M_2(a,M_1(a^\ast)))-Y(a^\ast,M_1(a^\ast),M_2(a,M_1(a^\ast)))\\
  & & -Y(a,M_1(a^\ast),M_2(a^\ast,M_1(a^\ast)))+Y(a^\ast,M_1(a^\ast),M_2(a^\ast,M_1(a^\ast)))\\
  & & +Y(a,M_1(a),M_2(a,M_1(a)))-Y(a^\ast,M_1(a),M_2(a,M_1(a)))\\
  & & -Y(a,M_1(a^\ast),M_2(a,M_1(a^\ast)))+Y(a^\ast,M_1(a^\ast),M_2(a,M_1(a^\ast)))\\
  & & -Y(a,M_1(a),M_2(a^\ast,M_1(a)))+Y(a^\ast,M_1(a),M_2(a^\ast,M_1(a)))\\
  & & +Y(a,M_1(a^\ast),M_2(a^\ast,M_1(a^\ast)))-Y(a^\ast,M_1(a^\ast),M_2(a^\ast,M_1(a^\ast)))
\end{eqnarray*}
where the second equality follows by adding and subtracting the same identifiable counterfactual formulas and the third equality follows by rearranging all the terms to satisfy the definition of the natural counterfactual interaction effects. 

Therefore, we have the following formulas satisfying Definition 3:
\begin{eqnarray*}
  NatINT_{AM_1} & = & Y(a,M_1(a),M_2(a^\ast,M_1(a)))-Y(a^\ast,M_1(a),M_2(a^\ast,M_1(a)))\\
  & & -Y(a,M_1(a^\ast),M_2(a^\ast,M_1(a^\ast)))+Y(a^\ast,M_1(a^\ast),M_2(a^\ast,M_1(a^\ast)))\\
  \\
  NatINT_{AM_2} & = & Y(a,M_1(a^\ast),M_2(a,M_1(a^\ast)))-Y(a^\ast,M_1(a^\ast),M_2(a,M_1(a^\ast)))\\
  & & -Y(a,M_1(a^\ast),M_2(a^\ast,M_1(a^\ast)))+Y(a^\ast,M_1(a^\ast),M_2(a^\ast,M_1(a^\ast)))\\
  \\
  NatINT_{AM_1M_2} & = & Y(a,M_1(a),M_2(a,M_1(a)))-Y(a^\ast,M_1(a),M_2(a,M_1(a)))\\
  & & -Y(a,M_1(a^\ast),M_2(a,M_1(a^\ast)))+Y(a^\ast,M_1(a^\ast),M_2(a,M_1(a^\ast)))\\
  & & -Y(a,M_1(a),M_2(a^\ast,M_1(a)))+Y(a^\ast,M_1(a),M_2(a^\ast,M_1(a)))\\
  & & +Y(a,M_1(a^\ast),M_2(a^\ast,M_1(a^\ast)))-Y(a^\ast,M_1(a^\ast),M_2(a^\ast,M_1(a^\ast))).
\end{eqnarray*}

Accordingly, $TDE$ can be decomposed into the following components:
\begin{eqnarray*}
  TDE = PDE + NatINT_{AM_1} + NatINT_{AM_2} + NatINT_{AM_1M_2}.
\end{eqnarray*}

We next focus on $PDE$ (path-specific effect) and decompose it into $CDE$ and reference interaction effects \cite{b,v4}:

\begin{eqnarray*}
  PDE & = & Y(a,M_1(a^\ast),M_2(a^\ast,M_1(a^\ast)))-Y(a^\ast,M_1(a^\ast),M_2(a^\ast,M_1(a^\ast)))\\
  \\
  & = & \sum_{m_2}\sum_{m_1} Y(a,m_1,m_2)\times I(M_1(a^\ast)=m_1)\times I(M_2(a^\ast,m_1)=m_2)\\
  & & -\sum_{m_2}\sum_{m_1} Y(a^\ast,m_1,m_2)\times I(M_1(a^\ast)=m_1)\times I(M_2(a^\ast,m_1)=m_2)\\
  \\
  & = &\sum_{m_2}\sum_{m_1}[Y(a,m_1,m_2)-Y(a^\ast,m_1,m_2)]\times I(M_1(a^\ast)=m_1)\times I(M_2(a^\ast,m_1)=m_2)\\
  \\
  & = &\sum_{m_2}\sum_{m_1}[Y(a,m_1,m_2)-Y(a^\ast,m_1,m_2)-Y(a,m_1^\ast,m_2^\ast)+Y(a^\ast,m_1^\ast,m_2^\ast)]\\
  & & \times I(M_1(a^\ast)=m_1)\times I(M_2(a^\ast,m_1)=m_2)\\
  & & +\sum_{m_2}\sum_{m_1}[Y(a,m_1^\ast,m_2^\ast)-Y(a^\ast,m_1^\ast,m_2^\ast)]\times I(M_1(a^\ast)=m_1)\times I(M_2(a^\ast,m_1)=m_2)\\
  \\
  & = &\sum_{m_2}\sum_{m_1}[Y(a,m_1,m_2)-Y(a^\ast,m_1,m_2)-Y(a,m_1^\ast,m_2^\ast)+Y(a^\ast,m_1^\ast,m_2^\ast)]\\
  & & \times I(M_1(a^\ast)=m_1)\times I(M_2(a^\ast,m_1)=m_2)\\
  & & +Y(a,m_1^\ast,m_2^\ast)-Y(a^\ast,m_1^\ast,m_2^\ast)\\
  \\
  & = & \sum_{m_2}\sum_{m_1}[Y(a,m_1,m_2)-Y(a^\ast,m_1,m_2)-Y(a,m_1^\ast,m_2^\ast)+Y(a^\ast,m_1^\ast,m_2^\ast)\\
  & & +Y(a^\ast,m_1,m_2^\ast)-Y(a^\ast,m_1,m_2^\ast)+Y(a,m_1,m_2^\ast)-Y(a,m_1,m_2^\ast)]\\
  & & \times I(M_1(a^\ast)=m_1)\times I(M_2(a^\ast,m_1)=m_2)\\
  & & +Y(a,m_1^\ast,m_2^\ast)-Y(a^\ast,m_1^\ast,m_2^\ast)\\
  \\
  & = & \sum_{m_2}\sum_{m_1} [Y(a,m_1,m_2^\ast)-Y(a^\ast,m_1,m_2^\ast)-Y(a,m_1^\ast,m_2^\ast)+Y(a^\ast,m_1^\ast,m_2^\ast)]\\
  & & \times I(M_1(a^\ast)=m_1)\times I(M_2(a^\ast,m_1)=m_2)\\
  & & + \sum_{m_2}\sum_{m_1} [Y(a,m_1,m_2)-Y(a,m_1,m_2^\ast)-Y(a^\ast,m_1,m_2)+Y(a^\ast,m_1,m_2^\ast)]\\
  & & \times I(M_1(a^\ast)=m_1)\times I(M_2(a^\ast,m_1)=m_2)\\
  & & + Y(a,m_1^\ast,m_2^\ast)-Y(a^\ast,m_1^\ast,m_2^\ast)\\
  \\
  & = & \sum_{m_1} [Y(a,m_1,m_2^\ast)-Y(a^\ast,m_1,m_2^\ast)-Y(a,m_1^\ast,m_2^\ast)+Y(a^\ast,m_1^\ast,m_2^\ast)]\\
  & & \times I(M_1(a^\ast)=m_1)\\
  & & + \sum_{m_2}\sum_{m_1} [Y(a,m_1,m_2)-Y(a,m_1,m_2^\ast)-Y(a^\ast,m_1,m_2)+Y(a^\ast,m_1,m_2^\ast)]\\
  & & \times I(M_1(a^\ast)=m_1)\times I(M_2(a^\ast,m_1)=m_2)\\
  & & + Y(a,m_1^\ast,m_2^\ast)-Y(a^\ast,m_1^\ast,m_2^\ast).
\end{eqnarray*}

According to the derivation above, the following formulas can be obtained:
\begin{eqnarray*}
  CDE(m_1^\ast,m_2^\ast) & = & Y(a,m_1^\ast,m_2^\ast)-Y(a^\ast,m_1^\ast,m_2^\ast)\\
  \\
  INT_{ref\mbox{-}AM_1}(m_1^\ast,m_2^\ast) & = & \sum_{m_1} [Y(a,m_1,m_2^\ast)-Y(a^\ast,m_1,m_2^\ast)-Y(a,m_1^\ast,m_2^\ast)+Y(a^\ast,m_1^\ast,m_2^\ast)]\\
  & & \times I(M_1(a^\ast)=m_1)\\
  \\
  INT_{ref\mbox{-}AM_2+AM_1M_2}(m_2^\ast) & = & \sum_{m_2}\sum_{m_1} [Y(a,m_1,m_2)-Y(a,m_1,m_2^\ast)-Y(a^\ast,m_1,m_2)+Y(a^\ast,m_1,m_2^\ast)]\\
  & & \times I(M_1(a^\ast)=m_1)\times I(M_2(a^\ast,m_1)=m_2).
\end{eqnarray*}

With a little mathematical derivation, $INT_{ref\mbox{-}AM_1}$ and $INT_{ref\mbox{-}AM_2+AM_1M_2}$ can be expressed in the form of the counterfactual formula:
\begin{eqnarray*}
  INT_{ref\mbox{-}AM_1}(m_1^\ast,m_2^\ast) & = & Y(a,M_1(a^\ast),m_2^\ast)-Y(a^\ast,M_1(a^\ast),m_2^\ast)-Y(a,m_1^\ast,m_2^\ast)+Y(a^\ast,m_1^\ast,m_2^\ast)\\
  \\
  INT_{ref\mbox{-}AM_2+AM_1M_2}(m_2^\ast) & = & Y(a,M_1(a^\ast),M_2(a^\ast,M_1(a^\ast)))-Y(a,M_1(a^\ast),m_2^\ast)\\
  & & - Y(a^\ast,M_1(a^\ast),M_2(a^\ast,M_1(a^\ast)))+Y(a^\ast,M_1(a^\ast),m_2^\ast)
\end{eqnarray*}

It is worth noting that $INT_{ref\mbox{-}AM_2+AM_1M_2}(m_2^\ast)$ cannot be separated into $INT_{ref\mbox{-}AM_2}(m_1^\ast,m_2^\ast)$ and $INT_{ref\mbox{-}AM_1M_2}(m_1^\ast,m_2^\ast)$ since both of the two terms are non-identifiable, which will be discussed in details in Appendix D. 

Therefore, $PDE$ can be decomposed into the following components:
\begin{eqnarray*}
  PDE = CDE(m_1^\ast,m_2^\ast) + INT_{ref\mbox{-}AM_1}(m_1^\ast,m_2^\ast) + INT_{ref\mbox{-}AM_2+AM_1M_2}(m_2^\ast).
\end{eqnarray*}

$TDE$ can be decomposed into the following components:
\begin{eqnarray*}
  TDE & = & PDE + NatINT_{AM_1} + NatINT_{AM_2}+ NatINT_{AM_1M_2}\\
  \\
      & = & CDE(m_1^\ast,m_2^\ast) + INT_{ref\mbox{-}AM_1}(m_1^\ast,m_2^\ast) + INT_{ref\mbox{-}AM_2+AM_1M_2}(m_2^\ast)\\
      & & + NatINT_{AM_1} + NatINT_{AM_2}+ NatINT_{AM_1M_2}.
\end{eqnarray*}

We next focus on $SIE_{M_1}$ and decompose it into $PIE_{M_1}$ and $NatINT_{M_1M_2}$ by subtracting $PIE_{M_1}$ from $SIE_{M_1}$:
\begin{eqnarray*}
  SIE_{M_1}-PIE_{M_1} & = & Y(a^\ast,M_1(a),M_2(a,M_1(a)))-Y(a^\ast,M_1(a^\ast),M_2(a,M_1(a^\ast)))\\
  & & -Y(a^\ast,M_1(a),M_2(a^\ast,M_1(a)))+Y(a^\ast,M_1(a^\ast),M_2(a^\ast,M_1(a^\ast)))\\
  \\
  & = & NatINT_{M_1M_2},
\end{eqnarray*}
where $NatINT_{M_1M_2}$ is listed in Definition 3.

Therefore, $SIE_{M_1}$ can be decomposed into the following components:
\begin{eqnarray*}
  SIE_{M_1} = PIE_{M_1} + NatINT_{M_1M_2}.
\end{eqnarray*}

Combining all the derivations above, we have the decomposition of total effect as follows:
\begin{eqnarray*}
  TE & = & CDE(m_1^\ast,m_2^\ast)+INT_{ref\mbox{-}AM_1}(m_1^\ast,m_2^\ast)+INT_{ref\mbox{-}AM_2+AM_1M_2}(m_2^\ast)\\
  & & + NatINT_{AM_1} + NatINT_{AM_2}+ NatINT_{AM_1M_2}+ NatINT_{M_1M_2}\\
  & & + PIE_{M_1} + PIE_{M_2}.
\end{eqnarray*}

We next present the interpretation for each component assuming binary $A$, $M_1$ and $M_2$ with the conditions $a=1$, $a^\ast=0$, $m_1^\ast=0$ and $m_2^\ast=0$ for illustration purpose. While other interpretations were proposed in the literature \cite{v4,b}, our work represent a different and more flexible interpretation from the perspective of population averages which accounts for the distribution of the mediators in the causal structure.

\subsection*{controlled direct effect}
With the specified conditions, the controlled direct effect can be written as:
\begin{eqnarray*}
CDE(m_1^\ast,m_2^\ast) & = & Y(a,m_1^\ast,m_2^\ast)-Y(a^\ast,m_1^\ast,m_2^\ast)\\
\\
\Rightarrow \quad\quad CDE(0,0) & = & Y(1,0,0)-Y(0,0,0).
\end{eqnarray*}

$CDE(m_1^\ast,m_2^\ast)$ can be interpreted as the effect due to neither mediation nor interaction.
\subsection*{reference interaction effects}
With the specified conditions, the reference interaction effect between $A$ and $M_1$ can be written as:
\begin{eqnarray*}
INT_{ref\mbox{-}AM_1}(m_1^\ast,m_2^\ast) & = & \sum_{m_1}[Y(a,m_1,m_2^\ast)-Y(a^\ast,m_1,m_2^\ast)-Y(a,m_1^\ast,m_2^\ast)+Y(a^\ast,m_1^\ast,m_2^\ast)]\\
& & \times I(M_1(a^\ast)=m_1)\\
\\
\Rightarrow INT_{ref\mbox{-}AM_1}(0,0) & = & \sum_{m_1}[Y(1,m_1,0)-Y(0,m_1,0)-Y(1,0,0)+Y(0,0,0)]\times I(M_1(0)=m_1)\\
\\
& = & [Y(1,0,0)-Y(0,0,0)-Y(1,0,0)+Y(0,0,0)]\times I(M_1(0)=0)\\
& & + [Y(1,1,0)-Y(0,1,0)-Y(1,0,0)+Y(0,0,0)]\times I(M_1(0)=1)\\
\\
& = & [Y(1,1,0)-Y(0,1,0)-Y(1,0,0)+Y(0,0,0)]\times I(M_1(0)=1)\\
\\
& = & [Y(1,1,0)-Y(0,1,0)-Y(1,0,0)+Y(0,0,0)]\times M_1(0).
\end{eqnarray*}

$INT_{ref\mbox{-}AM_1}(m_1^\ast,m_2^\ast)$ can be interpreted as the effect due to the interaction between $A$ and $M_1$ only.\\

The sum of reference interaction effect between $A$ and $M_2$ and reference interaction effect between $A$, $M_1$ and $M_2$ can be written as:
\begin{eqnarray*}
INT_{ref\mbox{-}AM_2+AM_1M_2}(m_2^\ast) & = & \sum_{m_2}\sum_{m_1}[Y(a,m_1,m_2)-Y(a,m_1,m_2^\ast)-Y(a^\ast,m_1,m_2)+Y(a^\ast,m_1,m_2^\ast)]\\
& & \times I(M_1(a^\ast)=m_1)\times I(M_2(a^\ast,m_1)=m_2)\\
\\
\Rightarrow INT_{ref\mbox{-}AM_2+AM_1M_2}(0) & = & \sum_{m_2}\sum_{m_1}[Y(1,m_1,m_2)-Y(1,m_1,0)-Y(0,m_1,m_2)+Y(0,m_1,0)]\\
& & \times I(M_1(0)=m_1)\times I(M_2(0,m_1)=m_2)\\
\\
& = & \sum_{M_2}[Y(1,0,m_2)-Y(1,0,0)-Y(0,0,m_2)+Y(0,0,0)]\\& & \times I(M_1(0)=0)\times I(M_2(0,0)=m_2)\\
& & + \sum_{M_2}[Y(1,1,m_2)-Y(1,1,0)-Y(0,1,m_2)+Y(0,1,0)]\\& & \times I(M_1(0)=1)\times I(M_2(0,1)=m_2)\\
\\
& = & [Y(1,0,0)-Y(1,0,0)-Y(0,0,0)+Y(0,0,0)]\\
& & \times I(M_1(0)=0)\times I(M_2(0,0)=0)\\
& & + [Y(1,0,1)-Y(1,0,0)-Y(0,0,1)+Y(0,0,0)]\\
& & \times I(M_1(0)=0)\times I(M_2(0,0)=1)\\
& & + [Y(1,1,0)-Y(1,1,0)-Y(0,1,0)+Y(0,1,0)]\\
& & \times I(M_1(0)=1)\times I(M_2(0,1)=0)\\
& & + [Y(1,1,1)-Y(1,1,0)-Y(0,1,1)+Y(0,1,0)]\\
& & \times I(M_1(0)=1)\times I(M_2(0,1)=1)\\
\\
& = & [Y(1,0,1)-Y(1,0,0)-Y(0,0,1)+Y(0,0,0)]\\
& & \times I(M_1(0)=0)\times I(M_2(0,0)=1)\\
& & + [Y(1,1,1)-Y(1,1,0)-Y(0,1,1)+Y(0,1,0)]\\
& & \times I(M_1(0)=1)\times I(M_2(0,1)=1)\\
\\
& = & [Y(1,0,1)-Y(1,0,0)-Y(0,0,1)+Y(0,0,0)]\\
& & \times [1-I(M_1(0)=1)]\times I(M_2(0,0)=1)\\
& & + [Y(1,1,1)-Y(1,1,0)-Y(0,1,1)+Y(0,1,0)]\\
& & \times I(M_1(0)=1)\times I(M_2(0,1)=1)\\
\\
& = & [Y(1,0,1)-Y(1,0,0)-Y(0,0,1)+Y(0,0,0)]\\
& & \times [1-M_1(0)]\times M_2(0,0)\\
& & + [Y(1,1,1)-Y(1,1,0)-Y(0,1,1)+Y(0,1,0)]\\
& & \times M_1(0)\times M_2(0,1).\\
\end{eqnarray*}

$INT_{ref\mbox{-}AM_2+AM_1M_2}(m_2^\ast)$ can be interpreted as the effect due to the interaction between $A$ and $M_2$ only, conditioning on the potential value of $M_1$ with the fixed reference level $a^\ast$.\\
\subsection*{natural counterfactual interaction effects}
The natural counterfactual interaction effect between $A$ and $M_1$ can be rewritten as: 
\begin{eqnarray*}
  NatINT_{AM_1} & = & Y(a,M_1(a),M_2(a^\ast,M_1(a)))-Y(a^\ast,M_1(a),M_2(a^\ast,M_1(a)))\\
  & & - Y(a,M_1(a^\ast),M_2(a^\ast,M_1(a^\ast)))+Y(a^\ast,M_1(a^\ast),M_2(a^\ast,M_1(a^\ast)))\\
\\
& = & \sum_{m_2}\sum_{m_1}Y(a,m_1,m_2)I(M_1(a)=m_1)I(M_2(a^\ast,m_1)=m_2)\\
& & - \sum_{m_2}\sum_{m_1}Y(a^\ast,m_1,m_2)I(M_1(a)=m_1)I(M_2(a^\ast,m_1)=m_2)\\
& & - \sum_{m_2}\sum_{m_1}Y(a,m_1,m_2)I(M_1(a^\ast)=m_1)I(M_2(a^\ast,m_1)=m_2)\\
& & + \sum_{m_2}\sum_{m_1}Y(a^\ast,m_1,m_2)I(M_1(a^\ast)=m_1)I(M_2(a^\ast,m_1)=m_2)\\
\\
& = & \sum_{m_2}\sum_{m_1}[Y(a,m_1,m_2)-Y(a^\ast,m_1,m_2)]I(M_1(a)=m_1)I(M_2(a^\ast,m_1)=m_2)\\
& & - \sum_{m_2}\sum_{m_1}[Y(a,m_1,m_2)-Y(a^\ast,m_1,m_2)]I(M_1(a^\ast)=m_1)I(M_2(a^\ast,m_1)=m_2)\\
\\
& = & \sum_{m_2}\sum_{m_1}[Y(a,m_1,m_2)-Y(a^\ast,m_1,m_2)]\times[I(M_1(a)=m_1)-I(M_1(a^\ast)=m_1)]\\
& & \times I(M_2(a^\ast,m_1)=m_2)\\
\\
& = & \sum_{m_2}\sum_{m_1}[Y(a,m_1,m_2)I(M_2(a^\ast,m_1)=m_2)-Y(a^\ast,m_1,m_2)I(M_2(a^\ast,m_1)=m_2)]\\
& & \times [I(M_1(a)=m_1)-I(M_1(a^\ast)=m_1)]\\
\\
& = & \sum_{m_2}\sum_{m_1}[Y(a,m_1,m_2)I(M_2(a^\ast,m_1)=m_2)-Y(a^\ast,m_1,m_2)I(M_2(a^\ast,m_1)=m_2)\\
& & - Y(a,m_1^\ast,m_2)I(M_2(a^\ast,m_1^\ast)=m_2)-Y(a^\ast,m_1^\ast,m_2)I(M_2(a^\ast,m_1^\ast)=m_2)]\\
& & \times [I(M_1(a)=m_1)-I(M_1(a^\ast)=m_1)],
\end{eqnarray*}
where the last equation follows by adding two extra terms which do not change the value of $NatINT_{AM_1}$.\\

With the specified conditions, $NatINT_{AM_1}$ can be written as:
\begin{eqnarray*}
  NatINT_{AM_1} & = & \sum_{m_2}\sum_{m_1}[Y(1,m_1,m_2)I(M_2(0,m_1)=m_2)-Y(0,m_1,m_2)I(M_2(0,m_1)=m_2)\\
& & - Y(1,0,m_2)I(M_2(0,0)=m_2)+Y(0,0,m_2)I(M_2(0,0)=m_2)]\\
& & \times[I(M_1(1)=m_1)-I(M_1(0)=m_1)]\\
\\
& = & \sum_{m_2}[Y(1,0,m_2)I(M_2(0,0)=m_2)-Y(0,0,m_2)I(M_2(0,0)=m_2)\\
& & - Y(1,0,m_2)I(M_2(0,0)=m_2)+Y(0,0,m_2)I(M_2(0,0)=m_2)]\\
& & \times[I(M_1(1)=0)-I(M_1(0)=0)]\\
& & + \sum_{m_2}[Y(1,1,m_2)I(M_2(0,1)=m_2)-Y(0,1,m_2)I(M_2(0,1)=m_2)\\
& & - Y(1,0,m_2)I(M_2(0,0)=m_2)+Y(0,0,m_2)I(M_2(0,0)=m_2)]\\
& & \times[I(M_1(1)=1)-I(M_1(0)=1)]\\
\\
& = & \sum_{m_2}[Y(1,1,m_2)I(M_2(0,1)=m_2)-Y(0,1,m_2)I(M_2(0,1)=m_2)\\
& & - Y(1,0,m_2)I(M_2(0,0)=m_2)+Y(0,0,m_2)I(M_2(0,0)=m_2)]\\
& & \times[I(M_1(1)=1)-I(M_1(0)=1)]\\
\\
& = & \sum_{m_2}[Y(1,1,m_2)I(M_2(0,1)=m_2)-Y(0,1,m_2)I(M_2(0,1)=m_2)\\
& & - Y(1,0,m_2)I(M_2(0,0)=m_2)+Y(0,0,m_2)I(M_2(0,0)=m_2)]\\
& & \times[M_1(1)-M_1(0)],
\end{eqnarray*}
where the indicator functions $I(M_2(0,1)=m_2)$ and $I(M_2(0,0)=m_2)$ indicate that $M_2$ is at its potential values $M_2(0,1)$ and $M_2(0,0)$ which may vary with respect to different individuals. \\

$NatINT_{AM_1}$ can be interpreted as the effect due to the mediation through $M_1$ and the interaction between $A$ and $M_1$ conditioning on the potential values of $M_2$ with the fixed reference level $a^\ast$.\\

The natural counterfactual interaction effect between $A$ and $M_2$ can be rewritten as: 
\begin{eqnarray*}  
  NatINT_{AM_2} & = & Y(a,M_1(a^\ast),M_2(a,M_1(a^\ast)))-Y(a^\ast,M_1(a^\ast),M_2(a,M_1(a^\ast)))\\
  & & -Y(a,M_1(a^\ast),M_2(a^\ast,M_1(a^\ast)))+Y(a^\ast,M_1(a^\ast),M_2(a^\ast,M_1(a^\ast)))\\
\\
& = & \sum_{m_2}\sum_{m_1}Y(a,m_1,m_2)I(M_1(a^\ast)=m_1)I(M_2(a,m_1)=m_2)\\
& & - \sum_{m_2}\sum_{m_1}Y(a^\ast,m_1,m_2)I(M_1(a^\ast)=m_1)I(M_2(a,m_1)=m_2)\\
& & - \sum_{m_2}\sum_{m_1}Y(a,m_1,m_2)I(M_1(a^\ast)=m_1)I(M_2(a^\ast,m_1)=m_2)\\
& & + \sum_{m_2}\sum_{m_1}Y(a^\ast,m_1,m_2)I(M_1(a^\ast)=m_1)I(M_2(a^\ast,m_1)=m_2)\\
\\
& = & \sum_{m_2}\sum_{m_1}[Y(a,m_1,m_2)-Y(a^\ast,m_1,m_2)]I(M_1(a^\ast)=m_1)I(M_2(a,m_1)=m_2)\\
& & - \sum_{m_2}\sum_{m_1}[Y(a,m_1,m_2)-Y(a^\ast,m_1,m_2)]I(M_1(a^\ast)=m_1)I(M_2(a^\ast,m_1)=m_2)\\
\\
& = & \sum_{m_2}\sum_{m_1}[Y(a,m_1,m_2)-Y(a^\ast,m_1,m_2)]I(M_1(a^\ast)=m_1)\\
& & \times[I(M_2(a,m_1)=m_2)-I(M_2(a^\ast,m_1)=m_2)]\\
\\
& = & \sum_{m_2}\sum_{m_1}[Y(a,m_1,m_2)I(M_1(a^\ast)=m_1)-Y(a^\ast,m_1,m_2)I(M_1(a^\ast)=m_1)]\\
& & \times[I(M_2(a,m_1)=m_2)-I(M_2(a^\ast,m_1)=m_2)]\\
\\
& = & \sum_{m_2}\sum_{m_1}[Y(a,m_1,m_2)I(M_1(a^\ast)=m_1)-Y(a^\ast,m_1,m_2)I(M_1(a^\ast)=m_1)\\
& & - Y(a,m_1,m_2^\ast)I(M_1(a^\ast)=m_1) + Y(a^\ast,m_1,m_2^\ast)I(M_1(a^\ast)=m_1)]\\
& & \times[I(M_2(a,m_1)=m_2)-I(M_2(a^\ast,m_1)=m_2)],
\end{eqnarray*}
where the last equation follows by adding two extra terms which do not change the value of $NatINT_{AM_2}$.\\

With the specified conditions, $NatINT_{AM_2}$ can be written as:
\begin{eqnarray*}
  NatINT_{AM_2} & = & \sum_{m_2}\sum_{m_1}[Y(1,m_1,m_2)I(M_1(0)=m_1)-Y(0,m_1,m_2)I(M_1(0)=m_1)\\
& & - Y(1,m_1,0)I(M_1(0)=m_1)+Y(0,m_1,0)I(M_1(0)=m_1)]\\
& & \times[I(M_2(1,m_1)=m_2)-I(M_2(0,m_1)=m_2)]\\
\\
& = & \sum_{m_1}[Y(1,m_1,0)I(M_1(0)=m_1)-Y(0,m_1,0)I(M_1(0)=m_1)\\
& & - Y(1,m_1,0)I(M_1(0)=m_1)+Y(0,m_1,0)I(M_1(0)=m_1)]\\
& & \times[I(M_2(1,m_1)=0)-I(M_2(0,m_1)=0)]\\
& & + \sum_{m_1}[Y(1,m_1,1)I(M_1(0)=m_1)-Y(0,m_1,1)I(M_1(0)=m_1)\\
& & - Y(1,m_1,0)I(M_1(0)=m_1)+Y(0,m_1,0)I(M_1(0)=m_1)]\\
& & \times[I(M_2(1,m_1)=1)-I(M_2(0,m_1)=1)]\\
\\
& = & \sum_{m_1}[Y(1,m_1,1)I(M_1(0)=m_1)-Y(0,m_1,1)I(M_1(0)=m_1)\\
& & - Y(1,m_1,0)I(M_1(0)=m_1)+Y(0,m_1,0)I(M_1(0)=m_1)]\\
& & \times[I(M_2(1,m_1)=1)-I(M_2(0,m_1)=1)]\\
\\
& = & \sum_{m_1}[Y(1,m_1,1)I(M_1(0)=m_1)-Y(0,m_1,1)I(M_1(0)=m_1)\\
& & - Y(1,m_1,0)I(M_1(0)=m_1)+Y(0,m_1,0)I(M_1(0)=m_1)]\\
& & \times[M_2(1,m_1)-M_2(0,m_1)],
\end{eqnarray*}
where the indicator function $I(M_1(0)=m_1)$ indicates that $M_1$ is at its potential value $M_1(0)$ which may vary with respect to different individuals. \\

$NatINT_{AM_2}$ can be interpreted as the effect due to the mediation through $M_2$ and the interaction between $A$ and $M_2$, conditioning on the potential value of $M_1$ with the fixed reference level $a^\ast$.\\

The natural counterfactual interaction effect between $A$, $M_1$ and $M_2$ can be rewritten as:
\begin{eqnarray*}
  NatINT_{AM_1M_2} & = & Y(a,M_1(a),M_2(a,M_1(a)))-Y(a^\ast,M_1(a),M_2(a,M_1(a)))\\
  & & - Y(a,M_1(a^\ast),M_2(a,M_1(a^\ast)))+Y(a^\ast,M_1(a^\ast),M_2(a,M_1(a^\ast)))\\
  & & -Y(a,M_1(a),M_2(a^\ast,M_1(a)))+Y(a^\ast,M_1(a),M_2(a^\ast,M_1(a)))\\
  & & +Y(a,M_1(a^\ast),M_2(a^\ast,M_1(a^\ast)))-Y(a^\ast,M_1(a^\ast),M_2(a^\ast,M_1(a^\ast)))\\
\\
& = & \sum_{m_2}\sum_{m_1}Y(a,m_1,m_2)I(M_1(a)=m_1)I(M_2(a,m_1)=m_2)\\
& & - \sum_{m_2}\sum_{m_1}Y(a^\ast,m_1,m_2)I(M_1(a)=m_1)I(M_2(a,m_1)=m_2)\\
& & - \sum_{m_2}\sum_{m_1}Y(a,m_1,m_2)I(M_1(a^\ast)=m_1)I(M_2(a,m_1)=m_2)\\
& & + \sum_{m_2}\sum_{m_1}Y(a^\ast,m_1,m_2)I(M_1(a^\ast)=m_1)I(M_2(a,m_1)=m_2)\\
& & - \sum_{m_2}\sum_{m_1}Y(a,m_1,m_2)I(M_1(a)=m_1)I(M_2(a^\ast,m_1)=m_2)\\
& & + \sum_{m_2}\sum_{m_1}Y(a^\ast,m_1,m_2)I(M_1(a)=m_1)I(M_2(a^\ast,m_1)=m_2)\\
& & + \sum_{m_2}\sum_{m_1}Y(a,m_1,m_2)I(M_1(a^\ast)=m_1)I(M_2(a^\ast,m_1)=m_2)\\
& & - \sum_{m_2}\sum_{m_1}Y(a^\ast,m_1,m_2)I(M_1(a^\ast)=m_1)I(M_2(a^\ast,m_1)=m_2)\\
\\
& = & \sum_{m_2}\sum_{m_1}[Y(a,m_1,m_2)-Y(a^\ast,m_1,m_2)]I(M_1(a)=m_1)I(M_2(a,m_1)=m_2)\\
& & - \sum_{m_2}\sum_{m_1}[Y(a,m_1,m_2)-Y(a^\ast,m_1,m_2)]I(M_1(a^\ast)=m_1)I(M_2(a,m_1)=m_2)\\
& & - \sum_{m_2}\sum_{m_1}[Y(a,m_1,m_2)-Y(a^\ast,m_1,m_2)]I(M_1(a)=m_1)I(M_2(a^\ast,m_1)=m_2)\\
& & + \sum_{m_2}\sum_{m_1}[Y(a,m_1,m_2)-Y(a^\ast,m_1,m_2)]I(M_1(a^\ast)=m_1)I(M_2(a^\ast,m_1)=m_2)\\
\\
& = & \sum_{m_2}\sum_{m_1}[Y(a,m_1,m_2)-Y(a^\ast,m_1,m_2)][I(M_1(a)=m_1)-I(M_1(a^\ast)=m_1)]\\
& & \times[I(M_2(a,m_1)=m_2)-I(M_2(a^\ast,m_1)=m_2)]\\
\\
& = & \sum_{m_2}\sum_{m_1}[Y(a,m_1,m_2)-Y(a^\ast,m_1,m_2)- Y(a,m_1,m_2^\ast)+Y(a^\ast,m_1,m_2^\ast)\\
& & + Y(a,m_1^\ast,m_2^\ast)-Y(a^\ast,m_1^\ast,m_2^\ast)]\\
& & \times[I(M_1(a)=m_1)-I(M_1(a^\ast)=m_1)]\\
& & \times[I(M_2(a,m_1)=m_2)-I(M_2(a^\ast,m_1)=m_2)]\\
& & + \sum_{m_2}\sum_{m_1}[-Y(a,m_1^\ast,m_2)+Y(a^\ast,m_1^\ast,m_2)]\\
& & \times[I(M_1(a)=m_1)-I(M_1(a^\ast)=m_1)]\\
& & \times[I(M_2(a,m_1^\ast)=m_2)-I(M_2(a^\ast,m_1^\ast)=m_2)],
\end{eqnarray*}
where the last equation follows by adding six extra terms which do not change the value of $NatINT_{AM_1M_2}$.\\

With the specified conditions, $NatINT_{AM_1M_2}$ can be written as: 
\begin{eqnarray*}
  NatINT_{AM_1M_2} & = & \sum_{m_2}\sum_{m_1}[Y(1,m_1,m_2)-Y(0,m_1,m_2)- Y(1,m_1,0)+Y(0,m_1,0) + Y(1,0,0)-Y(0,0,0)]\\
& & \times[I(M_1(1)=m_1)-I(M_1(0)=m_1)]\\
& & \times[I(M_2(1,m_1)=m_2)-I(M_2(0,m_1)=m_2)]\\
& & + \sum_{m_2}\sum_{m_1}[-Y(1,0,m_2)+Y(0,0,m_2)]\\
& & \times[I(M_1(1)=m_1)-I(M_1(0)=m_1)]\\
& & \times[I(M_2(1,0)=m_2)-I(M_2(0,0)=m_2)]\\
\\
& = & \sum_{m_2} [Y(1,0,m_2)-Y(0,0,m_2)- Y(1,0,0)+Y(0,0,0) + Y(1,0,0)-Y(0,0,0)]\\
& & \times[I(M_1(1)=0)-I(M_1(0)=0)]\\
& & \times[I(M_2(1,0)=m_2)-I(M_2(0,0)=m_2)]\\
& & + \sum_{m_2} [Y(1,1,m_2)-Y(0,1,m_2)- Y(1,1,0)+Y(0,1,0)+ Y(1,0,0)-Y(0,0,0)]\\
& & \times[I(M_1(1)=1)-I(M_1(0)=1)]\\
& & \times[I(M_2(1,1)=m_2)-I(M_2(0,1)=m_2)]\\
& & + \sum_{m_2}[-Y(1,0,m_2)+Y(0,0,m_2)]\\
& & \times[I(M_1(1)=0)-I(M_1(0)=0)]\\
& & \times[I(M_2(1,0)=m_2)-I(M_2(0,0)=m_2)]\\
& & + \sum_{m_2}[-Y(1,0,m_2)+Y(0,0,m_2)]\\
& & \times[I(M_1(1)=1)-I(M_1(0)=1)]\\
& & \times[I(M_2(1,0)=m_2)-I(M_2(0,0)=m_2)]\\
\\
& = & \sum_{m_2} [Y(1,1,m_2)-Y(0,1,m_2)- Y(1,1,0)+Y(0,1,0) + Y(1,0,0)-Y(0,0,0)]\\
& & \times[I(M_1(1)=1)-I(M_1(0)=1)]\\
& & \times[I(M_2(1,1)=m_2)-I(M_2(0,1)=m_2)]\\
& & + \sum_{m_2}[-Y(1,0,m_2)+Y(0,0,m_2)]\\
& & \times[I(M_1(1)=1)-I(M_1(0)=1)]\\
& & \times[I(M_2(1,0)=m_2)-I(M_2(0,0)=m_2)]\\
\\
& = & [Y(1,1,0)-Y(0,1,0)- Y(1,1,0)+Y(0,1,0) + Y(1,0,0)-Y(0,0,0)]\\
& & \times[I(M_1(1)=1)-I(M_1(0)=1)]\\
& & \times[I(M_2(1,1)=0)-I(M_2(0,1)=0)]\\
& & + [Y(1,1,1)-Y(0,1,1)- Y(1,1,0)+Y(0,1,0) + Y(1,0,0)-Y(0,0,0)]\\
& & \times[I(M_1(1)=1)-I(M_1(0)=1)]\\
& & \times[I(M_2(1,1)=1)-I(M_2(0,1)=1)]\\
& & + [-Y(1,0,0)+Y(0,0,0)]\\
& & \times[I(M_1(1)=1)-I(M_1(0)=1)]\\
& & \times[I(M_2(1,0)=0)-I(M_2(0,0)=0)]\\
& & + [-Y(1,0,1)+Y(0,0,1)]\\
& & \times[I(M_1(1)=1)-I(M_1(0)=1)]\\
& & \times[I(M_2(1,0)=1)-I(M_2(0,0)=1)]\\
\\
& = & [Y(1,0,0)-Y(0,0,0)]\\
& & \times[I(M_1(1)=1)-I(M_1(0)=1)]\\
& & \times[I(M_2(1,1)=0)-I(M_2(0,1)=0)]\\
& & + [Y(1,1,1)-Y(0,1,1)- Y(1,1,0)+Y(0,1,0) + Y(1,0,0)-Y(0,0,0)]\\
& & \times[I(M_1(1)=1)-I(M_1(0)=1)]\\
& & \times[I(M_2(1,1)=1)-I(M_2(0,1)=1)]\\
& & + [-Y(1,0,0)+Y(0,0,0)]\\
& & \times[I(M_1(1)=1)-I(M_1(0)=1)]\\
& & \times[I(M_2(1,0)=0)-I(M_2(0,0)=0)]\\
& & + [-Y(1,0,1)+Y(0,0,1)]\\
& & \times[I(M_1(1)=1)-I(M_1(0)=1)]\\
& & \times[I(M_2(1,0)=1)-I(M_2(0,0)=1)]\\
\\
& = & -[Y(1,0,0)-Y(0,0,0)]\\
& & \times[I(M_1(1)=1)-I(M_1(0)=1)]\\
& & \times[I(M_2(1,1)=1)-I(M_2(0,1)=1)]\\
& & + [Y(1,1,1)-Y(0,1,1)- Y(1,1,0)+Y(0,1,0) + Y(1,0,0)-Y(0,0,0)]\\
& & \times[I(M_1(1)=1)-I(M_1(0)=1)]\\
& & \times[I(M_2(1,1)=1)-I(M_2(0,1)=1)]\\
& & + [Y(1,0,0)-Y(0,0,0)]\\
& & \times[I(M_1(1)=1)-I(M_1(0)=1)]\\
& & \times[I(M_2(1,0)=1)-I(M_2(0,0)=1)]\\
& & + [-Y(1,0,1)+Y(0,0,1)]\\
& & \times[I(M_1(1)=1)-I(M_1(0)=1)]\\
& & \times[I(M_2(1,0)=1)-I(M_2(0,0)=1)]\\
\\
& = & [Y(1,1,1)-Y(0,1,1)- Y(1,1,0)+Y(0,1,0)]\\
& & \times[I(M_1(1)=1)-I(M_1(0)=1)]\\
& & \times[I(M_2(1,1)=1)-I(M_2(0,1)=1)]\\
& & + [Y(1,0,0)-Y(0,0,0)]\\
& & \times[I(M_1(1)=1)-I(M_1(0)=1)]\\
& & \times[I(M_2(1,0)=1)-I(M_2(0,0)=1)]\\
& & + [-Y(1,0,1)+Y(0,0,1)]\\
& & \times[I(M_1(1)=1)-I(M_1(0)=1)]\\
& & \times[I(M_2(1,0)=1)-I(M_2(0,0)=1)]\\
\\
& = & [Y(1,1,1)-Y(0,1,1)- Y(1,1,0)+Y(0,1,0)]\\
& & \times[I(M_1(1)=1)-I(M_1(0)=1)]\\
& & \times[I(M_2(1,1)=1)-I(M_2(0,1)=1)]\\
& & - [Y(1,0,1)-Y(0,0,1)-Y(1,0,0)+Y(0,0,0)]\\
& & \times[I(M_1(1)=1)-I(M_1(0)=1)]\\
& & \times[I(M_2(1,0)=1)-I(M_2(0,0)=1)]\\
\\
& = & [Y(1,1,1)-Y(0,1,1)- Y(1,1,0)+Y(0,1,0)]\\
& & \times[M_1(1)-M_1(0)]\\
& & \times[M_2(1,1)-M_2(0,1)]\\
& & + [-Y(1,0,1)+Y(0,0,1)+Y(1,0,0)-Y(0,0,0)]\\
& & \times[M_1(1)-M_1(0)]\\
& & \times[M_2(1,0)-M_2(0,0)],
\end{eqnarray*}
where the six equality follows by the facts that $I(M_2(1,1)=0)=1-I(M_2(1,1)=1)$ and $I(M_2(0,1)=0)=1-I(M_2(0,1)=1)$. \\

$NatINT_{AM_1M_2}$ can be interpreted as the effect due to mediation through both $M_1$ and $M_2$, and the interaction between $A$, $M_1$ and $M_2$. \\

The natural counterfactual interaction effect between $M_1$ and $M_2$ can be rewritten as: 
\begin{eqnarray*}
  NatINT_{M_1M_2} & = & Y(a^\ast,M_1(a),M_2(a,M_1(a)))-Y(a^\ast,M_1(a^\ast),M_2(a,M_1(a^\ast)))\\
  & & -Y(a^\ast,M_1(a),M_2(a^\ast,M_1(a)))+Y(a^\ast,M_1(a^\ast),M_2(a^\ast,M_1(a^\ast)))\\
  \\
  & = & \sum_{m_2}\sum_{m_1} Y(a^\ast,m_1,m_2)I(M_1(a)=m_1)I(M_2(a,m_1)=m_2)\\
  & & - \sum_{m_2}\sum_{m_1} Y(a^\ast,m_1,m_2)I(M_1(a^\ast)=m_1)I(M_2(a,m_1)=m_2)\\
  & & - \sum_{m_2}\sum_{m_1} Y(a^\ast,m_1,m_2)I(M_1(a)=m_1)I(M_2(a^\ast,m_1)=m_2)\\
  & & + \sum_{m_2}\sum_{m_1} Y(a^\ast,m_1,m_2)I(M_1(a^\ast)=m_1)I(M_2(a^\ast,m_1)=m_2)\\
\\
& = & \sum_{m_2}\sum_{m_1} Y(a^\ast,m_1,m_2)[I(M_1(a)=m_1)-I(M_1(a^\ast)=m_1)]I(M_2(a,m_1)=m_2)\\
& & - \sum_{m_2}\sum_{m_1} Y(a^\ast,m_1,m_2)[I(M_1(a)=m_1)-I(M_1(a^\ast)=m_1)]I(M_2(a^\ast,m_1)=m_2)\\
\\
& = & \sum_{m_2}\sum_{m_1} Y(a^\ast,m_1,m_2)[I(M_1(a)=m_1)-I(M_1(a^\ast)=m_1)]\\
& & \times [I(M_2(a,m_1)=m_2)-I(M_2(a^\ast,m_1)=m_2)]\\
\\
& = & \sum_{m_2}\sum_{m_1} [Y(a^\ast,m_1,m_2)-Y(a^\ast,m_1,m_2^\ast)+Y(a^\ast,m_1^\ast,m_2^\ast)]\\
& & \times [I(M_1(a)=m_1)-I(M_1(a^\ast)=m_1)]\\
& & \times [I(M_2(a,m_1)=m_2)-I(M_2(a^\ast,m_1)=m_2)]\\
& & + \sum_{m_2}\sum_{m_1} [-Y(a^\ast,m_1^\ast,m_2)]\\
& & \times [I(M_1(a)=m_1)-I(M_1(a^\ast)=m_1)]\\
& & \times [I(M_2(a,m_1^\ast)=m_2)-I(M_2(a^\ast,m_1^\ast)=m_2)],
\end{eqnarray*}
where the last equality follows by adding three extra terms which do not change the value of $NatINT_{M_1M_2}$.\\

With the specified conditions, $NatINT_{M_1M_2}$ can be written as:
\begin{eqnarray*}
  NatINT_{M_1M_2} & = & \sum_{m_2}\sum_{m_1} [Y(0,m_1,m_2)-Y(0,m_1,0)+Y(0,0,0)]\\
& & \times [I(M_1(1)=m_1)-I(M_1(0)=m_1)]\\
& & \times [I(M_2(1,m_1)=m_2)-I(M_2(0,m_1)=m_2)]\\
& & + \sum_{m_2}\sum_{m_1} [-Y(0,0,m_2)]\\
& & \times [I(M_1(1)=m_1)-I(M_1(0)=m_1)]\\
& & \times [I(M_2(1,0)=m_2)-I(M_2(0,0)=m_2)]\\
\\
& = & \sum_{m_2} [Y(0,0,m_2)-Y(0,0,0)+Y(0,0,0)]\\
& & \times [I(M_1(1)=0)-I(M_1(0)=0)]\\
& & \times [I(M_2(1,0)=m_2)-I(M_2(0,0)=m_2)]\\
& & + \sum_{m_2}[Y(0,1,m_2)-Y(0,1,0)+Y(0,0,0)]\\
& & \times [I(M_1(1)=1)-I(M_1(0)=1)]\\
& & \times [I(M_2(1,1)=m_2)-I(M_2(0,1)=m_2)]\\
& & + \sum_{m_2} [-Y(0,0,m_2)]\\
& & \times [I(M_1(1)=0)-I(M_1(0)=0)]\\
& & \times [I(M_2(1,0)=m_2)-I(M_2(0,0)=m_2)]\\
& & + \sum_{m_2} [-Y(0,0,m_2)]\\
& & \times [I(M_1(1)=1)-I(M_1(0)=1)]\\
& & \times [I(M_2(1,0)=m_2)-I(M_2(0,0)=m_2)]\\
\\
& & = \sum_{m_2}[Y(0,1,m_2)-Y(0,1,0)+Y(0,0,0)]\\
& & \times [I(M_1(1)=1)-I(M_1(0)=1)]\\
& & \times [I(M_2(1,1)=m_2)-I(M_2(0,1)=m_2)]\\
& & + \sum_{m_2} [-Y(0,0,m_2)]\\
& & \times [I(M_1(1)=1)-I(M_1(0)=1)]\\
& & \times [I(M_2(1,0)=m_2)-I(M_2(0,0)=m_2)]\\
\\
& = & [Y(0,1,0)-Y(0,1,0)+Y(0,0,0)]\\
& & \times [I(M_1(1)=1)-I(M_1(0)=1)]\\
& & \times [I(M_2(1,1)=0)-I(M_2(0,1)=0)]\\
& & + [Y(0,1,1)-Y(0,1,0)+Y(0,0,0)]\\
& & \times [I(M_1(1)=1)-I(M_1(0)=1)]\\
& & \times [I(M_2(1,1)=1)-I(M_2(0,1)=1)]\\
& & + [-Y(0,0,0)]\\
& & \times [I(M_1(1)=1)-I(M_1(0)=1)]\\
& & \times [I(M_2(1,0)=0)-I(M_2(0,0)=0)]\\
& & + [-Y(0,0,1)]\\
& & \times [I(M_1(1)=1)-I(M_1(0)=1)]\\
& & \times [I(M_2(1,0)=1)-I(M_2(0,0)=1)]\\
\\
& = & [-Y(0,0,0)]\\
& & \times [I(M_1(1)=1)-I(M_1(0)=1)]\\
& & \times [I(M_2(1,1)=1)-I(M_2(0,1)=1)]\\
& & + [Y(0,1,1)-Y(0,1,0)+Y(0,0,0)]\\
& & \times [I(M_1(1)=1)-I(M_1(0)=1)]\\
& & \times [I(M_2(1,1)=1)-I(M_2(0,1)=1)]\\
& & + [Y(0,0,0)]\\
& & \times [I(M_1(1)=1)-I(M_1(0)=1)]\\
& & \times [I(M_2(1,0)=1)-I(M_2(0,0)=1)]\\
& & + [-Y(0,0,1)]\\
& & \times [I(M_1(1)=1)-I(M_1(0)=1)]\\
& & \times [I(M_2(1,0)=1)-I(M_2(0,0)=1)]\\
\\
& = & [Y(0,1,1)-Y(0,1,0)]\\
& & \times [I(M_1(1)=1)-I(M_1(0)=1)]\\
& & \times [I(M_2(1,1)=1)-I(M_2(0,1)=1)]\\
& & + [-Y(0,0,1)+Y(0,0,0)]\\
& & \times [I(M_1(1)=1)-I(M_1(0)=1)]\\
& & \times [I(M_2(1,0)=1)-I(M_2(0,0)=1)]\\
\\
& = & [Y(0,1,1)-Y(0,1,0)]\times [M_1(1)-M_1(0)]\times [M_2(1,1)-M_2(0,1)]\\
& & + [-Y(0,0,1)+Y(0,0,0)] \times [M_1(1)-M_1(0)]\times [M_2(1,0)-M_2(0,0)],
\end{eqnarray*}
where the fifth equality follows by the facts that $I(M_2(1,1)=0)=1-I(M_2(1,1)=1)$ and $I(M_2(0,1)=0)=1-I(M_2(0,1)=1)$. \\

$NatINT_{M_1M_2}$ can be interpreted as the effect due to mediation through both $M_1$ and $M_2$, and the interaction between $M_1$ and $M_2$. Since the interaction is not involved with the change in exposure $A$, the interpretation can be simply put as the effect due to the mediation through both $M_1$ and $M_2$ only.\\
\subsection*{pure indirect effects}
The pure indirect effect through $M_1$ can be rewritten as:
\begin{eqnarray*}
  PIE_{M_1} & = & Y(a^\ast,M_1(a),M_2(a^\ast,M_1(a)))-Y(a^\ast,M_1(a^\ast),M_2(a^\ast,M_1(a^\ast)))\\
\\
& = & \sum_{m_2}\sum_{m_1}Y(a^\ast,m_1,m_2)I(M_1(a)=m_1)I(M_2(a^\ast,m_1)=m_2)\\
& & - \sum_{m_2}\sum_{m_1}Y(a^\ast,m_1,m_2)I(M_1(a^\ast)=m_1)I(M_2(a^\ast,m_1)=m_2)\\
\\
& = & \sum_{m_2}\sum_{m_1}Y(a^\ast,m_1,m_2)\times [I(M_1(a)=m_1)-I(M_1(a^\ast)=m_1)] \times I(M_2(a^\ast,m_1)=m_2).
\end{eqnarray*}

With the specified conditions, $PIE_{M_1}$ can be written as:
\begin{eqnarray*}
  PIE_{M_1} & = & \sum_{m_2}\sum_{m_1}Y(0,m_1,m_2)\times [I(M_1(1)=m_1)-I(M_1(0)=m_1)]\times I(M_2(0,m_1)=m_2)\\
\\
& = & \sum_{m_2} Y(0,0,m_2)\times [I(M_1(1)=0)-I(M_1(0)=0)] \times I(M_2(0,0)=m_2)\\
& & + \sum_{m_2} Y(0,1,m_2)\times [I(M_1(1)=1)-I(M_1(0)=1)]\times I(M_2(0,1)=m_2)\\
\\
& = & -\sum_{m_2} Y(0,0,m_2)\times [I(M_1(1)=1)-I(M_1(0)=1)]\times I(M_2(0,0)=m_2)\\
& & + \sum_{m_2} Y(0,1,m_2)\times [I(M_1(1)=1)-I(M_1(0)=1)]\times I(M_2(0,1)=m_2)\\
\\
& = & \sum_{m_2} [Y(0,1,m_2)I(M_2(0,1)=m_2)-Y(0,0,m_2)I(M_2(0,0)=m_2)]\\
& & \times [I(M_1(1)=1)-I(M_1(0)=1)]\\
\\
& = & \sum_{m_2} [Y(0,1,m_2)I(M_2(0,1)=m_2)-Y(0,0,m_2)I(M_2(0,0)=m_2)]\\
& & \times [M_1(1)-M_1(0)],
\end{eqnarray*}
where the third equation follows by the facts that $I(M_1(1)=0)=1-I(M_1(1)=1)$ and $I(M_1(0)=0)=1-I(M_1(0)=1)$ and the indicator functions, $I(M_2(0,1)=m_2)$ and  $I(M_2(0,0)=m_2)$, indicate that $M_2$ is at its potential values which may vary with respect to different individuals. \\

$PIE_{M_1}$ can be interpreted as the effect due to the mediation through $M_1$ only, conditioning on the potential values of $M_2$ with the fixed reference level $a^\ast$. \\

The pure indirect effect through $M_2$ can be rewritten as:
\begin{eqnarray*}
  PIE_{M_2} & = & Y(a^\ast,M_1(a^\ast),M_2(a,M_1(a^\ast)))-Y(a^\ast,M_1(a^\ast),M_2(a^\ast,M_1(a^\ast)))\\
\\
& = & \sum_{m_2}\sum_{m_1}Y(a^\ast,m_1,m_2)I(M_1(a^\ast)=m_1)I(M_2(a,m_1)=m_2)\\
& & - \sum_{m_2}\sum_{m_1}Y(a^\ast,m_1,m_2)I(M_1(a^\ast)=m_1)I(M_2(a^\ast,m_1)=m_2)\\
\\
& = & \sum_{m_2}\sum_{m_1}Y(a^\ast,m_1,m_2)\times I(M_1(a^\ast)=m_1) \times [I(M_2(a,m_1)=m_2)-I(M_2(a^\ast,m_1)=m_2)].
\end{eqnarray*}

With the specified conditions, $PIE_{M_2}$ can be written as:
\begin{eqnarray*}
  PIE_{M_1} & = & \sum_{m_2}\sum_{m_1}Y(0,m_1,m_2)\times I(M_1(0)=m_1)\times [I(M_2(1,m_1)=m_2)-I(M_2(0,m_1)=m_2)]\\
\\
& = & \sum_{m_1}Y(0,m_1,0)\times I(M_1(0)=m_1)\times [I(M_2(1,m_1)=0)-I(M_2(0,m_1)=0)]\\
& & + \sum_{m_1}Y(0,m_1,1)\times I(M_1(0)=m_1)\times [I(M_2(1,m_1)=1)-I(M_2(0,m_1)=1)]\\
\\
& = & -\sum_{m_1}Y(0,m_1,0)\times I(M_1(0)=m_1)\times [I(M_2(1,m_1)=1)-I(M_2(0,m_1)=1)]\\
& & + \sum_{m_1}Y(0,m_1,1)\times I(M_1(0)=m_1)\times [I(M_2(1,m_1)=1)-I(M_2(0,m_1)=1)]\\
\\
& = & \sum_{m_1}[Y(0,m_1,1)\times I(M_1(0)=m_1)-Y(0,m_1,0)\times I(M_1(0)=m_1)]\\
& & \times [I(M_2(1,m_1)=1)-I(M_2(0,m_1)=1)]\\
\\
& = & \sum_{m_1}[Y(0,m_1,1)\times I(M_1(0)=m_1)-Y(0,m_1,0)\times I(M_1(0)=m_1)]\\
& & \times [M_2(1,m_1)-M_2(0,m_1)],
\end{eqnarray*}
where the third equation follows by the facts that $I(M_2(1,m_1)=0)=1-I(M_2(1,m_1)=1)$ and $I(M_2(0,m_1)=0)=1-I(M_2(0,m_1)=1)$ and the indicator functions, $I(M_1(0)=m_1)$, indicates that $M_1$ is at its potential values which may vary with respect to different individuals. \\

$PIE_{M_2}$ can be interpreted as the effect due to the mediation through $M_2$ only, conditioning on the potential values of $M_1$ with the fixed reference level $a^\ast$. 

\clearpage
\section*{Appendix D. Non-identifiability issues of $INT_{ref\mbox{-}AM_2}(m_1^\ast,m_2^\ast)$ \\
and $INT_{ref\mbox{-}AM_1M_2}(m_1^\ast,m_2^\ast)$ in a sequential two-mediator scenario}

We show that the reference interaction effects, $INT_{ref\mbox{-}AM_2}(m_1^\ast,m_2^\ast)$ and $INT_{ref\mbox{-}AM_1M_2}(m_1^\ast,m_2^\ast)$, are non-identifiable in a sequential two-mediator scenario as shown in Figure \ref{fig5}.\\

\noindent
\emph{Proof}: 

We first decompose $INT_{ref\mbox{-}AM_2+AM_1M_2}(m_2^\ast)$ into $INT_{ref\mbox{-}AM_2}(m_1^\ast,m_2^\ast)$ and $INT_{ref\mbox{-}AM_1M_2}(m_1^\ast,m_2^\ast)$: 
\begin{eqnarray*}
INT_{ref\mbox{-}AM_2+AM_1M_2}(m_2^\ast) & = & \sum_{m_2}\sum_{m_1} [Y(a,m_1,m_2)-Y(a,m_1,m_2^\ast)-Y(a^\ast,m_1,m_2)+Y(a^\ast,m_1,m_2^\ast)]\\
  & & \times I(M_1(a^\ast)=m_1)\times I(M_2(a^\ast,m_1)=m_2)\\
  \\
  & = & \sum_{m_2}\sum_{m_1} [Y(a,m_1,m_2)-Y(a,m_1,m_2^\ast)-Y(a^\ast,m_1,m_2)+Y(a^\ast,m_1,m_2^\ast)\\
  & & + Y(a^\ast,m_1^\ast,m_2^\ast)-Y(a^\ast,m_1^\ast,m_2^\ast)+Y(a,m_1^\ast,m_2^\ast)-Y(a,m_1^\ast,m_2^\ast)\\
  & & +Y(a^\ast,m_1^\ast,m_2)-Y(a^\ast,m_1^\ast,m_2)+Y(a,m_1^\ast,m_2)-Y(a,m_1^\ast,m_2)]\\
  & & \times I(M_1(a^\ast)=m_1)\times I(M_2(a^\ast,m_1)=m_2)\\
  \\
  & = & \sum_{m_2}\sum_{m_1}[Y(a,m_1^\ast,m_2)-Y(a^\ast,m_1^\ast,m_2)-Y(a,m_1^\ast,m_2^\ast)+Y(a^\ast,m_1^\ast,m_2^\ast)]\\
  & & \times I(M_1(a^\ast)=m_1)\times I(M_2(a^\ast,m_1)=m_2)\\
  & & +\sum_{m_2}\sum_{m_1}[Y(a,m_1,m_2)-Y(a^\ast,m_1,m_2)-Y(a,m_1^\ast,m_2)+Y(a^\ast,m_1^\ast,m_2)\\
  & & -Y(a,m_1,m_2^\ast)+Y(a^\ast,m_1,m_2^\ast)+Y(a,m_1^\ast,m_2^\ast)-Y(a^\ast,m_1^\ast,m_2^\ast)]\\
  & & \times I(M_1(a^\ast)=m_1)\times I(M_2(a^\ast,m_1)=m_2).
\end{eqnarray*}

Therefore, we have the following formulas:
\begin{eqnarray*}
  INT_{ref\mbox{-}AM_2}(m_1^\ast,m_2^\ast) & = & \sum_{m_2}\sum_{m_1}[Y(a,m_1^\ast,m_2)-Y(a^\ast,m_1^\ast,m_2)-Y(a,m_1^\ast,m_2^\ast)+Y(a^\ast,m_1^\ast,m_2^\ast)]\\
  & & \times I(M_1(a^\ast)=m_1)\times I(M_2(a^\ast,m_1)=m_2)\\
  \\
  INT_{ref\mbox{-}AM_1M_2}(m_1^\ast,m_2^\ast) & = & \sum_{m_2}\sum_{m_1}[Y(a,m_1,m_2)-Y(a^\ast,m_1,m_2)-Y(a,m_1^\ast,m_2)+Y(a^\ast,m_1^\ast,m_2)\\
  & & -Y(a,m_1,m_2^\ast)+Y(a^\ast,m_1,m_2^\ast)+Y(a,m_1^\ast,m_2^\ast)-Y(a^\ast,m_1^\ast,m_2^\ast)]\\
  & & \times I(M_1(a^\ast)=m_1)\times I(M_2(a^\ast,m_1)=m_2).
\end{eqnarray*}

It can be seen that both formulas include the following term:
\begin{eqnarray*}
  \sum_{m_2}\sum_{m_1}Y(a,m_1^\ast,m_2)\times I(M_1(a^\ast)=m_1)\times I(M_2(a^\ast,m_1)=m_2),
\end{eqnarray*}
which can be rewritten as the counterfactual formula $Y(a,m_1^\ast,M_2(a^\ast,M_1(a^\ast)))$. 

Note that $m_1^\ast$ is an arbitrary reference level of $M_2$. Let us consider an instance that there exists $a^{\ast\ast}\neq a^\ast$ such that $M_1(a^{\ast\ast})=m_1^\ast$. In this case, the counterfactual formula can be rewritten as $Y(a,M_1(a^{\ast\ast}),M_2(a^\ast,M_1(a^\ast)))$, where $M_1$ is being activated by two different values of exposure $A$ in the kite graph formed up by the path $A\rightarrow M_1\rightarrow Y$ and the path $A\rightarrow M_1\rightarrow M_2 \rightarrow Y$ in Figure \ref{fig5}. Avin et al. \cite{a} showed that such counterfactual formulas are non-identifiable and referred to as problematic counterfactual formulas. Because the instance cannot be ruled out in any certain population, $Y(a,m_1^\ast,M_2(a^\ast,M_1(a^\ast)))$ is non-identifiable. Therefore, $INT_{ref\mbox{-}AM_2}(m_1^\ast,m_2^\ast)$ and $INT_{ref\mbox{-}AM_1M_2}(m_1^\ast,m_2^\ast)$ are non-identifiable.

\clearpage
\section*{Appendix E. Linear regression models with continuous outcome and continuous mediators in a sequential two-mediator scenario}

Suppose we have a directed acyclic graph as shown in Figure \ref{fig5}. Assume the following linear models for $Y$, $M_2$ and $M_1$ are correctly specified:
\begin{eqnarray*}
E[Y|A,M_1,M_2,C] & = & \theta_0 + \theta_1A + \theta_2M_1 + \theta_3M_2 + \theta_4AM_1 + \theta_5AM_2 + \theta_6M_1M_2\\ 
& & + \theta_7AM_1M_2 + \theta_8^\prime C\\
\\
E[M_2|A,M_1,C] & = & \beta_0 + \beta_1A + \beta_2M_1 + \beta_3AM_1 + \beta_4^\prime C\\
\\
E[M_1|A,C] & = & \gamma_0 + \gamma_1A + \gamma_2^\prime C, 
\end{eqnarray*}
where $C$ is a sufficient confounding set that satisfies the identification assumptions $(A1)$-$(A6)$; $\epsilon_Y$, $\epsilon_{M_2}$ and $\epsilon_{M_1}$ denote independent random error terms for $Y$, $M_2$ and $M_1$ and follow $N(0,\sigma_{Y}^2)$, $N(0,\sigma_{M_2}^2)$ and $N(0,\sigma_{M_1}^2)$, respectively. According to Appendix C, the total effect can be decomposed into the following components:
\begin{eqnarray*}
  TE & = & CDE(m_1^\ast,m_2^\ast)+INT_{ref\mbox{-}AM_1}(m_1^\ast,m_2^\ast)+INT_{ref\mbox{-}AM_2+AM_1M_2}(m_2^\ast)\\
  & & + NatINT_{AM_1} + NatINT_{AM_2}+ NatINT_{AM_1M_2}+ NatINT_{M_1M_2}\\
  & & + PIE_{M_1} + PIE_{M_2}.
\end{eqnarray*}

The expected value of each component conditional on the sufficient confounding set are presented in the following.

\subsection*{Controlled direct effect}
\begin{eqnarray*}
CDE(m_1^\ast,m_2^\ast) & = & Y(a,m_1^\ast,m_2^\ast)-Y(a^\ast,m_1^\ast,m_2^\ast)
\end{eqnarray*}
\begin{eqnarray*}
&\Rightarrow & E[CDE(m_1^\ast,m_2^\ast)|c]\\
\\
 & = & E[Y(a,m_1^\ast,m_2^\ast)-Y(a^\ast,m_1^\ast,m_2^\ast)|c]\\
 \\
 & = & E[Y(a,m_1^\ast,m_2^\ast)|c]-E[Y(a^\ast,m_1^\ast,m_2^\ast)|c]\\
 \\
 & = & E[Y(a,m_1^\ast,m_2^\ast)|a,c]-E[Y(a^\ast,m_1^\ast,m_2^\ast)|a^\ast,c]\quad by\; A1\\
 \\
 & = & E[Y(a,m_1^\ast,m_2^\ast)|a,m_1^\ast,m_2^\ast,c]-E[Y(a^\ast,m_1^\ast,m_2^\ast)|a^\ast,m_1^\ast,m_2^\ast,c]\quad by\; A2\\
 \\
 & = & E[Y|a,m_1^\ast,m_2^\ast,c]-E[Y|a^\ast,m_1^\ast,m_2^\ast,c]\quad by\; consistency\\
 \\
 &= & (\theta_0+\theta_1a+\theta_2m_1^\ast+\theta_3m_2^\ast+\theta_4am_1^\ast+\theta_5am_2^\ast+\theta_6m_1^\ast m_2^\ast+\theta_7am_1^\ast m_2^\ast+\theta_8^\prime c)\\
 & & -(\theta_0+\theta_1a^\ast+\theta_2m_1^\ast+\theta_3m_2^\ast+\theta_4a^\ast m_1^\ast+\theta_5a^\ast m_2^\ast+\theta_6m_1^\ast m_2^\ast+\theta_7a^\ast m_1^\ast m_2^\ast+\theta_8^\prime c)\\
 \\
 & = & (\theta_1a+\theta_4am_1^\ast+\theta_5am_2^\ast+\theta_7am_1^\ast m_2^\ast)-(\theta_1a^\ast+\theta_4a^\ast m_1^\ast+\theta_5a^\ast m_2^\ast+\theta_7a^\ast m_1^\ast m_2^\ast)\\
 \\
 & = & \theta_1\left(a-a^\ast\right)+\theta_4m_1^\ast\left(a-a^\ast\right)+\theta_5m_2^\ast\left(a-a^\ast\right)+\theta_7m_1^\ast m_2^\ast\left(a-a^\ast\right)\\
 \\
 & = & \left(\theta_1+\theta_4m_1^\ast+\theta_5m_2^\ast+\theta_7m_1^\ast m_2^\ast\right)\left(a-a^\ast\right).
\end{eqnarray*}
\\
\subsection*{Reference interaction effect between $A$ and $M_1$}
We first consider $M_1$ as a categorical random variable.
\begin{eqnarray*}
INT_{ref\mbox{-}AM_1}(m_1^\ast,m_2^\ast) & = & \sum_{m_1} [Y(a,m_1,m_2^\ast)-Y(a^\ast,m_1,m_2^\ast)-Y(a,m_1^\ast,m_2^\ast)+Y(a^\ast,m_1^\ast,m_2^\ast)]\\
  & & \times I(M_1(a^\ast)=m_1)\\
\end{eqnarray*}
\begin{eqnarray*}
&\Rightarrow & E[INT_{ref\mbox{-}AM_1}(m_1^\ast,m_2^\ast)|c]\\
\\
& = & E\left[\sum_{m_1} [Y(a,m_1,m_2^\ast)-Y(a^\ast,m_1,m_2^\ast)-Y(a,m_1^\ast,m_2^\ast)+Y(a^\ast,m_1^\ast,m_2^\ast)]\times I(M_1(a^\ast)=m_1)\bigg| c\right]\\
  \\
  & = & \sum_{m_1}E\left[[Y(a,m_1,m_2^\ast)-Y(a^\ast,m_1,m_2^\ast)-Y(a,m_1^\ast,m_2^\ast)+Y(a^\ast,m_1^\ast,m_2^\ast)]\times I(M_1(a^\ast)=m_1)|c\right]\\
  \\
  & = & \sum_{m_1}E\left[Y(a,m_1,m_2^\ast)-Y(a^\ast,m_1,m_2^\ast)-Y(a,m_1^\ast,m_2^\ast)+Y(a^\ast,m_1^\ast,m_2^\ast)|c\right]\\
  & &\times E\left[I(M_1(a^\ast)=m_1)|c\right]\quad by\; A4\\
  \\
  & = & \sum_{m_1}E\left[Y(a,m_1,m_2^\ast)-Y(a^\ast,m_1,m_2^\ast)-Y(a,m_1^\ast,m_2^\ast)+Y(a^\ast,m_1^\ast,m_2^\ast)|c\right]\\
  & &\times \Pr(M_1(a^\ast)=m_1|c)\\
  \\
  & = & \sum_{m_1}E\left[Y(a,m_1,m_2^\ast)-Y(a^\ast,m_1,m_2^\ast)-Y(a,m_1^\ast,m_2^\ast)+Y(a^\ast,m_1^\ast,m_2^\ast)|c\right]\\
  & &\times \Pr(M_1(a^\ast)=m_1|a^\ast,c)\quad by\; A3\\
  \\
  & = & \sum_{m_1}E\left[Y(a,m_1,m_2^\ast)-Y(a^\ast,m_1,m_2^\ast)-Y(a,m_1^\ast,m_2^\ast)+Y(a^\ast,m_1^\ast,m_2^\ast)|c\right]\\
  & &\times \Pr(M_1=m_1|a^\ast,c)\quad by\; consistency\\
  \\
  & = & \sum_{m_1}{E\left[Y\left(a,m_1,m_2^\ast\right)\middle| c\right]}\Pr{\left(M_1=m_1\middle| a^\ast,c\right)}-\sum_{m_1}{E\left[Y\left(a^\ast,m_1,m_2^\ast\right)\middle| c\right]}\Pr{\left(M_1=m_1\middle| a^\ast,c\right)}\\
  & & -\sum_{m_1}{E\left[Y\left(a,m_1^\ast,m_2^\ast\right)\middle| c\right]}\Pr{\left(M_1=m_1\middle| a^\ast,c\right)}+\sum_{m_1}{E\left[Y\left(a^\ast,m_1^\ast,m_2^\ast\right)\middle| c\right]}\Pr{\left(M_1=m_1\middle| a^\ast,c\right)}\\
  \\
  & = & \sum_{m_1}{E\left[Y\left(a,m_1,m_2^\ast\right)\middle|a,m_1,m_2^\ast,c\right]}\Pr{\left(M_1=m_1\middle| a^\ast,c\right)}\\
  & & -\sum_{m_1}{E\left[Y\left(a^\ast,m_1,m_2^\ast\right)\middle|a^\ast,m_1,m_2^\ast, c\right]}\Pr{\left(M_1=m_1\middle| a^\ast,c\right)}\\
  & & -\sum_{m_1}{E\left[Y\left(a,m_1^\ast,m_2^\ast\right)\middle|a,m_1^\ast,m_2^\ast, c\right]}\Pr{\left(M_1=m_1\middle| a^\ast,c\right)}\\
  & & +\sum_{m_1}{E\left[Y\left(a^\ast,m_1^\ast,m_2^\ast\right)\middle|a^\ast,m_1^\ast,m_2^\ast, c\right]}\Pr{\left(M_1=m_1\middle| a^\ast,c\right)}\quad by\; A1\;A2\\
  \\
  & = & \sum_{m_1}{E\left[Y|a,m_1,m_2^\ast,c\right]}\Pr{\left(M_1=m_1\middle| a^\ast,c\right)}-\sum_{m_1}{E\left[Y|a^\ast,m_1,m_2^\ast, c\right]}\Pr{\left(M_1=m_1\middle| a^\ast,c\right)}\\
  & & -\sum_{m_1}{E\left[Y|a,m_1^\ast,m_2^\ast\, c\right]}\Pr{\left(M_1=m_1\middle| a^\ast,c\right)}+\sum_{m_1}{E\left[Y|a^\ast,m_1^\ast,m_2^\ast, c\right]}\Pr{\left(M_1=m_1\middle| a^\ast,c\right)}\; by\; consistency
\end{eqnarray*}

We next extend the formula to consider a continuous $M_1$. 
\begin{eqnarray*}
& & E[INT_{ref\mbox{-}AM_1}(m_1^\ast,m_2^\ast)|c] \\
\\
& = & \int_{m_1}{E\left[Y\middle| a,m_1,m_2^\ast,c\right]d\Pr{\left(M_1=m_1\middle| a^\ast,c\right)}}\\
& & -\int_{m_1}{E\left[Y\middle| a^\ast,m_1,m_2^\ast,c\right]d\Pr{\left(M_1=m_1\middle| a^\ast,c\right)}}\\
& & -\int_{m_1}{E\left[Y\middle| a,m_1^\ast,m_2^\ast,c\right]d\Pr{\left(M_1=m_1\middle| a^\ast,c\right)}}\\
& & +\int_{m_1}{E\left[Y\middle| a^\ast,m_1^\ast,m_2^\ast,c\right]d\Pr{\left(M_1=m_1\middle| a^\ast,c\right)}}\\
\\
& = & \int_{m_1}(\theta_0+\theta_1a+\theta_2m_1+\theta_3m_2^\ast+\theta_4am_1+\theta_5am_2^\ast\\
& & +\theta_6m_1m_2^\ast+\theta_7am_1m_2^\ast+\theta_8^\prime c)d\Pr{\left(M_1=m_1\middle| a^\ast,c\right)})\\
& & -\int_{m_1}(\theta_0+\theta_1a^\ast+\theta_2m_1+\theta_3m_2^\ast+\theta_4a^\ast m_1+\theta_5a^\ast m_2^\ast\\
& & +\theta_6m_1m_2^\ast+\theta_7a^\ast m_1m_2^\ast+\theta_8^\prime c)d\Pr{\left(M_1=m_1\middle| a^\ast,c\right)}\\
& & -\int_{m_1}(\theta_0+\theta_1a+\theta_2m_1^\ast+\theta_3m_2^\ast+\theta_4am_1^\ast+\theta_5am_2^\ast\\
& & +\theta_6m_1^\ast m_2^\ast+\theta_7am_1^\ast m_2^\ast+\theta_8^\prime c)d\Pr{\left(M_1=m_1\middle| a^\ast,c\right)}\\
& & + \int_{m_1}(\theta_0+\theta_1a^\ast+\theta_2m_1^\ast+\theta_3m_2^\ast+\theta_4a^\ast m_1^\ast+\theta_5a^\ast m_2^\ast\\
& & +\theta_6m_1^\ast m_2^\ast+\theta_7a^\ast m_1^\ast m_2^\ast+\theta_8^\prime c)d\Pr{\left(M_1=m_1\middle| a^\ast,c\right)}\\
\\
& = & \left(\theta_0+\theta_1a+\theta_3m_2^\ast+\theta_5am_2^\ast+\theta_8^\prime c\right)+\left(\theta_2+\theta_4a+\theta_6m_2^\ast+\theta_7am_2^\ast\right)\\
& & \times\left(\gamma_0+\gamma_1a^\ast+\gamma_2^\prime c\right)\\
& & -\left(\theta_0+\theta_1a^\ast+\theta_3m_2^\ast+\theta_5a^\ast m_2^\ast+\theta_8^\prime c\right)-\left(\theta_2+\theta_4a^\ast+\theta_6m_2^\ast+\theta_7a^\ast m_2^\ast\right)\\
& & \times\left(\gamma_0+\gamma_1a^\ast+\gamma_2^\prime c\right)\\
& & -\left(\theta_0+\theta_1a+\theta_2m_1^\ast+\theta_3m_2^\ast+\theta_4am_1^\ast+\theta_5am_2^\ast+\theta_6m_1^\ast m_2^\ast+\theta_7am_1^\ast m_2^\ast+\theta_8^\prime c\right)\\
& & +\left(\theta_0+\theta_1a^\ast+\theta_2m_1^\ast+\theta_3m_2^\ast+\theta_4a^\ast m_1^\ast+\theta_5a^\ast m_2^\ast+\theta_6m_1^\ast m_2^\ast+\theta_7a^\ast m_1^\ast m_2^\ast+\theta_8^\prime c\right)\\
\\
& = & \left(\gamma_0+\gamma_1a^\ast+\gamma_2^\prime c-m_1^\ast\right)\times\left(\theta_4+\theta_7m_2^\ast\right)\times\left(a-a^\ast\right).
\end{eqnarray*}
\\
\subsection*{The sum of two reference interaction effects: $INT_{ref\mbox{-}AM_2+AM_1M_2}(m_2^\ast)$}
\begin{eqnarray*}
INT_{ref\mbox{-}AM_2+AM_1M_2}(m_2^\ast) & = & \sum_{m_2}\sum_{m_1} [Y(a,m_1,m_2)-Y(a,m_1,m_2^\ast)-Y(a^\ast,m_1,m_2)+Y(a^\ast,m_1,m_2^\ast)]\\
  & & \times I(M_1(a^\ast)=m_1)\times I(M_2(a^\ast,m_1)=m_2)
\end{eqnarray*}
\begin{eqnarray*}
&\Rightarrow & E[INT_{ref\mbox{-}AM_2+AM_1M_2}(m_2^\ast)|c]\\
\\
& = & E\left[\sum_{m_2}\sum_{m_1} [Y(a,m_1,m_2)-Y(a,m_1,m_2^\ast)-Y(a^\ast,m_1,m_2)+Y(a^\ast,m_1,m_2^\ast)]\right. \\
& & \left.\times I(M_1(a^\ast)=m_1)\times I(M_2(a^\ast,m_1)=m_2)\bigg|c\right]\\
\\
& = & \sum_{m_2}\sum_{m_1}E\left[[Y(a,m_1,m_2)-Y(a,m_1,m_2^\ast)-Y(a^\ast,m_1,m_2)+Y(a^\ast,m_1,m_2^\ast)]\right.\\
& &\left. \times I(M_1(a^\ast)=m_1)\times I(M_2(a^\ast,m_1)=m_2)|c\right]\\
\\
& = & \sum_{m_2}\sum_{m_1}E[Y(a,m_1,m_2)-Y(a,m_1,m_2^\ast)-Y(a^\ast,m_1,m_2)+Y(a^\ast,m_1,m_2^\ast)|c]\\
& & \times E[I(M_1(a^\ast)=m_1)\times I(M_2(a^\ast,m_1)=m_2)|c]\; by\; A4\;A6\\
\\
& = & \sum_{m_2}\sum_{m_1}E[Y(a,m_1,m_2)-Y(a,m_1,m_2^\ast)-Y(a^\ast,m_1,m_2)+Y(a^\ast,m_1,m_2^\ast)|c]\\
& & \times \Pr(M_1(a^\ast)=m_1|c)\times \Pr(M_2(a^\ast,m_1)=m_2)|c)\\
\\
& = & \sum_{m_2}\sum_{m_1}E[Y(a,m_1,m_2)-Y(a,m_1,m_2^\ast)-Y(a^\ast,m_1,m_2)+Y(a^\ast,m_1,m_2^\ast)|c]\\
& & \times \Pr(M_1(a^\ast)=m_1|a^\ast,c)\times \Pr(M_2(a^\ast,m_1)=m_2)|a^\ast,m_1,c)\; by\; A3\;A5\\
\\
& = & \sum_{m_2}\sum_{m_1}E[Y(a,m_1,m_2)-Y(a,m_1,m_2^\ast)-Y(a^\ast,m_1,m_2)+Y(a^\ast,m_1,m_2^\ast)|c]\\
& & \times \Pr(M_1=m_1|a^\ast,c)\times \Pr(M_2=m_2|a^\ast,m_1,c)\; by\;consistency\\
\\
& = & \sum_{m_2}\sum_{m_1}E[Y(a,m_1,m_2)|c]\times \Pr(M_1=m_1|a^\ast,c)\times \Pr(M_2=m_2|a^\ast,m_1,c)\\
& & - \sum_{m_2}\sum_{m_1}E[Y(a,m_1,m_2^\ast)|c]\times \Pr(M_1=m_1|a^\ast,c)\times \Pr(M_2=m_2|a^\ast,m_1,c)\\
& & - \sum_{m_2}\sum_{m_1}E[Y(a^\ast,m_1,m_2)|c]\times \Pr(M_1=m_1|a^\ast,c)\times \Pr(M_2=m_2|a^\ast,m_1,c)\\
& & + \sum_{m_2}\sum_{m_1}E[Y(a^\ast,m_1,m_2^\ast)|c]\times \Pr(M_1=m_1|a^\ast,c)\times \Pr(M_2=m_2|a^\ast,m_1,c)\\
\\
& = & \sum_{m_2}\sum_{m_1}E[Y(a,m_1,m_2)|a,m_1,m_2,c]\times \Pr(M_1=m_1|a^\ast,c)\times \Pr(M_2=m_2|a^\ast,m_1,c)\\
& & - \sum_{m_2}\sum_{m_1}E[Y(a,m_1,m_2^\ast)|a,m_1,m_2^\ast,c]\times \Pr(M_1=m_1|a^\ast,c)\times \Pr(M_2=m_2|a^\ast,m_1,c)\\
& & - \sum_{m_2}\sum_{m_1}E[Y(a^\ast,m_1,m_2)|a^\ast,m_1,m_2,c]\times \Pr(M_1=m_1|a^\ast,c)\times \Pr(M_2=m_2|a^\ast,m_1,c)\\
& & + \sum_{m_2}\sum_{m_1}E[Y(a^\ast,m_1,m_2^\ast)|a^\ast,m_1,m_2^\ast,c]\times \Pr(M_1=m_1|a^\ast,c)\times \Pr(M_2=m_2|a^\ast,m_1,c)\; by\;A1\;A2\\
\\
& = & \sum_{m_2}\sum_{m_1}E[Y|a,m_1,m_2,c]\times \Pr(M_1=m_1|a^\ast,c)\times \Pr(M_2=m_2|a^\ast,m_1,c)\\
& & - \sum_{m_2}\sum_{m_1}E[Y|a,m_1,m_2^\ast,c]\times \Pr(M_1=m_1|a^\ast,c)\times \Pr(M_2=m_2|a^\ast,m_1,c)\\
& & - \sum_{m_2}\sum_{m_1}E[Y|a^\ast,m_1,m_2,c]\times \Pr(M_1=m_1|a^\ast,c)\times \Pr(M_2=m_2|a^\ast,m_1,c)\\
& & + \sum_{m_2}\sum_{m_1}E[Y|a^\ast,m_1,m_2^\ast,c]\times \Pr(M_1=m_1|a^\ast,c)\times \Pr(M_2=m_2|a^\ast,m_1,c)\; by\;consistency\\
\\
& = & \int_{m_2}\int_{m_1}{E\left[Y\middle| a,m_1,m_2,c\right]}d\Pr{\left(M_1=m_1\middle| a^\ast,c\right)}d\Pr{\left(M_2=m_2\middle| a^\ast,m_1,c\right)}\\
& & - \int_{m_2}\int_{m_1}{E\left[Y\middle| a,m_1,m_2^\ast,c\right]}d\Pr{\left(M_1=m_1\middle| a^\ast,c\right)}d\Pr{\left(M_2=m_2\middle| a^\ast,m_1,c\right)}\\
& & - \int_{m_2}\int_{m_1}{E\left[Y\middle| a^\ast,m_1,m_2,c\right]}d\Pr{\left(M_1=m_1\middle| a^\ast,c\right)}d\Pr{\left(M_2=m_2\middle| a^\ast,m_1,c\right)}\\
& & + \int_{m_2}\int_{m_1}{E\left[Y\middle| a^\ast,m_1,m_2^\ast,c\right]}d\Pr{\left(M_1=m_1\middle| a^\ast,c\right)}d\Pr{\left(M_2=m_2\middle| a^\ast,m_1,c\right)}\\
\\
& = & \int_{m_1}\int_{m_2}{E\left[Y\middle| a,m_1,m_2,c\right]}d\Pr{\left(M_2=m_2\middle| a^\ast,m_1,c\right)}d\Pr{\left(M_1=m_1\middle| a^\ast,c\right)}\\
& & - \int_{m_1}\int_{m_2}{E\left[Y\middle| a,m_1,m_2^\ast,c\right]}d\Pr{\left(M_2=m_2\middle| a^\ast,m_1,c\right)}d\Pr{\left(M_1=m_1\middle| a^\ast,c\right)}\\
& & - \int_{m_1}\int_{m_2}{E\left[Y\middle| a^\ast,m_1,m_2,c\right]}d\Pr{\left(M_2=m_2\middle| a^\ast,m_1,c\right)}d\Pr{\left(M_1=m_1\middle| a^\ast,c\right)}\\
& & + \int_{m_1}\int_{m_2}{E\left[Y\middle| a^\ast,m_1,m_2^\ast,c\right]}d\Pr{\left(M_2=m_2\middle| a^\ast,m_1,c\right)}d\Pr{\left(M_1=m_1\middle| a^\ast,c\right)}\\
\\
& = & \int_{m_1}\int_{m_2}(\theta_0+\theta_1a+\theta_2m_1+\theta_3m_2+\theta_4am_1+\theta_5am_2\\
& & +\theta_6m_1m_2+\theta_7am_1m_2+\theta_8^\prime c) d\Pr{\left(M_2=m_2\middle| a^\ast,m_1,c\right)}d\Pr{\left(M_1=m_1\middle| a^\ast,c\right)}\\
& & - \int_{m_1}\int_{m_2}(\theta_0+\theta_1a+\theta_2m_1+\theta_3m_2^\ast+\theta_4am_1+\theta_5am_2^\ast\\
& & +\theta_6m_1m_2^\ast+\theta_7am_1m_2^\ast+\theta_8^\prime c)d\Pr{\left(M_2=m_2\middle| a^\ast,m_1,c\right)}d\Pr{\left(M_1=m_1\middle| a^\ast,c\right)}\\
& & - \int_{m_1}\int_{m_2}(\theta_0+\theta_1a^\ast+\theta_2m_1+\theta_3m_2+\theta_4a^\ast m_1+\theta_5a^\ast m_2\\
& & +\theta_6m_1m_2+\theta_7a^\ast m_1m_2+\theta_8^\prime c)d\Pr{\left(M_2=m_2\middle| a^\ast,m_1,c\right)}d\Pr{\left(M_1=m_1\middle| a^\ast,c\right)}\\
& & + \int_{m_1}\int_{m_2} (\theta_0+\theta_1a^\ast+\theta_2m_1+\theta_3m_2^\ast+\theta_4a^\ast m_1+\theta_5a^\ast m_2^\ast\\
& & +\theta_6m_1m_2^\ast+\theta_7a^\ast m_1m_2^\ast+\theta_8^\prime c)d\Pr{\left(M_2=m_2\middle| a^\ast,m_1,c\right)}d\Pr{\left(M_1=m_1\middle| a^\ast,c\right)}\\
\\
& = & \int_{m_1}\left[(\theta_0+\theta_1a+\theta_2m_1+\theta_4am_1+\theta_8^\prime c)\right.\\
& & +\left.\left(\theta_3+\theta_5a+\theta_6m_1+\theta_7am_1\right)\times\left(\beta_0+\beta_1a^\ast+\beta_2m_1+\beta_3a^\ast m_1+\beta_4^\prime c\right)\right]d\Pr{\left(M_1=m_1\middle| a^\ast,c\right)}\\
& & - \int_{m_1} (\theta_0+\theta_1a+\theta_2m_1+\theta_3m_2^\ast+\theta_4am_1+\theta_5am_2^\ast\\
& & +\theta_6m_1m_2^\ast+\theta_7am_1m_2^\ast+\theta_8^\prime c)d\Pr{\left(M_1=m_1\middle| a^\ast,c\right)}\\
& & - \int_{m_1} \left[(\theta_0+\theta_1a^\ast+\theta_2m_1+\theta_4a^\ast m_1+\theta_8^\prime c)\right.\\
& & +\left.\left(\theta_3+\theta_5a^\ast+\theta_6m_1+\theta_7a^\ast m_1\right)\times\left(\beta_0+\beta_1a^\ast+\beta_2m_1+\beta_3a^\ast m_1+\beta_4^\prime c\right)\right]d\Pr{\left(M_1=m_1\middle| a^\ast,c\right)}\\
& & + \int_{m_1} (\theta_0+\theta_1a^\ast+\theta_2m_1+\theta_3m_2^\ast+\theta_4a^\ast m_1+\theta_5a^\ast m_2^\ast\\
& & +\theta_6m_1m_2^\ast+\theta_7a^\ast m_1m_2^\ast+\theta_8^\prime c)d\Pr{\left(M_1=m_1\middle| a^\ast,c\right)}\\
\\
& = & \left(\theta_0+\theta_1a+\theta_8^\prime c\right)+\left(\theta_3+\theta_5a\right)\left(\beta_0+\beta_1a^\ast+\beta_4^\prime c\right)\\
& & +\left(\theta_2+\theta_4a\right)\left(\gamma_0+\gamma_1a^\ast+\gamma_2^\prime c\right)\\
& & +\left(\theta_6+\theta_7a\right)\left(\beta_0+\beta_1a^\ast+\beta_4^\prime c\right)\left(\gamma_0+\gamma_1a^\ast+\gamma_2^\prime c\right)\\
& & +\left(\theta_3+\theta_5a\right)\left(\beta_2+\beta_3a^\ast\right)\left(\gamma_0+\gamma_1a^\ast+\gamma_2^\prime c\right)\\
& & +\left(\theta_6+\theta_7a\right)\left(\beta_2+\beta_3a^\ast\right)\left[\sigma_{M_1}^2+\left(\gamma_0+\gamma_1a^\ast+\gamma_2^\prime c\right)^2\right]\\
& & -\left(\theta_0+\theta_1a+\theta_3m_2^\ast+\theta_5am_2^\ast+\theta_8^\prime c\right)\\
& & -\left(\theta_2+\theta_4a+\theta_6m_2^\ast+\theta_7am_2^\ast\right)\left(\gamma_0+\gamma_1a^\ast+\gamma_2^\prime c\right)\\
& & -\left(\theta_0+\theta_1a^\ast+\theta_8^\prime c\right)-\left(\theta_3+\theta_5a^\ast\right)\left(\beta_0+\beta_1a^\ast+\beta_4^\prime c\right)\\
& & -\left(\theta_2+\theta_4a^\ast\right)\left(\gamma_0+\gamma_1a^\ast+\gamma_2^\prime c\right)\\
& & -\left(\theta_6+\theta_7a^\ast\right)\left(\beta_0+\beta_1a^\ast+\beta_4^\prime c\right)\left(\gamma_0+\gamma_1a^\ast+\gamma_2^\prime c\right)\\
& & -\left(\theta_3+\theta_5a^\ast\right)\left(\beta_2+\beta_3a^\ast\right)\left(\gamma_0+\gamma_1a^\ast+\gamma_2^\prime c\right)\\
& & -\left(\theta_6+\theta_7a^\ast\right)\left(\beta_2+\beta_3a^\ast\right)\left[\sigma_{M_1}^2+\left(\gamma_0+\gamma_1a^\ast+\gamma_2^\prime c\right)^2\right]\\
& & +\left(\theta_0+\theta_1a^\ast+\theta_3m_2^\ast+\theta_5a^\ast m_2^\ast+\theta_8^\prime c\right)\\
& & +\left(\theta_2+\theta_4a^\ast+\theta_6m_2^\ast+\theta_7a^\ast m_2^\ast\right)\left(\gamma_0+\gamma_1a^\ast+\gamma_2^\prime c\right)\\
\\
& = & \theta_1\left(a-a^\ast\right)+\theta_5\left(\beta_0+\beta_1a^\ast+\beta_4^\prime c\right)\left(a-a^\ast\right)\\
& & +\theta_4\left(\gamma_0+\gamma_1a^\ast+\gamma_2^\prime c\right)\left(a-a^\ast\right)\\
& & +\theta_7\left(\beta_0+\beta_1a^\ast+\beta_4^\prime c\right)\left(\gamma_0+\gamma_1a^\ast+\gamma_2^\prime c\right)\left(a-a^\ast\right)\\
& & +\theta_5\left(\beta_2+\beta_3a^\ast\right)\left(\gamma_0+\gamma_1a^\ast+\gamma_2^\prime c\right)\left(a-a^\ast\right)\\
& & +\theta_7\left(\beta_2+\beta_3a^\ast\right)\left[\sigma_{M_1}^2+\left(\gamma_0+\gamma_1a^\ast+\gamma_2^\prime c\right)^2\right]\left(a-a^\ast\right)\\
& & -\left(\theta_1+\theta_5m_2^\ast\right)\left(a-a^\ast\right)-(\theta_4+\theta_7m_2^\ast)\left(\gamma_0+\gamma_1a^\ast+\gamma_2^\prime c\right)(a-a^\ast)\\
\\
& = & \left\{\theta_1+\theta_5\left(\beta_0+\beta_1a^\ast+\beta_4^\prime c\right)\right.\\
& & \left.+\theta_7\left(\beta_0+\beta_1a^\ast+\beta_4^\prime c\right)\left(\gamma_0+\gamma_1a^\ast+\gamma_2^\prime c\right)\right. \\
& & \left.+\theta_5\left(\beta_2+\beta_3a^\ast\right)\left(\gamma_0+\gamma_1a^\ast+\gamma_2^\prime c\right)\right.\\
& & \left.+\theta_7\left(\beta_2+\beta_3a^\ast\right)\left[\sigma_{M_1}^2+\left(\gamma_0+\gamma_1a^\ast+\gamma_2^\prime c\right)^2\right]\right.\\
& & \left. -\left(\theta_1+\theta_5m_2^\ast\right)-\theta_7m_2^\ast\left(\gamma_0+\gamma_1a^\ast+\gamma_2^\prime c\right) \right\}(a-a^\ast).
\end{eqnarray*}
\\
\subsection*{Natural counterfactual interaction effects}
We derive the the expected value of each interaction effect. 

\begin{eqnarray*}
Y(a,M_1(a),M_2(a,M_1(a))) = \sum_{m_2}\sum_{m_1}Y(a,m_1,m_2)\times I(M_1(a)=m_1)\times I(M_2(a,m_1)=m_2)
\end{eqnarray*}
\begin{eqnarray*}
&\Rightarrow& E[Y(a,M_1(a),M_2(a,M_1(a)))|c]\\
\\
& = & E\left[\sum_{m_2}\sum_{m_1}Y(a,m_1,m_2)\times I(M_1(a)=m_1)\times I(M_2(a,m_1)=m_2) \bigg|c\right]\\
\\
& = & \sum_{m_2}\sum_{m_1}{E\left[Y\left(a,m_1,m_2\right)\times I\left(M_1\left(a\right)=m_1\right)\times I\left(M_2\left(a,m_1\right)=m_2\right)\middle| c\right]}\\
\\
& = & \sum_{m_2}\sum_{m_1}{E\left[Y\left(a,m_1,m_2\right)\middle| c\right]E\left[I\left(M_1\left(a\right)=m_1\right)\middle| c\right]E\left[I\left(M_2\left(a,m_1\right)=m_2\right)\middle| c\right]}\; by\; A4\;A6\\
\\
& = & \sum_{m_2}\sum_{m_1}{E\left[Y\left(a,m_1,m_2\right)\middle| c\right]\Pr{\left(M_1\left(a\right)=m_1\middle| c\right)}\Pr{\left(M_2\left(a,m_1\right)=m_2\middle| c\right)}}\\
\\
& = & \sum_{m_2}\sum_{m_1}{E\left[Y\left(a,m_1,m_2\right)\middle| c\right]\Pr{\left(M_1\left(a\right)=m_1\middle| a,c\right)}\Pr{\left(M_2\left(a,m_1\right)=m_2\middle| a,m_1,c\right)}}\; by\; A3\;A5\\
\\
& = & \sum_{m_2}\sum_{m_1}{E\left[Y\left(a,m_1,m_2\right)\middle| c\right]\Pr{\left(M_1=m_1\middle| a,c\right)}\Pr{\left(M_2=m_2\middle| a,m_1,c\right)}}\; by\; consistency\\
\\
& = & \sum_{m_2}\sum_{m_1}{E\left[Y\left(a,m_1,m_2\right)\middle| a,m_1,m_2,c\right]\Pr{\left(M_1=m_1\middle| a,c\right)}\Pr{\left(M_2=m_2\middle| a,m_1,c\right)}}\; by\; A1\;A2\\
\\
& = & \sum_{m_2}\sum_{m_1}{E\left[Y\middle| a,m_1,m_2,c\right]\Pr{\left(M_1=m_1\middle| a,c\right)}\Pr{\left(M_2=m_2\middle| a,m_1,c\right)}}\; by\; consistency\\
\\
& = & \int_{m_2}\int_{m_1}{E\left[Y\middle| a,m_1,m_2,c\right]d}\Pr{\left(M_1=m_1\middle| a,c\right)}d\Pr{\left(M_2=m_2\middle| a,m_1,c\right)}\\
\\
& = & \int_{m_2}\int_{m_1}(\theta_0+\theta_1a+\theta_2m_1+\theta_3m_2+\theta_4am_1+\theta_5am_2\\
& & +\theta_6m_1m_2+\theta_7am_1m_2+\theta_8^\prime c)d\Pr{\left(M_1=m_1\middle| a,c\right)}d\Pr{\left(M_2=m_2\middle| a,m_1,c\right)}\\
\\
& = & \int_{m_1}\int_{m_2}(\theta_0+\theta_1a+\theta_2m_1+\theta_3m_2+\theta_4am_1+\theta_5am_2\\
& & +\theta_6m_1m_2+\theta_7am_1m_2+\theta_8^\prime c)d\Pr{\left(M_2=m_2\middle| a,m_1,c\right)}d\Pr{\left(M_1=m_1\middle| a,c\right)}\\
\\
& = & \int_{m_1}\int_{m_2} [(\theta_0+\theta_1a+\theta_2m_1+\theta_4am_1+\theta_8^\prime c)\\
& & +\left(\theta_3+\theta_5a+\theta_6m_1+\theta_7am_1\right)m_2]d\Pr{\left(M_2=m_2\middle| a,m_1,c\right)}d\Pr{\left(M_1=m_1\middle| a,c\right)}\\
\\
& = & \int_{m_1} \left[\left(\theta_0+\theta_1a+\theta_2m_1+\theta_4am_1+\theta_8^\prime c\right)\right.\\
& & +\left.\left(\theta_3+\theta_5a+\theta_6m_1+\theta_7am_1\right)\left(\beta_0+\beta_1a+\beta_2m_1+\beta_3am_1+\beta_4^\prime c\right)\right]d\Pr\left(M_1=m_1\middle| a,c\right)\\
\\
& = & \int_{m_1} \left[\left(\theta_0+\theta_1a+\theta_8^\prime c\right)+\left(\theta_2+\theta_4a\right)m_1\right.\\
& & +\left.\left(\theta_3+\theta_5a+\left(\theta_6+\theta_7a\right)m_1\right)\left(\beta_0+\beta_1a+\beta_4^\prime c+\left(\beta_2+\beta_3a\right)m_1\right)\right]d\Pr\left(M_1=m_1\middle| a,c\right)\\
\\
& = & \left(\theta_0+\theta_1a+\theta_8^\prime c\right)+\left(\theta_3+\theta_5a\right)\left(\beta_0+\beta_1a+\beta_4^\prime c\right)\\
& & +\left(\theta_2+\theta_4a\right)\left(\gamma_0+\gamma_1a+\gamma_2^\prime c\right)+\left(\theta_6+\theta_7a\right)\left(\beta_0+\beta_1a+\beta_4^\prime c\right)\left(\gamma_0+\gamma_1a+\gamma_2^\prime c\right)\\
& & +\left(\theta_3+\theta_5a\right)\left(\beta_2+\beta_3a\right)\left(\gamma_0+\gamma_1a+\gamma_2^\prime c\right)\\
& & +\left(\theta_6+\theta_7a\right)\left(\beta_2+\beta_3a\right)\left[\sigma_{M_1}^2+\left(\gamma_0+\gamma_1a+\gamma_2^\prime c\right)^2\right].\qquad\qquad(W1)
\end{eqnarray*}

Similarly, we can obtain the following expected values for the rest of the counterfactual formulas. 
\begin{eqnarray*}
& & E[Y(a,M_1(a),M_2(a^\ast,M_1(a)))|c]\\
\\
& = & \left(\theta_0+\theta_1a+\theta_8^\prime c\right)+\left(\theta_3+\theta_5a\right)\left(\beta_0+\beta_1a^\ast+\beta_4^\prime c\right)\\
& & +\left(\theta_2+\theta_4a\right)\left(\gamma_0+\gamma_1a+\gamma_2^\prime c\right)+\left(\theta_6+\theta_7a\right)\left(\beta_0+\beta_1a^\ast+\beta_4^\prime c\right)\left(\gamma_0+\gamma_1a+\gamma_2^\prime c\right)\\
& & +\left(\theta_3+\theta_5a\right)\left(\beta_2+\beta_3a^\ast\right)\left(\gamma_0+\gamma_1a+\gamma_2^\prime c\right)\\
& & +\left(\theta_6+\theta_7a\right)\left(\beta_2+\beta_3a^\ast\right)\left[\sigma_{M_1}^2+\left(\gamma_0+\gamma_1a+\gamma_2^\prime c\right)^2\right]\qquad\qquad(W2)
\end{eqnarray*}

\begin{eqnarray*}
& & E[Y(a,M_1(a^\ast),M_2(a,M_1(a^\ast)))|c]\\
\\
& = & \left(\theta_0+\theta_1a+\theta_8^\prime c\right)+\left(\theta_3+\theta_5a\right)\left(\beta_0+\beta_1a+\beta_4^\prime c\right)\\
& & +\left(\theta_2+\theta_4a\right)\left(\gamma_0+\gamma_1a^\ast+\gamma_2^\prime c\right)+\left(\theta_6+\theta_7a\right)\left(\beta_0+\beta_1a+\beta_4^\prime c\right)\left(\gamma_0+\gamma_1a^\ast+\gamma_2^\prime c\right)\\
& & +\left(\theta_3+\theta_5a\right)\left(\beta_2+\beta_3a\right)\left(\gamma_0+\gamma_1a^\ast+\gamma_2^\prime c\right)\\
& & +\left(\theta_6+\theta_7a\right)\left(\beta_2+\beta_3a\right)\left[\sigma_{M_1}^2+\left(\gamma_0+\gamma_1a^\ast+\gamma_2^\prime c\right)^2\right]\qquad\qquad(W3)
\end{eqnarray*}

\begin{eqnarray*}
& & E[Y(a^\ast,M_1(a),M_2(a,M_1(a)))|c]\\
\\
& = & \left(\theta_0+\theta_1a^\ast+\theta_8^\prime c\right)+\left(\theta_3+\theta_5a^\ast\right)\left(\beta_0+\beta_1a+\beta_4^\prime c\right)\\
& & +\left(\theta_2+\theta_4a^\ast\right)\left(\gamma_0+\gamma_1a+\gamma_2^\prime c\right)+\left(\theta_6+\theta_7a^\ast\right)\left(\beta_0+\beta_1a+\beta_4^\prime c\right)\left(\gamma_0+\gamma_1a+\gamma_2^\prime c\right)\\
& & +\left(\theta_3+\theta_5a^\ast\right)\left(\beta_2+\beta_3a\right)\left(\gamma_0+\gamma_1a+\gamma_2^\prime c\right)\\
& & +\left(\theta_6+\theta_7a^\ast\right)\left(\beta_2+\beta_3a\right)\left[\sigma_{M_1}^2+\left(\gamma_0+\gamma_1a+\gamma_2^\prime c\right)^2\right]\qquad\qquad(W4)
\end{eqnarray*}

\begin{eqnarray*}
& & E[Y(a^\ast,M_1(a^\ast),M_2(a,M_1(a^\ast)))|c]\\
\\
& = & \left(\theta_0+\theta_1a^\ast+\theta_8^\prime c\right)+\left(\theta_3+\theta_5a^\ast\right)\left(\beta_0+\beta_1a+\beta_4^\prime c\right)\\
& & +\left(\theta_2+\theta_4a^\ast\right)\left(\gamma_0+\gamma_1a^\ast+\gamma_2^\prime c\right)\\
& & +\left(\theta_6+\theta_7a^\ast\right)\left(\beta_0+\beta_1a+\beta_4^\prime c\right)\left(\gamma_0+\gamma_1a^\ast+\gamma_2^\prime c\right)\\
& & +\left(\theta_3+\theta_5a^\ast\right)\left(\beta_2+\beta_3a\right)\left(\gamma_0+\gamma_1a^\ast+\gamma_2^\prime c\right)\\
& & +\left(\theta_6+\theta_7a^\ast\right)\left(\beta_2+\beta_3a\right)\left[\sigma_{M_1}^2+\left(\gamma_0+\gamma_1a^\ast+\gamma_2^\prime c\right)^2\right]\qquad\qquad(W5)
\end{eqnarray*}

\begin{eqnarray*}
& & E[Y(a^\ast,M_1(a),M_2(a^\ast,M_1(a)))|c]\\
\\
& = & \left(\theta_0+\theta_1a^\ast+\theta_8^\prime c\right)+\left(\theta_3+\theta_5a^\ast\right)\left(\beta_0+\beta_1a^\ast+\beta_4^\prime c\right)\\
& & +\left(\theta_2+\theta_4a^\ast\right)\left(\gamma_0+\gamma_1a+\gamma_2^\prime c\right)\\
& & +\left(\theta_6+\theta_7a^\ast\right)\left(\beta_0+\beta_1a^\ast+\beta_4^\prime c\right)\left(\gamma_0+\gamma_1a+\gamma_2^\prime c\right)\\
& & +\left(\theta_3+\theta_5a^\ast\right)\left(\beta_2+\beta_3a^\ast\right)\left(\gamma_0+\gamma_1a+\gamma_2^\prime c\right)\\
& & +\left(\theta_6+\theta_7a^\ast\right)\left(\beta_2+\beta_3a^\ast\right)\left[\sigma_{M_1}^2+\left(\gamma_0+\gamma_1a+\gamma_2^\prime c\right)^2\right]\qquad\qquad(W6)
\end{eqnarray*}

\begin{eqnarray*}
& & E[Y(a,M_1(a^\ast),M_2(a^\ast,M_1(a^\ast)))|c]\\
\\
& = & \left(\theta_0+\theta_1a+\theta_8^\prime c\right)+\left(\theta_3+\theta_5a\right)\left(\beta_0+\beta_1a^\ast+\beta_4^\prime c\right)\\
& & +\left(\theta_2+\theta_4a\right)\left(\gamma_0+\gamma_1a^\ast+\gamma_2^\prime c\right)\\
& & +\left(\theta_6+\theta_7a\right)\left(\beta_0+\beta_1a^\ast+\beta_4^\prime c\right)\left(\gamma_0+\gamma_1a^\ast+\gamma_2^\prime c\right)\\
& & +\left(\theta_3+\theta_5a\right)\left(\beta_2+\beta_3a^\ast\right)\left(\gamma_0+\gamma_1a^\ast+\gamma_2^\prime c\right)\\
& & +\left(\theta_6+\theta_7a\right)\left(\beta_2+\beta_3a^\ast\right)\left[\sigma_{M_1}^2+\left(\gamma_0+\gamma_1a^\ast+\gamma_2^\prime c\right)^2\right]\qquad\qquad(W7)
\end{eqnarray*}

\begin{eqnarray*}
& & E[Y(a^\ast,M_1(a^\ast),M_2(a^\ast,M_1(a^\ast)))|c]\\
\\
& = & \left(\theta_0+\theta_1a^\ast+\theta_8^\prime c\right)+\left(\theta_3+\theta_5a^\ast\right)\left(\beta_0+\beta_1a^\ast+\beta_4^\prime c\right)\\
& & +\left(\theta_2+\theta_4a^\ast\right)\left(\gamma_0+\gamma_1a^\ast+\gamma_2^\prime c\right)\\
& & +\left(\theta_6+\theta_7a^\ast\right)\left(\beta_0+\beta_1a^\ast+\beta_4^\prime c\right)\left(\gamma_0+\gamma_1a^\ast+\gamma_2^\prime c\right)\\
& & +\left(\theta_3+\theta_5a^\ast\right)\left(\beta_2+\beta_3a^\ast\right)\left(\gamma_0+\gamma_1a^\ast+\gamma_2^\prime c\right)\\
& & +\left(\theta_6+\theta_7a^\ast\right)\left(\beta_2+\beta_3a^\ast\right)\left[\sigma_{M_1}^2+\left(\gamma_0+\gamma_1a^\ast+\gamma_2^\prime c\right)^2\right].\qquad\qquad(W8)
\end{eqnarray*}

The formulas of natural counterfactual interaction effects can be obtained as follows:
\begin{eqnarray*}
& & E[NatINT_{AM_1}|c]\\
\\
& = & (W2)-(W6)-(W7)+(W8)\\
\\
& = & \left[\theta_4\gamma_1+\theta_7\gamma_1\left(\beta_0+\beta_1a^\ast+\beta_4^\prime c\right)+\theta_5\gamma_1\left(\beta_2+\beta_3a^\ast\right)\right.\\
& & +2\theta_7\gamma_1\left(\beta_2+\beta_3a^\ast\right)\left(\gamma_0+\gamma_2^\prime c\right)\\
& & +\left.\theta_7\gamma_1^2\left(\beta_2+\beta_3a^\ast\right)\left(a+a^\ast\right)\right](a-a^\ast)^2
\end{eqnarray*}

\begin{eqnarray*}
& & E[NatINT_{AM_2}|c]\\
\\
& = & (W3)-(W5)-(W7)+(W8)\\
\\
& = & \left[\theta_5\beta_1+\theta_7\beta_1\left(\gamma_0+\gamma_1a^\ast+\gamma_2^\prime c\right)+\theta_5\beta_3\left(\gamma_0+\gamma_1a^\ast+\gamma_2^\prime c\right)\right.\\
& & \left.+\theta_7\beta_3\left[\sigma_{M_1}^2+\left(\gamma_0+\gamma_1a^\ast+\gamma_2^\prime c\right)^2\right] \right](a-a^\ast)^2
\end{eqnarray*}

\begin{eqnarray*}
& & E[NatINT_{AM_1M_2}|c]\\
\\
& = & (W1)-(W4)-(W3)+(W5)-(W2)+(W6)+(W7)-(W8)\\
\\
& = & \left[\theta_7\beta_1\gamma_1+\theta_5\beta_3\gamma_1+2\theta_7\beta_3\gamma_1\left(\gamma_0+\gamma_2^\prime c\right)+\theta_7\beta_3\gamma_1^2\left(a+a^\ast\right) \right](a-a^\ast)^3
\end{eqnarray*}

\begin{eqnarray*}
& & E[NatINT_{M_1M_2}|c]\\
\\
& = & (W4)-(W5)-(W6)+(W8)\\
\\
& = & \left[\beta_1\gamma_1\left(\theta_6+\theta_7a^\ast\right)+\beta_3\gamma_1\left(\theta_3+\theta_5a^\ast\right)\right.\\
& & +2\beta_3\gamma_1\left(\theta_6+\theta_7a^\ast\right)\left(\gamma_0+\gamma_2^\prime c\right)\\
& & \left. +\beta_3\gamma_1^2\left(\theta_6+\theta_7a^\ast\right)\left(a+a^\ast\right)\right](a-a^\ast)^2.
\end{eqnarray*}
\\
\subsection*{Pure indirect effects}
The pure indirect effect through $M_1$ can be obtained by the following derivation:
\begin{eqnarray*}
PIE_{M_1} = Y(a^\ast,M_1(a),M_2(a^\ast,M_1(a)))-Y(a^\ast,M_1(a^\ast),M_2(a^\ast,M_1(a^\ast)))
\end{eqnarray*}
\begin{eqnarray*}
&\Rightarrow& E[PIE_{M_1}|c]\\
\\
& = & E[Y(a^\ast,M_1(a),M_2(a^\ast,M_1(a)))-Y(a^\ast,M_1(a^\ast),M_2(a^\ast,M_1(a^\ast)))|c]\\
\\
& = & E[Y(a^\ast,M_1(a),M_2(a^\ast,M_1(a)))|c]-E[Y(a^\ast,M_1(a^\ast),M_2(a^\ast,M_1(a^\ast)))|c]\\
\\
& = & (W6)-(W8)
\\
& = & \left[\gamma_1\left(\theta_2+\theta_4a^\ast\right)+\gamma_1\left(\theta_6+\theta_7a^\ast\right)\left(\beta_0+\beta_1a^\ast+\beta_4^\prime c\right)\right.\\
& & +\gamma_1\left(\theta_3+\theta_5a^\ast\right)\left(\beta_2+\beta_3a^\ast\right)\\
& & +2\gamma_1\left(\theta_6+\theta_7a^\ast\right)\left(\beta_2+\beta_3a^\ast\right)\left(\gamma_0+\gamma_2^\prime c\right)\\
& & \left.+\gamma_1^2\left(\theta_6+\theta_7a^\ast\right)\left(\beta_2+\beta_3a^\ast\right)\left(a+a^\ast\right)\right](a-a^\ast).
\end{eqnarray*}

Similarly, the pure indirect effect through $M_2$ can be obtained by the following derivation:
\begin{eqnarray*}
PIE_{M_2} = Y(a^\ast,M_1(a^\ast),M_2(a,M_1(a^\ast)))-Y(a^\ast,M_1(a^\ast),M_2(a^\ast,M_1(a^\ast)))
\end{eqnarray*}
\begin{eqnarray*}
&\Rightarrow& E[PIE_{M_2}|c]\\
\\
& = & E[Y(a^\ast,M_1(a^\ast),M_2(a,M_1(a^\ast)))-Y(a^\ast,M_1(a^\ast),M_2(a^\ast,M_1(a^\ast)))|c]\\
\\
& = & E[Y(a^\ast,M_1(a^\ast),M_2(a,M_1(a^\ast)))|c]-E[Y(a^\ast,M_1(a^\ast),M_2(a^\ast,M_1(a^\ast)))|c]\\
\\
& = & (W5)-(W8)
\\
& = & \left[\beta_1\left(\theta_3+\theta_5a^\ast\right)+\beta_1\left(\theta_6+\theta_7a^\ast\right)\left(\gamma_0+\gamma_1a^\ast+\gamma_2^\prime c\right)\right.\\
& & +\beta_3\left(\theta_3+\theta_5a^\ast\right)\left(\gamma_0+\gamma_1a^\ast+\gamma_2^\prime c\right)\\
& & \left.+\beta_3\left(\theta_6+\theta_7a^\ast\right)\left[\sigma_{M_1}^2+\left(\gamma_0+\gamma_1a^\ast+\gamma_2^\prime c\right)^2\right]\right](a-a^\ast).
\end{eqnarray*}
\\
\subsection*{Total effect}
\begin{eqnarray*}
TE = Y(a,M_1(a),M_2(a,M_1(a)))-Y(a^\ast,M_1(a^\ast),M_2(a^\ast,M_1(a^\ast)))
\end{eqnarray*}
\begin{eqnarray*}
&\Rightarrow& E[TE|c]\\
\\
& = & E[Y(a,M_1(a),M_2(a,M_1(a)))-Y(a^\ast,M_1(a^\ast),M_2(a^\ast,M_1(a^\ast)))|c]\\
\\
& = & E[Y(a,M_1(a),M_2(a,M_1(a)))|c]-E[Y(a^\ast,M_1(a^\ast),M_2(a^\ast,M_1(a^\ast)))|c]\\
\\
& = & (W1)-(W8)
\\
& = & \left[\theta_1+\theta_5\left(\beta_0+\beta_4^\prime c\right)+\beta_1\theta_3+\theta_4\left(\gamma_0+\gamma_2^\prime c\right)+\gamma_1\theta_2\right.\\
& & +\theta_7\left(\beta_0+\beta_4^\prime c\right)\left(\gamma_0+\gamma_2^\prime c\right)+\beta_1\theta_6\left(\gamma_0+\gamma_2^\prime c\right)\\
& & +\gamma_1\theta_6\left(\beta_0+\beta_4^\prime c\right)+\theta_5\beta_2\left(\gamma_0+\gamma_2^\prime\right)+\theta_3\beta_3\left(\gamma_0+\gamma_2^\prime c\right)\\
& & +\theta_3\beta_2\gamma_1+\theta_7\beta_2\sigma_{M_1}^2+\theta_6\beta_3\sigma_{M_1}^2+\theta_7\beta_2\left(\gamma_0+\gamma_2^\prime c\right)^2\\
& & \left. +\theta_6\beta_3\left(\gamma_0+\gamma_2^\prime c\right)^2+2\gamma_1\theta_6\beta_2\left(\gamma_0+\gamma_2^\prime c\right)\right](a-a^\ast)\\
\\
& & +\left[\beta_1\theta_5+\gamma_1\theta_4+\beta_1\theta_7\left(\gamma_0+\gamma_2^\prime c\right)\right.\\
& & +\gamma_1\theta_7\left(\beta_0+\beta_4^\prime c\right)+\gamma_1\beta_1\theta_6+\theta_5\beta_3\left(\gamma_0+\gamma_2^\prime c\right)\\
& & +\theta_5\beta_2\gamma_1+\theta_3\beta_3\gamma_1+\theta_7\beta_3\sigma_{M_1}^2+\theta_7\beta_3\left(\gamma_0+\gamma_2^\prime c\right)^2\\
& & \left.+2\gamma_1\theta_7\beta_2\left(\gamma_0+\gamma_2^\prime c\right)+2\gamma_1\theta_6\beta_3\left(\gamma_0+\gamma_2^\prime c\right)+\theta_6\beta_2\gamma_1^2\right]\left(a^2-{a^\ast}^2\right)\\
\\
& & + \left[\gamma_1\beta_1\theta_7+\theta_5\beta_3\gamma_1+2\gamma_1\theta_7\beta_3\left(\gamma_0+\gamma_2^\prime c\right)+\theta_7\beta_2\gamma_1^2+\theta_6\beta_3\gamma_1^2\right]\left(a^3-{a^\ast}^3\right)\\
\\
& & + \theta_7\beta_3\gamma_1^2\left(a^4-{a^\ast}^4\right).
\end{eqnarray*}
\\
\subsection*{Summary of Results}
In this subsection, the results of all components are listed below for a quick reference for the readers. 
\begin{eqnarray*}
 E[CDE(m_1^\ast,m_2^\ast)|c]
 & = & \left(\theta_1+\theta_4m_1^\ast+\theta_5m_2^\ast+\theta_7m_1^\ast m_2^\ast\right)\left(a-a^\ast\right)\\
 \\
 E[INT_{ref\mbox{-}AM_1}(m_1^\ast,m_2^\ast)|c] 
& = & \left(\gamma_0+\gamma_1a^\ast+\gamma_2^\prime c-m_1^\ast\right)\times\left(\theta_4+\theta_7m_2^\ast\right)\times\left(a-a^\ast\right)\\
\\
 E[INT_{ref\mbox{-}AM_2+AM_1M_2}(m_2^\ast)|c]
& = & \left\{\theta_1+\theta_5\left(\beta_0+\beta_1a^\ast+\beta_4^\prime c\right)\right.\\
& & \left.+\theta_7\left(\beta_0+\beta_1a^\ast+\beta_4^\prime c\right)\left(\gamma_0+\gamma_1a^\ast+\gamma_2^\prime c\right)\right. \\
& & \left.+\theta_5\left(\beta_2+\beta_3a^\ast\right)\left(\gamma_0+\gamma_1a^\ast+\gamma_2^\prime c\right)\right.\\
& & \left.+\theta_7\left(\beta_2+\beta_3a^\ast\right)\left[\sigma_{M_1}^2+\left(\gamma_0+\gamma_1a^\ast+\gamma_2^\prime c\right)^2\right]\right.\\
& & \left. -\left(\theta_1+\theta_5m_2^\ast\right)-\theta_7m_2^\ast\left(\gamma_0+\gamma_1a^\ast+\gamma_2^\prime c\right) \right\}(a-a^\ast)\\
\\
 E[NatINT_{AM_1}|c]
& = & \left[\theta_4\gamma_1+\theta_7\gamma_1\left(\beta_0+\beta_1a^\ast+\beta_4^\prime c\right)+\theta_5\gamma_1\left(\beta_2+\beta_3a^\ast\right)\right.\\
& & +2\theta_7\gamma_1\left(\beta_2+\beta_3a^\ast\right)\left(\gamma_0+\gamma_2^\prime c\right)\\
& & +\left.\theta_7\gamma_1^2\left(\beta_2+\beta_3a^\ast\right)\left(a+a^\ast\right)\right](a-a^\ast)^2\\
\\
 E[NatINT_{AM_2}|c]
& = & \left[\theta_5\beta_1+\theta_7\beta_1\left(\gamma_0+\gamma_1a^\ast+\gamma_2^\prime c\right)+\theta_5\beta_3\left(\gamma_0+\gamma_1a^\ast+\gamma_2^\prime c\right)\right.\\
& & \left.+\theta_7\beta_3\left[\sigma_{M_1}^2+\left(\gamma_0+\gamma_1a^\ast+\gamma_2^\prime c\right)^2\right] \right](a-a^\ast)^2\\
\\
 E[NatINT_{AM_1M_2}|c]
& = & \left[\theta_7\beta_1\gamma_1+\theta_5\beta_3\gamma_1+2\theta_7\beta_3\gamma_1\left(\gamma_0+\gamma_2^\prime c\right)+\theta_7\beta_3\gamma_1^2\left(a+a^\ast\right) \right](a-a^\ast)^3\\
\\
 E[NatINT_{M_1M_2}|c]
& = & \left[\beta_1\gamma_1\left(\theta_6+\theta_7a^\ast\right)+\beta_3\gamma_1\left(\theta_3+\theta_5a^\ast\right)\right.\\
& & +2\beta_3\gamma_1\left(\theta_6+\theta_7a^\ast\right)\left(\gamma_0+\gamma_2^\prime c\right)\\
& & \left. +\beta_3\gamma_1^2\left(\theta_6+\theta_7a^\ast\right)\left(a+a^\ast\right)\right](a-a^\ast)^2\\
\\
 E[PIE_{M_1}|c]
& = & \left[\gamma_1\left(\theta_2+\theta_4a^\ast\right)+\gamma_1\left(\theta_6+\theta_7a^\ast\right)\left(\beta_0+\beta_1a^\ast+\beta_4^\prime c\right)\right.\\
& & +\gamma_1\left(\theta_3+\theta_5a^\ast\right)\left(\beta_2+\beta_3a^\ast\right)\\
& & +2\gamma_1\left(\theta_6+\theta_7a^\ast\right)\left(\beta_2+\beta_3a^\ast\right)\left(\gamma_0+\gamma_2^\prime c\right)\\
& & \left.+\gamma_1^2\left(\theta_6+\theta_7a^\ast\right)\left(\beta_2+\beta_3a^\ast\right)\left(a+a^\ast\right)\right](a-a^\ast)\\
\\
 E[PIE_{M_2}|c]
& = & \left[\beta_1\left(\theta_3+\theta_5a^\ast\right)+\beta_1\left(\theta_6+\theta_7a^\ast\right)\left(\gamma_0+\gamma_1a^\ast+\gamma_2^\prime c\right)\right.\\
& & +\beta_3\left(\theta_3+\theta_5a^\ast\right)\left(\gamma_0+\gamma_1a^\ast+\gamma_2^\prime c\right)\\
& & \left.+\beta_3\left(\theta_6+\theta_7a^\ast\right)\left[\sigma_{M_1}^2+\left(\gamma_0+\gamma_1a^\ast+\gamma_2^\prime c\right)^2\right]\right](a-a^\ast)\\
\\
 E[TE|c]
& = & \left[\theta_1+\theta_5\left(\beta_0+\beta_4^\prime c\right)+\beta_1\theta_3+\theta_4\left(\gamma_0+\gamma_2^\prime c\right)+\gamma_1\theta_2\right.\\
& & +\theta_7\left(\beta_0+\beta_4^\prime c\right)\left(\gamma_0+\gamma_2^\prime c\right)+\beta_1\theta_6\left(\gamma_0+\gamma_2^\prime c\right)\\
& & +\gamma_1\theta_6\left(\beta_0+\beta_4^\prime c\right)+\theta_5\beta_2\left(\gamma_0+\gamma_2^\prime\right)+\theta_3\beta_3\left(\gamma_0+\gamma_2^\prime c\right)\\
& & +\theta_3\beta_2\gamma_1+\theta_7\beta_2\sigma_{M_1}^2+\theta_6\beta_3\sigma_{M_1}^2+\theta_7\beta_2\left(\gamma_0+\gamma_2^\prime c\right)^2\\
& & \left. +\theta_6\beta_3\left(\gamma_0+\gamma_2^\prime c\right)^2+2\gamma_1\theta_6\beta_2\left(\gamma_0+\gamma_2^\prime c\right)\right](a-a^\ast)\\
\\
& & +\left[\beta_1\theta_5+\gamma_1\theta_4+\beta_1\theta_7\left(\gamma_0+\gamma_2^\prime c\right)\right.\\
& & +\gamma_1\theta_7\left(\beta_0+\beta_4^\prime c\right)+\gamma_1\beta_1\theta_6+\theta_5\beta_3\left(\gamma_0+\gamma_2^\prime c\right)\\
& & +\theta_5\beta_2\gamma_1+\theta_3\beta_3\gamma_1+\theta_7\beta_3\sigma_{M_1}^2+\theta_7\beta_3\left(\gamma_0+\gamma_2^\prime c\right)^2\\
& & \left.+2\gamma_1\theta_7\beta_2\left(\gamma_0+\gamma_2^\prime c\right)+2\gamma_1\theta_6\beta_3\left(\gamma_0+\gamma_2^\prime c\right)+\theta_6\beta_2\gamma_1^2\right]\left(a^2-{a^\ast}^2\right)\\
\\
& & + \left[\gamma_1\beta_1\theta_7+\theta_5\beta_3\gamma_1+2\gamma_1\theta_7\beta_3\left(\gamma_0+\gamma_2^\prime c\right)+\theta_7\beta_2\gamma_1^2+\theta_6\beta_3\gamma_1^2\right]\left(a^3-{a^\ast}^3\right)\\
\\
& & + \theta_7\beta_3\gamma_1^2\left(a^4-{a^\ast}^4\right).
\end{eqnarray*}


\begin{thebibliography}{99}

\bibitem{vmul} VanderWeele TJ, Vansteelandt S. Mediation analysis with multiple mediators. Epidemiol Methods. 2014;2:95-115.
\\
\bibitem{vpre} VanderWeele TJ, Vansteelandt S, Robins JM. Effect decomposition in the presence of an exposure-induced mediator-outcome confounder. Epidemiology. 2014;25:300-306.
\\
\bibitem{d} Daniel RM, De Stavola BL, Cousens SN, et al. Causal mediation analysis with multiple mediators. Biometrics. 2015;71:1-14.
\\
\bibitem{s} Steen J, Loeys T, Moerkerke B, et al. Flexible mediation analysis with multiple mediators. Am J Epidemiol. 2017;182:184-193. 
\\
\bibitem{m} Mittinty MN, Lynch JW, Forbes AB, et al. Effect decomposition through multiple causally nonordered mediators in the presence of exposure-induced mediator-outcome confounding. Stat Med. 2019;38:5085-5102. 
\\
\bibitem{v3} VanderWeele TJ. A three-way decomposition of a total effect into direct, indirect, and interactive effects. Epidemiology. 2013;24:224-232.
\\
\bibitem{v4} VanderWeele TJ. A unification of mediation and interaction: a 4-way decomposition. Epidemiology. 2014;25:749-761.
\\
\bibitem{vbook} VanderWeele TJ. Explanation in Causal Inference: Methods for Mediation and Interaction. New York: Oxford University Press; 2015.
\\
\bibitem{b} Bellavia A, Valeri L. Decomposition of the total effect in the presence of multiple mediators and interactions. Am J Epidemiol. 2018;187:1311-1318. 
\\
\bibitem{Rothman1} Rothman KJ. Modern Epidemiology. 1st ed. Boston, Mass: Little, Brown and Company; 1986.
\\
\bibitem{Rothman2} Rothman KJ, Greenland S, Lash TL. Concepts of interaction. Chapter 5, 3rd ed. In: Modern Epidemiology. Philadelphia, PA: Lippincott Williams and Wilkins; 2008:71-84.
\\
\bibitem{Hosmer}  Hosmer DW, Lemeshow S. Confidence interval estimation of interaction. Epidemiology. 1992; 3:452-56.
\\
\bibitem{riden} Robins JM, Greenland S. Identifiability and exchangeability for direct and indirect effects. Epidemiology. 1992;3:143-155.
\\
\bibitem{p01} Pearl J. Direct and indirect effects. In: Proceedings of the Seventeenth Conference on Uncertainty in Artificial Intelligence. San Francisco, CA: Morgan Kaufmann Publishers Inc; 2001.p.411-420.
\\
\bibitem{rsem} Robins JM. Semantics of causal DAG models and the identification of direct and indirect effects. In: Green JP, Hjort NL, Richardson S, eds. Highly Structured Stochastic Systems. New York: Oxford University Press; 2003:70-81. 
\\
\bibitem{a} Avin C, Shpitser I, Pearl J. Identifiability of path-specific effects. In: Proceedings of the International Joint Conferences on Artificial Intelligence. Edinburgh, Schotland:2005.p.357-363.  
\\
\bibitem{vcon} VanderWeele TJ, Vansteelandt S. Conceptual issues concerning mediation, interventions and composition. Statistics and Its Interface. 2009;2:457-468. 
\\
\bibitem{ralt} Robins JM, Richardson TS. Alternative graphical causal models and the identification of direct effects. In: Shrout P, eds. Causality and Psychopathology: Finding the Determinants of Disorders and Their Cures. New York: Oxford University Press; 2010.
\\
\bibitem{l} Leon DA, Saburova L, Tomkins S, et al. Hazardous alcohol drinking and premature mortality in Russia: a population based case-control study. Lancet. 2007;369:2001-2009. 
\\
\bibitem{v} Valeri L, VanderWeele TJ. Mediation analysis allowing for exposure-mediator interactions and causal interpretation: theoretical assumptions and implementation with SAS and SPSS macros. Psychol Methods. 2013;18:137-150. 
\\
\bibitem{p14} Pearl J. Interpretation and identification of causal mediation. Psychol Methods. 2014;19:459-481. 
\end{thebibliography}
\end{document}